\PassOptionsToPackage{numbers,sort&compress}{natbib}
\documentclass[review,11pt]{elsarticle}

\usepackage{lineno,hyperref}
\usepackage{bm}
\usepackage{algorithm}
\usepackage{algorithmicx}
\usepackage[noend]{algpseudocode}
\let\Algorithm\algorithm
\usepackage{caption}
\usepackage{hyperref}
\usepackage[utf8]{inputenc}
\usepackage{lscape}
\usepackage{tikz}
\usepackage{float}
\usepackage{graphicx}
\usepackage{listings}%
\usepackage{amsmath,amsfonts}
\usepackage{wrapfig}
\usepackage{amssymb}
\usepackage{alltt}%
\usepackage{siunitx}
\usepackage{setspace}
\usepackage{booktabs}
\usepackage{eqparbox}
\usepackage{mathtools}
\usepackage{chemformula}
\usepackage[margin=2cm]{geometry}
\renewcommand\algorithm[1][]{\Algorithm[#1]\setstretch{1.4}}

\DeclareMathOperator*{\diag}{\textrm{diag}}

\usepackage[labelformat=simple]{subcaption} 

\DeclareCaptionLabelFormat{subcaptionlabel}{\normalfont(\textbf{#2}\normalfont)}
\modulolinenumbers[5]

\usepackage{xspace}
\newcommand{\NP}{\emph{Neural Physics}\xspace}
\usepackage{algorithmicx}
\usepackage{algpseudocode}
\usepackage{float}

\newcommand{\wdiffxi}{{\bf{w}}_{\scriptscriptstyle \nabla^2_\xi}}

\newcommand{\fdiff}{\bm{f}^{\scriptscriptstyle \nabla^2}}
\newcommand{\half}{\frac{1}{2}}

\begin{document}

\author[IC]{Boyang Chen} 
\author[IC,IX]{Claire E. Heaney}
\author[IC,IX,DA]{Christopher C. Pain} 

\address[IC]{Applied Modelling and Computation Group, Department of Earth Science and Engineering, Imperial College London, SW7 2AZ, UK}
\address[IX]{Centre for AI-Physics Modelling, Imperial-X, Imperial College London, W12 7SL}
\address[DA]{Data Assimilation Laboratory, Data Science Institute, Imperial College London, SW7 2AZ, UK}

\begin{frontmatter}
\title{
Solving the Discretised Shallow Water Equations 
on Distorted Structured Grids with \emph{Neural Physics}}
\begin{abstract}
This paper presents a method for solving the shallow water equations on highly distorted structured grids using convolution-based operators following the \NP approach. By mapping distorted physical grids to a uniform computational domain with Jacobian transformations, the governing advection and diffusion operators are reformulated in a structured convolutional form. This enables the use of efficient stencil-based operations and, following the \NP concept, we do this through convolutional layers whose weights are determined by the chosen numerical methods rather than by training.

The method introduces consistent mass-lumped formulations, conservative and non-conservative advection forms, and a diffusion operator, all expressed as convolutions. A nonlinear Petrov-Galerkin stabilisation is incorporated to control numerical oscillations on distorted meshes. Two approximations for the Laplacian operator are also proposed to accelerate pressure and diffusion solves.  The framework is validated using the Carlisle 2005 flood event, demonstrating accurate predictions of water depth and velocity on distorted meshes for both linear and quadratic discretisations.

The results show that distorted structured grids can retain conservation, stability and accuracy, whilst adapting the mesh statically. The convolutional formulation makes the method well suited to \NP implementations and GPU acceleration, offering a scalable approach for flood modelling and large-scale geophysical simulations.
\end{abstract}

\end{frontmatter}


\section{Introduction}

The shallow water equations (SWEs) are one of the most widely used mathematical models for describing depth-averaged free-surface flows in rivers, floodplains, estuaries and coastal environments. Derived from the incompressible Navier-Stokes equations under hydrostatic assumptions, the SWEs capture the dominant balance between gravity, inertia and bed friction in shallow flows~\citep{Stoker1957,Toro2001}. As a result, they form the basis of many hydrodynamic models used in flood prediction, coastal engineering and environmental management \cite{teng2017flood,Zanardo2022}. Such models are essential for flood risk assessment, disaster mitigation and infrastructure planning in increasingly urbanised environments \cite{Spiekermann2015,Henonin2013}.

Numerical solution of the shallow water equations presents several challenges due to nonlinear advection, moving wet–dry interfaces and discontinuities, such as hydraulic jumps and complex bathymetry. A wide range of numerical discretisation methods have therefore been developed for SWE simulations. These include finite difference schemes~\citep{CASULLI199056,lundgren2020efficient}, finite volume methods~\citep{LeVeque2002}, finite element methods~\citep{donea2003finite} and discontinuous Galerkin formulations~\cite{Cockburn2000,lee2021discontinuous}. Finite volume methods are particularly attractive due to their conservative properties and their ability to capture discontinuities using Riemann solvers~\citep{Toro2001,lefloch2011godunov,bernetti2008exact}. High-order discontinuous Galerkin methods combine conservation with geometric flexibility and have been successfully applied to coastal and flood modelling problems \cite{shaw2021lisflood,bunya2009wetting}.

A major difficulty in environmental flow modelling lies in the efficient representation of complex geometries and large spatial domains. Flood modelling applications often require high resolution near river channels, hydraulic structures or urban infrastructure, while large surrounding regions may require only coarse resolution~\citep{teng2017flood}. Uniform structured grids are computationally efficient and well suited to high-performance computing architectures, including GPUs~\citep{xia2016gpu,smith2013towards,lacasta2015gpu}. However, they can lead to inefficient use of computational resources when fine resolution is needed only in localised regions.

To address this issue, numerous adaptive mesh strategies have been proposed. Adaptive mesh refinement techniques dynamically increase spatial resolution in regions with strong gradients while maintaining coarse resolution elsewhere~\citep{Berger1989}. Unstructured mesh approaches allow flexible resolution near complex boundaries and have been widely used in CFD and geophysical modelling~\citep{Pain2001,Estrem2020}. Dynamic mesh movement and $r$-adaptation methods redistribute nodes in response to evolving solution features~\citep{Huang2011,Alauzet2010,Loseille2011}. Recent work has also explored multi-resolution approaches such as wavelet-based adaptive flood models~\citep{Kesserwani2023}. 

An alternative approach can be found by distorting structured grids, so the connectivity of a Cartesian mesh is retained whilst the nodes are located where required~\citep{Thompson1985,Thompson1998}. As the grid remains structured, differential operators can be evaluated using compact stencil operations with the deformation of the nodes being handled through coordinate transformations. Such techniques are widely used in atmospheric and ocean modelling through terrain-following coordinate systems that represent complex topography while preserving structured solvers~\citep{Staniforth2012,Miura2013}. Similar strategies have been applied in hydrodynamic models to concentrate resolution in regions with strong gradients or complex geometry.

Several recent studies have demonstrated that classical numerical methods can be reformulated within machine learning frameworks. The so-called \NP approach has been applied to a range of computational physics problems including neutron transport~\citep{Phillips2023diffusion}, incompressible flow~\citep{Chen2026}, multiphase CFD simulations~\citep{Chen2024CMA} and the shallow water equations~\citep{Chen2025,Nadimy2026}. This approach exploits the correspondence between the numerical stencils used to discretise equations and the convolutional operations commonly used in deep learning architectures. The resulting implementations can leverage high-performance hardware whilst maintaining the mathematical accuracy of the traditional discretisations. As well as platform agnosticism of the code, the \NP approach enables close coupling of physics-based models with trained surrogate models and is naturally suited to solving optimisation tasks as the code is differentiable \cite{Huang2026,jiao2026np4buoyancy}.

In this work we develop a \NP framework for solving the shallow water equations on distorted structured grids. By mapping the distorted physical grid to a uniform computational domain, the governing equations can be reformulated in terms of discretised operators that retain a structured convolutional representation, enabling efficient GPU implementation.

The contributions of this paper are twofold. First, we extend the existing ConvFEM-based \NP framework to distorted structured grids, enabling the governing equations to be discretised directly on geometry-adapted meshes through Jacobian-based coordinate transformations. Second, nonlinear Petrov-Galerkin stabilisation is incorporated to control oscillations arising from advection-dominated flows on distorted meshes. The resulting framework enables static mesh adaptation through diffusion-based grid distortion, allowing resolution to be concentrated in regions of high interpolation error whilst retaining the underlying structured computational topology. This provides a practical compromise between the flexibility of adaptive meshes and the computational efficiency of structured solvers.

The remainder of the paper is organised as follows. Section~\ref{sec:method} introduces implementation of \NP and the mapping from physical (distorted) coordinate system to the structured coordinate system. Section~\ref{sec:discretisation_SWE} introduces the discretisation of the SWEs. Section~\ref{static-grid-opt} describes the optimisation procedure for generating distorted grids. Numerical examples and validation cases are presented in Section~\ref{results}. Finally, Section~\ref{dis-conclusions} discusses the computational characteristics and broader implications of the method for large-scale environmental flow modelling within the \NP framework.

\section{Methodology}\label{sec:method}

This section describes the numerical framework used to solve the governing equations on statically distorted structured grids. It introduces the Jacobian transformation and the convolutional kernels used within \NP framework.

\subsection{Discretisation using Convolutional Filters} \label{sec:discretisation_NN4PDEs}
The idea behind the \NP approach is that the matrix-vector multiplications associated with numerical discretisations can be written as discrete convolutions. Such operations are commonly used in the convolutional layers of neural networks and can be expressed using functions within machine learning libraries. 

\subsubsection{Notation}
The action of a discretisation stencil on a tensor $\bm{C}^n$ can be written as
\begin{equation}
\bm{f}(\bm{C}^n;\bm{w})\big|_{i,j}
=
\sum_{ii=-\ell}^{\ell}\sum_{jj=-\ell}^{\ell}
\bm{w}_{ii,jj}\,\bm{C}^n_{i+ii,j+jj},
\end{equation}
where $\bm{w}$ is a $(2\ell+1)\times(2\ell+1)$ array containing the coefficients of the numerical stencil. This operation is equivalent to the application of a convolutional filter, where $\bm{w}$ represents the kernel weights. Throughout this work, the tensor $\bm{f}(\bm{C}^n;\bm{w})$ is used to denote the application of a filter $\bm{w}$ to the tensor $\bm{C}^n$. 

To evaluate the stencil near domain boundaries, a halo region (padding) of width $\ell$ is included in the computational domain, giving a solution array of size $(N_x+2\ell)\times(N_y+2\ell)$. The resulting output field is defined on the physical domain and therefore has dimensions $N_x\times N_y$. The stencil radius $\ell$ depends on the discretisation scheme used (see Section~\ref{sec:bcs}).

\subsection{Non-uniform distorted grid for advection term}
The variables $\xi$ and $\eta$ represent coordinates on the uniform structured grid, which, without loss of generality, can be assumed to have a grid spacing of~1. In this case, the lumped mass matrix on this grid has a coefficient of $m_{\xi L} = 1$ in the convolutional notation. The Jacobian associated with the transformation from the distorted grid to the uniform structured grid is,
\begin{equation}
J = 
\left(\begin{array}{cc}
\frac{\partial x}{\partial \xi} & \frac{\partial y}{\partial \xi} \\
\frac{\partial x}{\partial \eta}  & \frac{\partial y}{\partial \eta} \\
\end{array}\right), \;\;\; 
J^{-1} = \frac{1}{\det(J)}
\left(\begin{array}{rr}
\frac{\partial y}{\partial \eta} & -\frac{\partial y}{\partial \xi} \\
-\frac{\partial x}{\partial \eta}  & \frac{\partial x}{\partial \xi} \\
\end{array}\right)
= 
\left(\begin{array}{cc}
(J^{-1})_{x x} & (J^{-1})_{x y}  \\
(J^{-1})_{y x}  & (J^{-1})_{y y} \\
\end{array}\right) 
\label{Jacobian}
\end{equation}
where $\det(J)=\frac{\partial x}{\partial \xi} \frac{\partial y}{\partial \eta} - \frac{\partial x}{\partial \eta} \frac{\partial y}{\partial \xi}$. Based on the following relationship between coordinate systems
\begin{equation}
\left(\begin{array}{c}
\frac{\partial C}{\partial \xi} \\
\frac{\partial C}{\partial \eta} \\
\end{array}\right) = J \left(\begin{array}{c}
\frac{\partial C}{\partial x} \\
\frac{\partial C}{\partial y} \\
\end{array}\right),
\end{equation}
the advection terms on the distorted grid can be written as
\begin{equation}
\left(\begin{array}{c}
\frac{\partial C}{\partial x} \\
\frac{\partial C}{\partial y} \\
\end{array}\right) 
= 
J^{-1} 
\left(\begin{array}{c}
\frac{\partial C}{\partial \xi} \\
\frac{\partial C}{\partial \eta} \\
\end{array}\right).
\label{Jacobian}
\end{equation}

Moreover, the coordinates $x,y$ with nodal coordinate arrays $\bm{x},\bm{y}$ on the distorted grid can be differentiated,
\begin{equation}
\begin{split}
\frac{\partial \bm{x}}{\partial \xi} = \frac{1}{m_{\xi L}} \bm{f}(\bm{x}, \bm{w_\xi}) , \quad \frac{\partial \bm{y}}{\partial \xi} = \frac{1}{m_{\xi L}} \bm{f}(\bm{y};\bm{w_\xi}) , \quad \frac{\partial \bm{x}}{\partial \eta} = \frac{1}{m_{\xi L}}\bm{f}(\bm{x},\bm{w_\eta}) , \quad \frac{\partial \bm{y}}{\partial \eta} = \frac{1}{m_{\xi L}} \bm{f}(\bm{y};\bm{w_\eta}), 
\end{split}   
\label{x-derive-xi}
\end{equation}
where $\bm{w_\xi}$ and $\bm{w_\eta}$ are the weights of the convolutional filters that represent derivatives in $\xi$ and $\eta$ directions on the uniform structured grid. 
The derivation of the system in Equation~\eqref{x-derive-xi} can be seen from the FEM discretisation: 
\begin{equation}
\int_{V_\xi} N_{i,j} \frac{\partial {x}}{\partial \xi} d\xi d\eta  =
\int_{V_\xi} N_{i,j} 
\sum_{ii,jj} \frac{\partial (N_{ii,jj}\bm{x}_{ii,jj})}{\partial \xi} d\xi d\eta ,
\end{equation}
with the FEM expansion  of $x=\sum_{ii,jj} N_{ii,jj} \bm{x}_{ii,jj}$
and then substituting the lumped mass for the consistent mass approximation to form the equation on the left-hand-side of Equations~\eqref{x-derive-xi}. 
For anything but linear elements we follow the procedure outlined in \citep{Phillips2024} to form the ConvFEM discretisation. 
For the advection equation on the distorted grid,

\begin{equation}
   \det(J)  \odot \frac{\partial C}{\partial t} + \tilde{u} \odot\frac{\partial C}{\partial \xi} + \tilde{v} \odot\frac{\partial C}{\partial \eta} = \det(J)  \odot s
\end{equation}
where 
\begin{equation}
\tilde{u} = \frac{1}{m_{\xi L}} \det(J)  \odot \left( u \odot J_{x x}^{-1} + v \odot J_{y x}^{-1}\right), \qquad 
\tilde{v} = \frac{1}{m_{\xi L}}\det(J)  \odot \left( u \odot J_{x y}^{-1} + v \odot J_{y y}^{-1}\right),
\end{equation}

which, using mass lumping of the mass terms and one point quadrature at the nodes for the other terms, becomes: 
\begin{equation}
  \bm{\det}(\bm{\underline{\underline{J}}})  \odot \frac{\partial \bm{C}}{\partial t} + \bm{\tilde{u}} \odot\frac{\partial \bm{C}}{\partial \xi} + \bm{\tilde{v}} \odot\frac{\partial \bm{C}}{\partial \eta} = \bm{\det}(\bm{\underline{\underline{J}}})  \odot \bm{s}, 
\end{equation}
in which 
\begin{equation}
\bm{\tilde{u}} = \frac{1}{m_{\xi L}} \bm{\det}(\bm{\underline{\underline{J}}})  \odot \left(\bm{u} \odot \bm{J_{x x}^{-1}} + \bm{v} \odot \bm{J_{y x}^{-1}}\right), 
\end{equation}
\begin{equation}
\bm{\tilde{v}} = \frac{1}{m_{\xi L}} \bm{\det}(\bm{\underline{\underline{J}}})  \odot \left( \bm{u} \odot \bm{J_{x y}^{-1}} + \bm{v} \odot \bm{J_{y y}^{-1}}\right).
\end{equation}
Thus the non-conservative form can be written as follows,
\begin{eqnarray}
    \bm{M_L}\odot \frac{\partial \bm{C}}{\partial t} &+& \frac{1}{m_{\xi L}} \bm{\det}(\bm{\underline{\underline{J}}}) \odot \left( \bm{u} \odot \bm{J_{x x}^{-1}} + \bm{v} \odot \bm{J_{y x}^{-1}}\right)  \odot \bm{f}(\bm{C};\bm{w_{\xi}}) \nonumber \\ &+& \frac{1}{m_{\xi L}} \bm{\det}(\bm{\underline{\underline{J}}})  \odot \left( \bm{u} \odot \bm{J_{x y}^{-1}} + \bm{v} \odot \bm{J_{y y}^{-1}}\right)  \odot \bm{f}(\bm{C};\bm{w_\eta})  = \bm{M_L} \odot \bm{s},
\end{eqnarray}
or
\begin{equation}
   \bm{M_L}\odot  \frac{\partial \bm{C}}{\partial t} +   \bm{\tilde u}   \odot \bm{f}(\bm{C};\bm{w_\xi}) + \bm{\tilde v}  \odot \bm{f}(\bm{C};\bm{w_\eta}) + \gamma_c \bm{C}\odot (\bm{f}(\bm{\tilde u};\bm{w_\xi}) + \bm{f}(\bm{\tilde v};\bm{w_\eta}) ) = \bm{M_L} \odot \bm{s},
\end{equation}

in which $\gamma_c=1$ corresponds to conservative from and $\gamma_c=0$ non-conservative form and the lumped 
mass matrix is: 
\begin{equation}
\bm{M_L}=\bm{\det(\underline{\underline{J}})}.
\end{equation}
When $\gamma_c=1$, the term $\bm{C}\odot (\bm{f}(\bm{\tilde u};\bm{w_\xi}) + \bm{f}(\bm{\tilde v};\bm{w_\eta}) )$ can be smoothed out using mass tensor $\bm{m_{\xi}}$, which is on the uniform structured grid and can be expressed as a convolutional filter as below for linear and quadratic with $\Delta \xi = \Delta \eta = 1$. 
In addition, to ensure mass conservation, the conservative form can also be expressed as 
\begin{equation}
   \bm{M_L}\odot  \frac{\partial \bm{C}}{\partial t} +  (  \bm{f}(\bm{\tilde u}   \odot \bm{C};\bm{w_\xi}) +  \bm{f}(\bm{\tilde v}  \odot \bm{C};\bm{w_\eta}) ) = \bm{M_L}\odot \bm{s}.
\end{equation}

\subsection{Non-linear Petrov-Galerkin Diffusion on Distorted Grids}
\label{Petrov-distort-2d}
The diffusion based discrete residual $R_{i}$ can be written as below using the mixed mass approximation to the residual,
 \begin{equation}
\bm{R_c} =  \alpha_r 
\bm{f} (\,\tilde{\bm{u}} \odot \bm{f}(\bm{C};\bm{w_{\xi}}) +\tilde{\bm{v}} \odot \bm{f}(\bm{C};\bm{w_{\eta}}) 
\,; \, \frac{1}{m_{\xi L}} \bm{m_\xi} - \bm{I} ),
\end{equation}
where $m_{\xi L}=\Delta \xi \Delta \eta$ is the lumped mass term, $\bm{m_\xi}$ is the mass filter associated with the finite element matrices and $\alpha_r$ is set to~2 as default. Thus, the resulting diffusion term is $\nabla_{\xi} \cdot (k_{\xi} \nabla_{\xi}C)$ (that is, in the uniform structured grid) with the diffusion coefficient $k_{\xi}$ calculated as,
\begin{equation}
   \bm{k}_{\bm{\xi\eta}} = {\alpha_{k}|\bm{R_c}|   }
\oslash \left(\bm{\epsilon_k}+\sqrt{(\bm{f}(\bm{C};\bm{w_{\xi}^{}})^2+\bm{f}(\bm{C};\bm{w_{\eta}^{}})^2)} \right) \, , 
\end{equation}
in which the operator $\oslash$ is defined $\bm{a}\oslash\bm{b}=(a_1/b_1, a_2/b_2, \ldots)^T$, $\sqrt{\bm{a}}=(\sqrt{a_1},\sqrt{a_2}, \ldots)^T$, 
 ${\bm{a}}^2=(({a_1})^2,({a_2})^2,\ldots)^T$. 
The final Petrov–Galerkin diffusion coefficient is given by:
\begin{equation}
     \bm{k_\xi} = \bm{\det}(\bm{\underline{\underline{J}}})\odot \min\{ \bm{k}_{\bm{\xi\eta max}}, \bm{k}_{\bm{\xi\eta}}  \}, 
\end{equation}
in which $\min\{ \bm{a}, \bm{b}\}=(\min\{ a_1,b_1\}, \min\{a_2,b_2\},\ldots)^T$ and 
with $\bm{k_{\xi \eta  max}}=(0.25/\Delta t,0.25/\Delta t,0.25/\Delta t,\ldots)^T $ is introduced for 
numerical stability of the explicit predictor-corrector time stepping.  

The final diffusion term can be calculated as follows: 
\begin{equation}\label{final_diffusion_regular}
    \nabla_{\xi} \cdot (k_{\xi} \nabla_{\xi}C) \sim \frac{1}{2} \left(\bm{k_{\xi}} \odot \bm{f}(\bm{C};\bm{w_{\nabla\xi^{2}}}) + \bm{f}(  \bm{k_{\xi}}\odot \bm{C};\bm{w_{\nabla\xi^{2}}}) - \bm{C}\odot\bm{f}(  \bm{k_{\xi}};\bm{w_{\nabla_\xi^{2}}})\right),
\end{equation}
where $\bm{w_{\nabla^{2}_{\xi}}}$ is the Laplacian operation on the uniform regular grid, see \cite{Chen2026}.

\section{Discretisation and Solution of the Shallow Water Equations}\label{sec:discretisation_SWE}

\subsection{Discretisation in Time}\label{sec:disc_gov_eq_time}
\subsubsection{Discretisation of the Continuity Equation in Time}
The continuity equation, part of the Shallow Water Equations, which governs the water depth, $\eta$, is discretised in time using a central difference approximation about $t^{n+\half}$:

\begin{equation}\label{gov1-eta}
    \frac{h^{n+1}-h^{n}}{\Delta t} 
    + \frac{\partial({u}^{n+\frac{1}{2}} {\eta}^{n+\frac{1}{2}})}{\partial x} + \frac{\partial({v}^{n+\frac{1}{2}} {\eta}^{n+\frac{1}{2}})}{\partial y} 
    = s_h^{n+1},
\end{equation}

in which $h^n$ is the free surface height at time level~$n$, $\Delta t$ is the time step and the source, $s_h$, is treated implicitly. 

\subsubsection{Discretisation of the Momentum Equations in Time}\label{Momentum Equations section}
The momentum equations part of the Shallow Water Equations (written in non-conservative form and divided through by the water depth~$\eta$) are also discretised in time using a central difference approximation about $t^{n+\half}$ resulting in:  
\begin{equation}\label{NS-eqn-disc}
\begin{split}
\rho_g \left(\frac{\bm{q}^{n+1} -\bm{q}^{n} }{\Delta t} 
+ 
u^{n+\frac{1}{2}}\frac{\partial \bm{q}^{n+\frac{1}{2}} }{\partial x} +  v^{n+\frac{1}{2}}\frac{\partial \bm{q}^{n+\frac{1}{2}} }{\partial y} 
\right) + \sigma_q^{n+1} \bm{q}^{n+1}  
=  -\nabla h^{n+\frac{1}{2}} 
\,, 
\end{split}
\end{equation}
where $\rho_g=\frac{1}{g}$, $\bm{q}^{n} = (u^n\; v^n)^T$ denotes the velocity and the bed friction term is expressed as follows:

\begin{equation}\label{manning-cor}
\sigma_q^{n+1}=\frac{\hat\sigma_q^{n+1}}{g\eta^{n+1}}
=\frac{ g n^2_M (\eta^{n+1})^{-1/3} \sqrt{(u^{n+1})^2+(v^{n+1})^2}}{g\eta^{n+1}}\approx\frac{n^2_M  \sqrt{(u^{n+1})^2+(v^{n+1})^2}}
{(\max\{\epsilon_\eta,\eta^{n+1}\})^{4/3} }
\, , 
\end{equation}
in which a limit on how small $\eta$ can be in the denominator is applied to avoid dividing by a number close to zero and here $\epsilon_\eta=10^{-4}$. 
The source term from the continuity equation, $s_h$, would appear in the non-conservative form of the momentum equations, Equation~\eqref{NS-eqn-disc}. However, if the momentum equations, themselves, had a source, which had the same velocity as the fluid, the sources would cancel. This is assumed to be the case here.

\subsubsection{Discretisation of the Continuity Equation in Space}  \label{adv-section}
After applying the product rule to the first order derivatives in the continuity equation, Equation~\eqref{gov1-eta}, the FE spatial discretisation (written as convolutional filters that are derived in Section~\ref{sec:discretisation_NN4PDEs}) is applied. This produces a conservative (in ${\eta}^n$) discretised system of equations. For stabilisation, a diffusion operator (analytically $\nabla\cdot(k\nabla\eta)$) is added to the right-hand side of Equation~\eqref{gov1-eta} (see \ref{Petrov-distort-2d} for details of its discretisation). The resulting equation can be written as follows:
\begin{eqnarray} \bm{r_h}^{n+\frac{1}{2}} &=& \bm{M_L}\odot\left(\frac{\bm{h}^{n+1}-\bm{h}^{n} }{\Delta t}\right) 
-\bm{M_L}\odot \bm{s_h}^{n+1}   \nonumber \\
 &+&
  \tilde{\bm{u}}^{n+\frac{1}{2}}  \odot  \bm{f}\left(\bm{\eta}^{n+\frac{1}{2}} ;{\bf{w_{\xi}}}\right)
+
  \tilde{\bm{v}}^{n+\frac{1}{2}}  \odot \bm{f}\left( \bm{\eta}^{n+\frac{1}{2}} ;{\bf{w_{\eta}}}\right) \nonumber \\
 &+&
 \bm{\eta}^{n+\frac{1}{2}}   \odot  \left( \bm{f}\left(\tilde{\bm{u}}^{n+\frac{1}{2}} ;{\bf{w_{\xi}}}\right)
+
  \bm{f}\left( \tilde{\bm{v}}^{n+\frac{1}{2}}  ;{\bf{w_{\eta}}}\right) \right)
 \nonumber \\
&+&  
 \fdiff\left(\bm{\eta} ^{n+\frac{1}{2}}, \bm{k}_{\eta}^{n+\frac{1}{2}};\wdiffxi^{\rm{\bf{PG}}}\right)
 ,  %
\label{Petrov-h}
\end{eqnarray} 
where $\bm{r}_h$ is the residual associated with the free surface height, $\bm{h}^n$, $\bm{\eta}^n$, $\tilde{\bm{u}}^n$, $\tilde{\bm{v}}^n$ and $\bm{s_h}^n$ are tensors containing the nodal values at time level~$n$ of the free surface height, water depth, $\xi$ and $\eta$ components of velocity and source respectively. The entry-wise multiplication of tensors is represented by the Hadamard product, which is denoted by~$\odot$. The function $\bm{f}(\bm{\cdot}\,;\bm{\cdot})$ represents a convolution of the first argument with the kernel or filter given in the second argument, see Section~\ref{sec:discretisation_NN4PDEs}. The terms $\bf{w_{\xi}}$ and $\bf{w_{\eta}}$ represent the filters (whose entries make up the weights) of the neural network that correspond to the discretisations of the first derivative with respect to $\xi$ and $\eta$. Similarly, the term $\wdiffxi^{\rm{\bf{PG}}}$ represent filters corresponding to coefficients associated with the Petrov-Galerkin diffusion applied to stabilise the free surface height. See Section~\ref{sec:discretisation_NN4PDEs} for derivations of the convolutional filters. Petrov-Galerkin stabilisation~\citep{donea2003finite} is used here to control oscillations in the free surface height that often occur with central-difference or FEM-based discretisations of the advection term. The stabilisation introduces a diffusion term (with diffusion coefficients $\bm{k}_{\eta}^{n+\frac{1}{2}}$) and provides robustness for demanding problems, such as those which include wetting and drying. It can be convenient to use a linear FEM filter for $\wdiffxi^{\rm{\bf{PG}}}$ rather than a higher order ConvFEM filter, as the latter has a greater tendency to introduce oscillations due to its use of high order FEM polynomials. For idealised problems, the stabilisation term $\wdiffxi^{\rm{\bf{PG}}}$ was discretised with the same order of ConvFEM filters as the other spatial derivatives (that is $\wdiffxi^{\rm{\bf{PG}}}= \wdiffxi$), as the form of this term was found to have little influence on the results. However, for the more demanding test cases, the stabilisation term was discretised with linear elements irrespective of the order of discretisation used for the other terms.  

\subsubsection{Discretisation of the Momentum Equations in Space} 
The momentum equations include a term which depends on the bed friction, Equation~\eqref{manning-cor}. To discretise this term, we first need to distribute the value of $\bm{\eta}^{n+1}$ over the nodes using 

\begin{equation}\label{eta-spread}
\bm{\hat{\eta}}^{n+1} = \bm{f}\left( \bm{{\eta}}^{n+1}; {\bf{w_{ml\xi}}}^{-1} {\bf{w_{m\xi}}} \right)\,,
\end{equation}

where the filter ${\bf{w_{ml\xi}}}^{-1} {\bf{w_{m\xi}}}$ represents the product of two filters associated with the inverse lumped mass matrix 
(e.g. ${\bf{w_{ml\xi}}}^{-1} =\bf{I}$) and the consistent mass matrix in the non-distorted domain, respectively. Distributing the values of water depth over the nodes in this way enables the water to move from the wet to dry areas without being inhibited by excessive bottom drag. On discretising Equation~\eqref{manning-cor} using the expression in Equation~\eqref{eta-spread}, we obtain

\begin{equation}\label{manning-cor}
\bm{\sigma_q}^{n+1}
= 
(\bm{n_M})^2 \odot \bm{c}^{n+1} \oslash 
(\max\{ \bm{1}\epsilon_\eta , 
(1-\alpha_\eta ) \bm{\eta}^{n+1} + \alpha_\eta \bm{\hat{\eta}}^{n+1} \} )^{4/3} ,  
\end{equation}

in which the fluid speed tensor $\bm{c}^{n+1}$ 
is such that $c_{i,j}^{n+1} = \sqrt{ (u^{n+1}_{i,j})^2+(v^{n+1}_{i,j})^2}$. The tensor of nodal values of the Manning coefficient, $\bm{n_M}$, is assumed to be constant and $\alpha_\eta=0.1$ is a heuristically obtained parameter. 

When discretising the momentum equations (Equation~\eqref{NS-eqn-disc}), we adopt a Petrov-Galerkin stabilisation similar to that used for the free surface height. However, for the momentum equations, the stabilisation terms have the same order elements as the other terms in the equations, hence the weights are denoted by $\wdiffxi$ and not $\wdiffxi^{\textbf{PG}}$. Mass lumping is used for the advection, time and absorption terms. Applying the FE spatial discretisations through convolutional filters leads to:
\begin{eqnarray}
 \bm{r_u}^{n+\frac{1}{2}} =  \bm{M_L}\odot 
 \rho_g \left(\frac{\bm{u}^{n+1} -\bm{u}^n}{\Delta t}\right)
   +
 \bm{M_L}\odot \left( \bm{\sigma_q}^{n+1}\odot\bm{u}^{n+1}
  \right) \nonumber \\
 +
 \rho_g \left(
 \tilde{\bm{u}}^{n+\frac{1}{2}} \odot \bm{f}( \bm{u}^{n+\frac{1}{2}};{\bf{w_{\xi}}} )
+
 \tilde{\bm{v}}^{n+\frac{1}{2}} \odot \bm{f}( \bm{u}^{n+\frac{1}{2}};{\bf{w_{\eta}}}) 
 \right) \nonumber \\
+ 
 \fdiff (\bm{u}^{n+\frac{1}{2}},\bm{{k}}_u^{n+\frac{1}{2}};\wdiffxi )
 + \bm{J_{xx}^{-1}} \odot \bm{f}( \bm{h}^{n+\frac{1}{2}};{\bf{w_{\xi}}}) 
 +  \bm{J_{xy}^{-1}} \odot \bm{f}( \bm{h}^{n+\frac{1}{2}};{\bf{w_{\eta}}}) 
 \,,
 \label{Petrov-u} \\[3mm]
 \bm{r_v}^{n+\frac{1}{2}} = \bm{M_L}\odot 
 \rho_g \left(\frac{\bm{v}^{n+1} -\bm{v}^n}{\Delta t}\right)
  +
 \bm{M_L}\odot \left( \bm{\sigma_q}^{n+1}\odot\bm{v}^{n+1}\right)  \nonumber \\
 + 
 \rho_g \left(
 \tilde{\bm{u}}^{n+\frac{1}{2}} \odot \bm{f}( \bm{v}^{n+\frac{1}{2}};{\bf{w_{x}}})
+
 \tilde{\bm{v}}^{n+\frac{1}{2}} \odot \bm{f}( \bm{v}^{n+\frac{1}{2}};{\bf{w_{y}}}) 
 \right) \nonumber \\
+ 
 \fdiff (\bm{v}^{n+\frac{1}{2}},\bm{{k}}_v^{n+\frac{1}{2}};\wdiffxi )
 + \bm{J_{yx}^{-1}} \odot \bm{f}( \bm{h}^{n+\frac{1}{2}};{\bf{w_{\xi}}}) 
 +  \bm{J_{yy}^{-1}} \odot \bm{f}( \bm{h}^{n+\frac{1}{2}};{\bf{w_{\eta}}}) 
 \,,
 \label{Petrov-v}
\end{eqnarray}

where $\bm{k}_u$ and $\bm{k}_v$ are diffusion coefficients in $x$ and $y$~directions, respectively. 

In this work, the momentum equations are solved in non-conservative form. This choice is motivated by several considerations. First, it avoids the need to divide by the fluid depth $\eta$ after solving for the conservative variables $(\eta u)^n$ and $(\eta v)^n$ in order to recover the velocities $u^n$ and $v^n$. Second, when bed friction is included, momentum is no longer strictly conserved and the advantages of a conservative formulation are reduced, particularly in flows where frictional effects dominate. Third, solving directly for the velocity fields allows numerical diffusion arising from the Petrov-Galerkin stabilisation to be applied naturally to the velocity variables in the discretised momentum equations (Equation~\eqref{NS-eqn-disc}). 
Fourth, non-conservative formulations facilitate the construction of schemes that preserve motionless steady states. This desirable property, commonly referred to as the \emph{C-property}, ensures that a stationary solution remains stationary under the numerical discretisation~\citep{VANZO20161}. 

Finally, the non-conservative formulation leads to a simpler numerical implementation, which can improve computational efficiency in large-scale simulations.


\subsubsection{Application of Boundary Conditions}
\label{sec:bcs}

\begin{figure}[H]
\hspace{-0.1cm}
    \includegraphics[width=5.8cm]{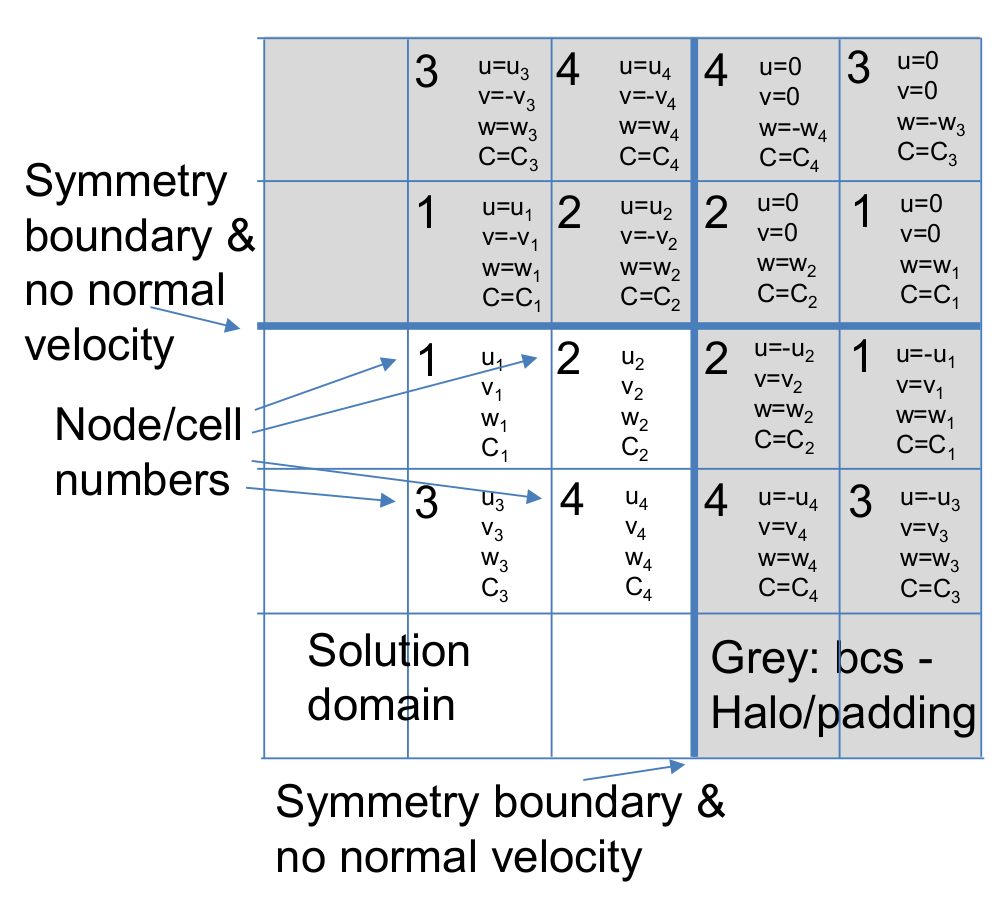}
\hspace{-0.2cm}
    \includegraphics[width=5.5cm]{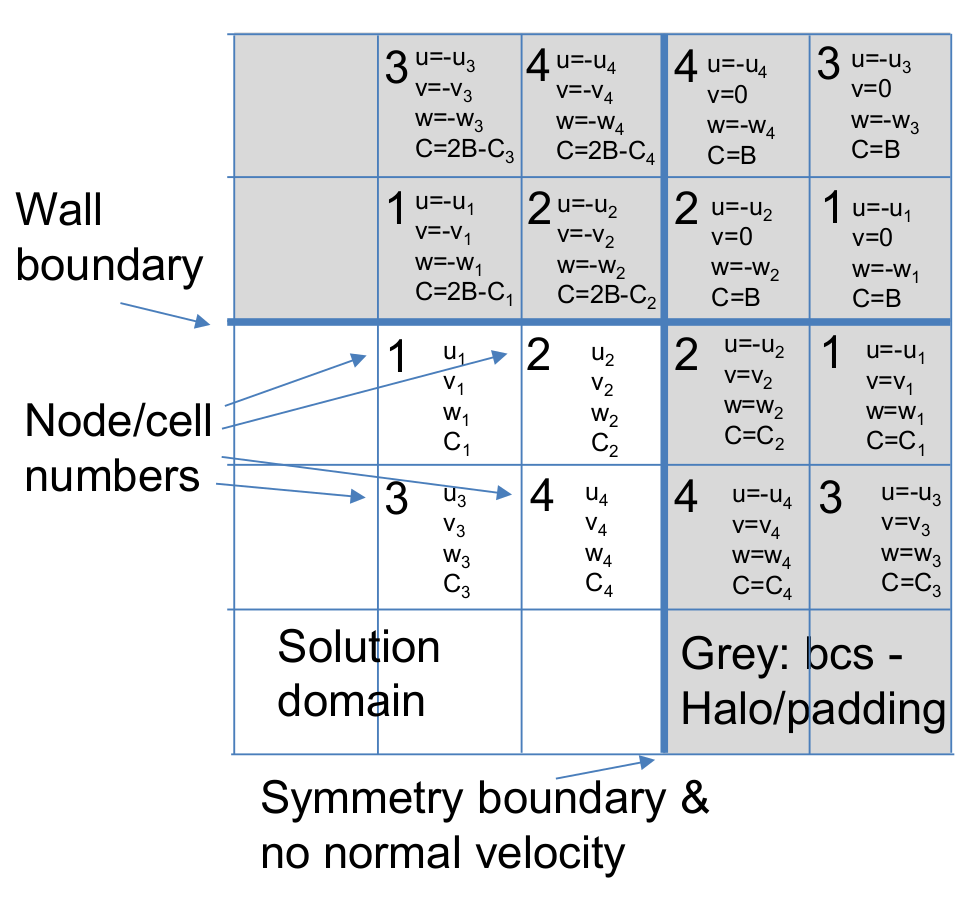}
\hspace{-0.2cm}
    \includegraphics[width=5.6cm]{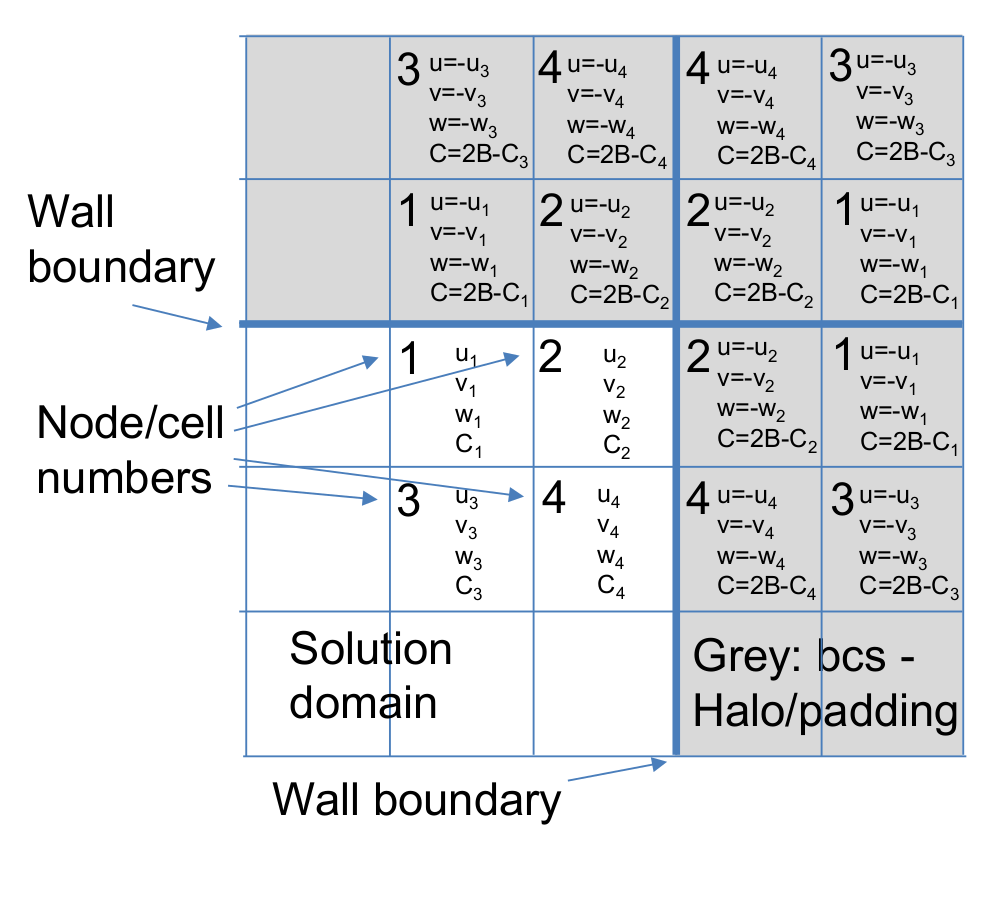}
    \caption{Schematic diagram showing how the boundary conditions work for symmetry boundaries and specified velocity and free surface height boundaries. The boundary conditions are enforced between the cells/nodes in the diagram by specifying suitable values of the halo values (in grey). }
    \label{fig:symm-bcs}
\end{figure}

Boundary conditions are imposed using halo nodes (also referred to as ghost or virtual cells) located around the perimeter of the computational domain. These halo nodes are implemented through padding, as illustrated in Figure~\ref{fig:symm-bcs}. Using halo nodes allows the same discretisation stencil to be applied at interior and boundary nodes, simplifying the numerical implementation.

The width of the halo region depends on the order of the numerical discretisation. For example, a halo width of one node is sufficient for linear elements, two nodes for quadratic ConvFEM elements, three nodes for cubic ConvFEM elements and so forth.

Dirichlet boundary conditions are imposed by assigning the prescribed value directly to the halo nodes. 
When a specified normal derivative is required, the halo values are obtained by extrapolating from the nearest interior node using the prescribed gradient. A common special case is the zero-gradient condition, which corresponds to symmetry or slip boundary conditions (see Figure~\ref{fig:symm-bcs}). In this case, the halo node values are simply set equal to the value at the nearest interior node.

\subsubsection{Stage 1: Velocity Solution}\label{sec:stage1}
To solve for velocity, we wish to reduce the residuals written in Equations~\eqref{Petrov-u} and~\eqref{Petrov-v} to zero. As the future velocity and the bed friction are the dominant terms in the residuals, we treat these terms implicitly by forming the equations below:

\begin{eqnarray}
\bm{u}^{n+1} &\leftarrow& -\bm{M_L^{-1}}\odot \bm{r_u}^{n+\frac{1}{2}} \oslash \left(\frac{\rho_g}{\Delta t}\bm{1} +\bm{\sigma_q}^{n+1}\right)+ {\bm{u}}^{n+1}, \label{velocity-1} \\ 
\bm{v}^{n+1} &\leftarrow& -\bm{M_L^{-1}}\odot \bm{r_v}^{n+\frac{1}{2}} \oslash \left(\frac{\rho_g}{\Delta t}\bm{1} +\bm{\sigma_q}^{n+1}\right) + {\bm{v}}^{n+1},
\label{velocity-2}
\end{eqnarray} 
    
where $\bm{1}$ is a tensor, in which each entry is unity $\left(\text{that is } \bm{1}\left\vert\right. _{i,j}=1\right)$. Hadamard entrywise division of tensors of the same size is denoted by $\oslash$ and defined as follows, 
    
\begin{equation*}
    \hat{\bm{c}}=\hat{\bm{a}}\oslash\hat{\bm{b}} \;\;\;\;\Longleftrightarrow \;\;\;\;\hat{c}_{i,j}=\frac{\hat{a}_{i,j}}{\hat{b}_{i,j}}.
\end{equation*}

Having solved for $\bm{u}^{n+1}$ and $\bm{v}^{n+1}$, the velocities $\bm{u}^{n+\frac{1}{2}}$ and $\bm{v}^{n+\frac{1}{2}}$ are updated. On the right-hand side of Equations~\eqref{velocity-1} and~\eqref{velocity-2}, the variables $\bm{u}^{n+1}$ and $\bm{v}^{n+1}$ refer to the most up-to-date approximation that is available --- for the predictor step this will be $\bm{u}^{n}$ and $\bm{v}^{n}$.  Writing Equations~\eqref{velocity-1} and~\eqref{velocity-2} in a residual form, enables one to use either the consistent mass matrix (defined by the convolutional filter $\bf{w}_{m}$) or the lumped mass matrix ($\bf{w}_{ml}$) for the time term within the residual. As, if the iteration process converges, the residuals will have been forced to zero whichever mass matrix is used. Choosing a diagonal lumped mass matrix has the benefit of being easy to invert, making the iterative process efficient. 
    
\subsubsection{Stage 2: Free Surface Height Correction}\label{sec:stage2}
The free surface correction equation is derived by subtracting out the velocity equation (Equation~\eqref{NS-eqn-disc}) with the previous free surface height at time level~$n$ and the new free surface height at time level~$n+1$ that leads to the satisfaction of the continuity equation (Equation~\eqref{gov1-eta}).   
The divergence of the resulting momentum equation is formed to produce the free surface height correction equation similar to~\cite{Pavlidis2016}. In continuous form, the resulting iterative correction $\Delta h$ to the free surface height can be determined by solving:

\begin{equation} \label{nonhydro-pressure1---} 
\left(- \nabla^2 +\frac{\beta_h \rho_g}{(\Delta t)^2\eta^{n+\frac{1}{2}}}\right) \Delta h 
= \frac{\beta_h \rho_g}{\Delta t \eta^{n+\frac{1}{2}}} \left(  \frac{ {h}^{n+1}-h^{n}}{\Delta t}
+\nabla \cdot ( \bm{q}^{n+\frac{1}{2}}\eta^{n+\frac{1}{2}})  -s_h^{n+1}
\right) \,,
\end{equation}
    
in which all the variables on the right-hand side of this expression use the best estimate of the variables ${h}^{n+1}$, ${\bm{q}}^{n+1}$. Here a parameter $\beta_h=4$ has been included. If large time steps are being used and the non-linear convergence is struggling, $\beta_h$ can be reduced. $\beta_h=1$ corresponds to a more stable, but less accurate, backward Euler time discretisation rather than the predictor-corrector method used here. The discretised form of Equation~\eqref{nonhydro-pressure1---} is 
\begin{equation}
 \beta_c \bm{f}(\bm{\Delta {h}} ;\wdiffxi) 
+ \frac{\beta_h\rho_g}{(\Delta t)^2}
\left( {\bf{1}}\oslash \max\{\epsilon_\eta \bm{1}, \bm{\eta}^{n+\frac{1}{2}} \}  \right)\odot \bm{M_L} \odot  \bm{\Delta {h}} 
=
-\frac{\beta_h \rho_g}{\Delta t}  
\bm{r_h}^{n+\frac{1}{2}} 
 \oslash \max\{\epsilon_\eta \bm{1}, \bm{\eta}^{n+\frac{1}{2}} \}, 
  \label{ph_disc1}
\end{equation}
in which $\bm{\eta}^{n+\frac{1}{2}}$ has been replaced by $\max\{\epsilon_\eta \bm{1}, \bm{\eta}^{n+\frac{1}{2}} \}$ in the above to aid stability for wetting and drying occurring at small water depths. Also to increase the speed of the algorithm the 
discretised Laplacian has been approximated by the discretised 
Laplacian in the uniform grid domain, that is the first term of 
Equation~\eqref{ph_disc1} and $\beta_c\in\left[0,1\right]$ is a relaxation 
coefficient $\beta_c=1$ is used here. 
For idealised problems, $\epsilon_\eta=0$, and for non-idealised problems,
$\epsilon_\eta=\SI{0.0001}{m}$.  
Equation~\eqref{ph_disc1} is solved for $\bm{\Delta {h}}$ using a Jacobi iteration, which is sufficient for Courant numbers (based on the wave speed) of less than $\mathcal{O}(10)$, say. For larger Courant numbers, a multigrid method could also be used to calculate the free surface height correction tensor. See~\citep{Phillips2023diffusion} for a description of how the multigrid method can be implemented through neural networks with the U-Net architecture~\citep{Ronneberger2015}. Having solved for $\bm{\Delta h}$, the free surface height is updated according to 
\begin{equation*}
    \bm{h}^{n+1} \leftarrow {\bm{h}}^{n+1} + \bm{\Delta h} \,. 
\end{equation*} 

\subsubsection{Stage 3: Velocity Correction}\label{sec:stage3}
The velocity correction $\Delta \bm{q}$ is then calculated, which, in continuous form, is given by

\begin{equation}
     \Delta \bm{q} = -\frac{\Delta t}{2\rho_g}  \nabla \Delta h\,.
\end{equation}

In discretised form, this becomes

\begin{equation}
    \bm{\Delta u} = -\frac{\Delta t}{2\rho_g}  \bm{M_L^{-1}}\odot 
    \left(
    \bm{J_{xx}^{-1}} \odot \bm{f}( \bm{\Delta h};{\bf{w_{\xi}}}) 
 +  \bm{J_{xy}^{-1}} \odot \bm{f}( \bm{\Delta h};{\bf{w_{\eta}}}) \right) ,
\label{vel-u-cor}
\end{equation}
\begin{equation}
    \bm{\Delta v} = -\frac{\Delta t}{2\rho_g}  \bm{M_L^{-1}}\odot 
    \left(
    \bm{J_{yx}^{-1}} \odot \bm{f}( \bm{\Delta h};{\bf{w_{\xi}}}) 
 +  \bm{J_{yy}^{-1}} \odot \bm{f}( \bm{\Delta h};{\bf{w_{\eta}}}) \right) .
\label{vel-v-cor}
\end{equation}

Once $\bm{\Delta u}$ and $\bm{\Delta v}$ are obtained, the tensors representing the discrete velocity components can be updated: 

\begin{equation}
      \bm{u}^{n+1} \leftarrow \bm{u}^{n+1} + \bm{\Delta u}, \quad
      \bm{v}^{n+1} \leftarrow \bm{v}^{n+1} + \bm{\Delta v}.
\end{equation}

\section{Static grid optimisation}
\label{static-grid-opt}
Static grid optimisation is achieved by first estimating a desired element size field from interpolation theory, where the local truncation error is related to solution curvature. To ensure robustness for time-dependent problems, the maximum interpolation error over all time levels is taken, producing a conservative measure of where resolution is required. This error is then converted into a target grid spacing or element size, and subsequently a diffusion coefficient, which is typically defined as the inverse of the target grid spacing. The diffusion coefficient is initially computed on the uniform computational grid and  interpolated onto the physical mesh. A diffusion equation is solved to redistribute the mesh nodes, driving them toward regions requiring higher resolution while smoothing the overall distribution. This process results in a mesh that equilibrates according to the prescribed spacing field, concentrating nodes in high-error regions and coarsening elsewhere, while maintaining stability and mesh quality through the diffusive smoothing mechanism.

When adapting the resolution, we also take account of element quality, and attempt to limit the movement of nodes to avoid badly conditioned elements. We express the redistribution of nodes and the element quality as two terms in a functional, $F$, which we will minimise: 

\begin{equation}
F = \int_{\Omega_\xi} \left(  \frac{1}{2} (\nabla_\xi \bm{x}  )^T \bm{k} (\nabla_\xi  \bm{x}  )  
 + \frac{1}{2} ( (\bm{x}-\bm{x}_0)^T \bm{\sigma} (\bm{x}-\bm{x}_0)  ) \right) \,dV_\xi .
\label{F-dff-diffusion}
\end{equation}
in which $\bm{k}$ is a tensor containing diffusion coefficients, $\bm{\sigma}$ is a tensor of relaxation coefficients, $\bm{x}=(x,y)^T$ are the 

coordinates and $\bm{x}_0$ contains the grid locations from a previous iteration that we relax towards. This expression balances relocating the nodes according to a diffusion mechanism whilst, at the same time, not moving the nodes too far from their previous locations. If time stepping is used to solve this equation, then one can use the second term to help find a solution by setting ${\bf x}_0$ to the previous time step value $\bm{x}_0 = \bm{x}^{(n)}$ and $\bm{\sigma}=\diag( \frac{1}{\Delta t^*})$, in which $n$ is the time level and $\Delta t^*$ the pseudo time-step size. 
We integrate over the entire uniform grid represented by $\Omega_\xi$. Differentiating Equation~\eqref{F-dff-diffusion}, assuming $\bm{k}$ is independent of $\bm{x}$, and setting this to zero results in: 
\begin{equation}
\frac{\partial F}{\partial \bm{x}} = 
 \int_{\Omega_\xi} \left( \; - (\nabla_\xi^T (\bm{k} \nabla_\xi  \bm{x}  ) +  \bm{\sigma} (\bm{x}-\bm{x}_0 )  \; \right) dV_\xi =0, 
\label{F-differentiated}
\end{equation}
which occurs at the extremum (the minimum) of the quadratic functional~$F$. The diffusion coefficient(s) can be set in a number of ways. A simple way to produce isotropic non-uniform resolution is to use a relative length scale $L_{ref}$ that varies non-uniformly across the grid. In this case the diffusion coefficient becomes:

\begin{equation}
\label{k-dff-diffusion}
k=\frac{1}{L_{ref}}. 
\end{equation}

The tensor $\bm{k}$ can also generate anisotropic grid spacing by defining multiple components:
\begin{equation}
\label{k-dff-diffusion}
\bm{k}={\bm{L_{ref}}^{-1}}. 
\end{equation}

If, alongside element quality, we wish to include solution error when redistributing the nodes, we can use interpolation theory and adjust the above definitions of $\bm{k}$. For a given direction across the undistorted system, we can estimate the solution error by $\vert\nabla_\xi \bm{a}^T \bm{a} \nabla_\xi  \psi\vert$ along a unit direction $\bm{a}$ in the undistorted grid. This term estimates the curvature of the solution field in the direction $a$. Assuming $\epsilon$ is the desired interpolation error of the solution variables, then a way to introduce anisotropy is to represent the resulting diffusion coefficient as a tensor ${\bm{k}}$, which can be thought of as a hyperellipse representing the strength of diffusion in different directions. Another way to represent this diffusion anisotropy is directly through the discretisation. Suppose we have a 3$\times$3$\times$3 stencil in 3D then a way to represent the anisotropy is to effectively have 13 directions across the stencil which are discretised separately with their own diffusion coefficient associated with them. Suppose the index for a node is $i,j$ then the diffusion coefficients across node $i,j$ is then

\begin{equation}
k^{i,j}_{i+li,j+lj}, \ \forall\ li,lj \in\{-1,0,1\} \quad {\text{other than}} \quad li=lj =0. 
\end{equation}
As the diffusion tensor is symmetric,  $k^{i,j}_{i+li,j+lj}=k^{i,j}_{i-li,j-lj}$. 

Interpolation theory suggests the absolute value of interpolation error is, in 1D: 
\begin{equation}
\epsilon_0 = \gamma h^2 \left| \frac{\partial^2 \psi}{\partial x^2}\right|. 
\end{equation}
in which $\epsilon_0$ is the specified interpolation error; $h$ is the desired edge length (distance between neighbouring nodes); $\gamma$ is an $\mathcal{O}(1)$ scalar (here $\gamma=1$ is used); and $\Delta x$ is the actual edge length. In a 1D non-distorted space represented by $\xi$, this equation is modified and can be rearranged for $h$: 

\begin{equation}
\epsilon_0 = \gamma \frac{h^2}{(\Delta x)^2} \left| \frac{\partial^2 \psi}{\partial \xi^2} \right| \quad\implies\quad h =   \sqrt{  \frac{(\Delta x)^2 \epsilon_0 }{  {\gamma  \left| \frac{\partial^2 \psi}{\partial \xi^2} \right|}  } }.   
\end{equation} 

The diffusion coefficient is $k=1/h$ in which $h$ is the desired node spacing. This can be seen from the discretised in 1D diffusion equation of the $x$ coordinate: $ k_{i+\frac{1}{2}} (x_{i+1} - x_i)/(\Delta\xi)^2 - k_{i-\frac{1}{2}} (x_{i} - x_{i-1})/(\Delta\xi)^2=0$ which re-arranging becomes $ k_{i-\frac{1}{2}}/k_{i+\frac{1}{2}} = \Delta x_{i+\frac{1}{2}}/\Delta x_{i-\frac{1}{2}}$ which is realised with $k_{i+\frac{1}{2}}=1/h_{i+\frac{1}{2}}$ and in which $\Delta\xi$ is the uniform grid spacing in the non-distorted domain.  
Now, since the diffusion coefficient is: 
\begin{equation}
k=\frac{1}{h}, 
\end{equation} 
then
\begin{equation}
k=
\sqrt{    \frac{\gamma \left|\frac{\partial^2 \psi}{\partial \xi^2}\right|}{(\Delta x)^2 \epsilon_0 }  }. 
\end{equation} 
For time dependent problem in order to produce a diffusion coefficient 
that reflects the length scales needed over all time we effectively use the infinity norm or:
\begin{equation}
k=
\sqrt{    \frac{\gamma \, \max_n\left\{ \left|\frac{\partial^2 \psi^n}{\partial \xi^2}\right|\right\}}{(\Delta x)^2 \,\epsilon_0 }  }. 
\end{equation} 
Now, introducing the maximum and minimum edge lengths $h_{max}$ and $h_{min}$ one obtains:
\begin{equation}
k=\min\left\{ 
\max\left\{
\sqrt{    \frac{\gamma \,\max_n\left\{ \left| \frac{\partial^2 \psi^n}{\partial \xi^2} \right| \right\}}{(\Delta x)^2 \,\epsilon_0 }  }, 
\frac{1}{h_{max}} 
\right\}, 
\frac{1}{h_{min}} 
\right\}
. 
\end{equation} 
Now using the approximations in 1D: 
\begin{equation}
\frac{\partial^2 \psi^n}{\partial \xi^2}
\approx 
\frac{2\psi_{i,j}^n -\psi_{i+1}^n -\psi_{i-1}^n}
{(\Delta \xi^{i}_{i+1})^2}.
\end{equation} 
In 2D, for the shallow water equations, this becomes: 
\begin{eqnarray}
k_{i,j}^{li,lj}&=& 
\sqrt{ 
\frac{\gamma \max_n\{\vert\nabla^2_\xi \psi^n\vert_{i,j}^{li,lj}\} }{ \frac{1}{2}( (\Delta x_{i,j}^{li,lj})^2+ (\Delta x_{i,j}^{-li,-lj})^2) \,  \epsilon_{0i,j}^{li,lj}  }
} \nonumber\\[2mm]
&\approx& 
\sqrt{ 
\frac{\gamma\max_n\{\vert 2\psi_{i,j}^n -\psi_{i+li,j+lj}^n -\psi_{i-li,j-lj}^n\vert\}/(\Delta \xi^{li,lj})^2}{ \frac{1}{2}( (\Delta x_{i,j}^{li,lj})^2+ (\Delta x_{i,j}^{-li,-lj})^2) \epsilon_{0i,j}^{li,lj}  } 
} \; \; \; \forall \, li,lj \in\{-1,0,1\}, \label{k-dff-diffusion-error-discretised-2}
\end{eqnarray}
in which $li,lj$ represents the direction of diffusion in $k_{i,j}^{li,lj}$ and $i,j$ the 
node on which the diffusion coefficient is centred. Also, 
\begin{eqnarray}
(\Delta \xi^{li,lj})^2=1, && \text{if}\; \vert li\vert+\vert lj\vert=1; \\
(\Delta \xi^{li,lj})^2=2, && \text{if}\; \vert li\vert+\vert lj\vert=2,
\end{eqnarray}
and $\Delta x_{i,j}^{li,lj}$ and $\Delta \xi^{li,lj}$ are the edge 
length scales (grid spacing) in the real (distorted) and local (uniform non-distorted) coordinated system 
in the direction associated with the stencil $li,lj$. 
$\epsilon_{0i,j}^{li,lj}$ represents the desired interpolation error. 
In addition, the distance between neighbouring nodes is calculated as follows: 
\begin{equation}
\Delta x_{i,j}^{li,lj} = \vert\vert \bm{x}_{i,j} - \bm{x}_{i+li,j+lj}\vert\vert_p,
\end{equation}
in which the distance or $p=2$ is generally used; however, $p=1$ or $p=\infty$ are more
computationally efficient. The $p$-norms are calculated using:
\begin{eqnarray}
\vert\vert \bm{x}_{i,j} - \bm{x}_{i+li,j+lj}\vert\vert_1 
&=& 
\vert {x}_{i,j} - {x}_{i+li,j+lj}\vert 
+ \vert {y}_{i,j} - {y}_{i+li,j+lj}\vert , 
\label{1-norm}\\
\vert\vert \bm{x}_{i,j} - \bm{x}_{i+li,j+lj}\vert\vert_2 
&=& 
\sqrt{ 
( {x}_{i,j} - {x}_{i+li,j+lj})^2 
+ ( {y}_{i,j} - {y}_{i+li,j+lj})^2 }, 
\label{2-norm}\\
\vert\vert \bm{x}_{i,j} - \bm{x}_{i+li,j+lj}\vert\vert_\infty 
&=& 
\max\{ \vert {x}_{i,j} - {x}_{i+li,j+lj}\vert ,
 \vert {y}_{i,j} - {y}_{i+li,j+lj}\vert \}. 
\label{inf-norm}
\end{eqnarray}
With the inclusion of a maximum and minimum length scale (${{h_{max}}_{i,j}^{li,lj}}$, ${{h_{min}}_{i,j}^{li,lj}}$ 
respectively), the approximation for $k_{i,j}^{li,lj}$ from Equation~\eqref{k-dff-diffusion-error-discretised-2} can be written as 
\begin{equation}
\min\left\{
\max\left\{ \quad
\sqrt{
\frac{ \gamma 
\max_n\{\vert 2\psi_{i,j}^n -\psi_{i+li,j+lj}^n -\psi_{i-li,j-lj}^n\vert\}/ (\Delta \xi_{i,j}^{li,lj})^2}{ \max\{ 
\frac{1}{2}( (\Delta x_{i,j}^{li,lj})^2+ (\Delta x_{i,j}^{-li,-lj})^2)
, (\Delta x_{min})^2 \}  \, \epsilon_{0i,j}^{li,lj}   }  } \;\;\; , 
\frac{1}{{h_{max}}_{i,j}^{li,lj}} \right\},  \frac{1}{{h_{min}}_{i,j}^{li,lj}} \right\}  , 
\label{final_error_diffusivity}
\end{equation}
in which we have introduced $\Delta x_{min}$ as the minimum element length or node to the spacing of the neighbouring nodes that we want in the grid. This stops the method from producing increasingly fine grids. 
After calculating these values of $k_{i,j}^{li,lj}$ they are smoothed using the consistent mass matrix $S_m = M_{\xi L}^{-1} M_\xi$ (the row sum lumped and consistent mass matrices $M_{\xi L}$,  $M_\xi$ are evaluated in the uniform domain $\Omega_\xi$ and thus these matrices do not depend on the position $i,j$ due to the grid being uniform), that is: 
\begin{equation}
k_{i,j}^{li,lj} \leftarrow \sum_{jj=-1}^{1}\sum_{ii=-1}^{1}  {S_m}^{ii,jj}  k_{i+ii,j+jj}^{li,lj}.
\end{equation}
This represents a single iteration of the smoothing. We typically use two smoothing iterations.  
For multiple solution variables for which we want to adapt the mesh, we typically repeat the 
above operations to calculate the diffusion coefficient and simply take the maximum diffusion coefficient 
at each node $i,j$ and in each direction $li,lj$. This will then produce a distorted grid that is able 
to represent all the solution variables as best it can. 

For simplicity, the desired interpolation error $\epsilon^{i,j}_{0 i+li,j+lj}$ can be defined isotropically (the same in each direction) and uniform across the whole domain, for example (although many variants 
of this simplification exist) with: 
\begin{equation}
{\epsilon_0}_{i,j}^{li,lj}  = \epsilon_0, 
\end{equation}
in which $\epsilon_0$ is a scalar. 
The same is true of the minimum and maximum edge length scale  scalars $h_{max}$ and $h_{min}$ 
and thus: 
\begin{equation}
{h_{min}}_{i,j}^{li,lj}  = h_{min},
\end{equation}
\begin{equation}
{h_{max}}_{i,j}^{li,lj}  = h_{max}.
\end{equation}

In addition, the curvature of the coordinate $x$ can be approximated as (as used in Equation~\eqref{final_error_diffusivity}):
\begin{equation}
\begin{split}
\nabla_\xi^2 x \vert_{i,j}^{li,lj} = -(2x_{i,j} -x_{i+li,j+lj} -x_{i-li,j-lj})/ (\Delta \xi_{i,j}^{li,lj})^2,  \\
 \;\;\; \forall \in\{-1,0,1\} \;\;\; {\text{other than}} \;\;\; li=lj =0. 
\end{split}
\end{equation}
In order to evaluate $\nabla_\xi k \nabla_\xi x \vert_{i,j}^{li,lj}$ we use the identity: 
\begin{equation}
\begin{split}
(\nabla_\xi k \nabla_\xi x )\vert_{i,j}^{li,lj} = 
\frac{1}{2} 
\left( 
k_{i,j}^{li,lj} \nabla_\xi^2 x \vert_{i,j}^{li,lj} 
+ \nabla_\xi^2( k x ) \vert_{i,j}^{li,lj}
- x_{i,j}  \nabla_\xi^2 k \vert_{i,j}^{li,lj}
\right), \\
 \;\;\; \forall \in\{-1,0,1\} \;\;\; {\text{other than}} \;\;\; li=lj =0. 
\end{split}
\end{equation}
The diffusion operator $\nabla_\xi k \nabla_\xi$ at node $i,j$ is then discretised using:
\begin{equation}
\nabla_\xi k \nabla_\xi x\vert_{i,j} \approx 
\sum_{li=-1}^{1} \; \sum_{lj=-1}^{1} 
(\nabla_\xi k \nabla_\xi x )\vert_{i,j}^{li,lj} .
\end{equation}
An alternative, computationally efficient  version of this operator 
can also be used:
\begin{equation}
\nabla_\xi k \nabla_\xi x\vert_{i,j} \approx 
(\nabla_\xi k \nabla_\xi x )\vert_{i,j}^{1,0} +
(\nabla_\xi k \nabla_\xi x )\vert_{i,j}^{0,1} .
\label{comp-efficient-diff} 
\end{equation}
This has the disadvantage that the cross terms that are used for controlling 
the element quality then require additional computational overheads to introduce, 
so we generally use this efficient approach without the element quality terms. 

 The resulting pseudo time-stepping approach discretised using forward Euler in time then becomes 
for each of the nodal/cell coordinates:  
\begin{equation}
x_{i,j} \leftarrow x_{i,j}
+ \Delta t^* \sum_{li=-1}^{1} \; \sum_{lj=-1}^{1} 
(\nabla_\xi k \nabla_\xi x )\vert_{i,j}^{li,lj} ,
\end{equation}
\begin{equation}
y_{i,j} \leftarrow y_{i,j}
+ \Delta t^* \sum_{li=-1}^{1} \; \sum_{lj=-1}^{1} 
(\nabla_\xi k \nabla_\xi y )\vert_{i,j}^{li,lj} ,
\end{equation}
which forms an iterative method within a time step that starts 
from $x_{i,j} = x_{i,j}^n$, $y_{i,j} = y_{i,j}^n$.  
The $\Delta t^*$ controls the pseudo time step size or step length 
associated with grid movement. For stability we limit this pseudo 
time step such that:
\begin{equation}
{k_{max-stable}}_{i,j}^{li,lj} = \gamma_{stab} \frac{(\Delta x_{i,j}^{li,lj})^2 }{\Delta t^*} , 
\end{equation}
in which the factor $\gamma_{stab}=0.25$ is typically used for safety rather than the strict value of 1 
and ${k_{max-stable}}_{i,j}^{li,lj}$ is the maximum stable diffusion value for a given 
pseudo time step $\Delta t^*$. Thus for stability the maximum stable diffusion size is: 
\begin{equation}
 \Delta t^* \leqslant \gamma_{stab} \frac{(\Delta x_{i,j}^{li,lj})^2 }{k_{i,j}^{li,lj}}, 
 \; \forall i\in\{1,2,...,N_x\}, \forall j\in\{1,2,...,N_y\} 
\end{equation}
\begin{equation}
 \text{and} \; 
 \forall li\in\{-1,0,1\}, \forall lj\in\{-1,0,1\}. 
\end{equation}
To form the most efficient algorithm, it may be advantageous to have a non-uniform 
step size with: 
\begin{equation}
 \Delta {t^*}_{i,j}^{li,lj} = \gamma_{stab} \frac{(\Delta x_{i,j}^{li,lj})^2 }{k_{i,j}^{li,lj}}.
 \label{d t*}
\end{equation}
It is convenient to have a square in $(\Delta x_{i,j}^{li,lj})^2$ as it avoids an additional 
square route in the calculation of $\Delta x_{i,j}^{li,lj}$ is a 2-norm or distance norm is used 
in its calculation. 
If one wants to take small pseudo-time steps because it helps with the non-linearity or produce a small 
grid velocity then one can use a much smaller $\gamma_{stab}$ (e.g. $\gamma_{stab}=0.01$) and not 
have too many iterations per time step. 
In this case the nodal updates become:
\begin{equation}
x_{i,j} \leftarrow x_{i,j}
+  \sum_{li=-1}^{1} \; \sum_{lj=-1}^{1} 
\Delta {t^*}_{i,j}^{li,lj} (\nabla_\xi k \nabla_\xi x )\vert_{i,j}^{li,lj} ,
\end{equation}
\begin{equation}
y_{i,j} \leftarrow y_{i,j}
+  \sum_{li=-1}^{1} \; \sum_{lj=-1}^{1} 
(\Delta {t^*}_{i,j}^{li,lj}) (\nabla_\xi k \nabla_\xi y )\vert_{i,j}^{li,lj} ,
\end{equation}
in which this iteration starts  
from initial guesses for $x_{i,j} $, $y_{i,j} $ from a uniform grid spacing, say. 
The main problem with the above approach is the calculation of 
$ \Delta {t^*}_{i,j}^{li,lj})^n$ in Equation~\eqref{d t*} as it has a the diffusion coefficient 
on the denominator which can result in too large a time step size. The next section 
fixes this issue.

\par\noindent
{\it{Choosing the time pseudo time step size using the analogy with Jacobi iteration method }} 
\nopagebreak
\par
\nopagebreak
Here it is computationally more efficient to 
use the discretisation of the diffusion operator based on a 
simple 1D finite difference scheme to obtain the nodal coordinate updates:


\begin{align}
x_{i,j} &\leftarrow x_{i,j}
-  
\frac{1}{D_{i,j} }
\sum_{li=-1}^{1} \; \sum_{lj=-1}^{1} 
\left(
\frac{1}{2} (k_{i,j} + k_{i+li,j+lj} ) \frac{(x_{i,j} - x_{i+li,j+lj})}{(\Delta\xi_{i,j}^{li,lj})^2} \right. \notag\\
& \qquad \qquad \qquad \qquad \qquad \qquad \qquad + \left. \frac{1}{2} (k_{i,j} + k_{i-li,j-lj} ) \frac{(x_{i,j} - x_{i-li,j-lj})}{(\Delta\xi_{i,j}^{-li,-lj})^2} \right) ,\\
y_{i,j} &\leftarrow y_{i,j}
-  
\frac{1}{D_{i,j} }
\sum_{li=-1}^{1} \; \sum_{lj=-1}^{1} 
\left(
\frac{1}{2} (k_{i,j} + k_{i+li,j+lj} ) \frac{(y_{i,j} - y_{i+li,j+lj})}{(\Delta\xi_{i,j}^{li,lj})^2} \right.\notag\\
& \qquad \qquad \qquad \qquad \qquad \qquad \qquad + \left. \frac{1}{2} (k_{i,j} + k_{i-li,j-lj} ) \frac{(y_{i,j} - y_{i-li,j-lj})}{(\Delta\xi_{i,j}^{-li,-lj})^2} 
\right) ,
\end{align}
with
\begin{equation}
D_{i,j} = w_d 
\sum_{li=-1}^{1} \; \sum_{lj=-1}^{1} 
\left( \frac{1}{2} 
\frac{(k_{i,j} + k_{i+li,j+lj} )}{(\Delta\xi_{i,j}^{li,lj})^2}  + 
\frac{1}{2} \frac{(k_{i,j} + k_{i-li,j-lj} )}{(\Delta\xi_{i,j}^{-li,-lj})^2} \right),
\end{equation}
in which $w_d>1$ under relaxes the Jacobi iteration and $w_d=1$ is the Jacobi method. 
This corresponds to a pseudo time step of 
\begin{equation}
(\Delta t^*_{i,j}) = 
\frac{1}{D_{i,j}} . 
\end{equation}

The more computationally efficient version of the equations above using 
Equation~\eqref{comp-efficient-diff} as the basis are:
\begin{align}
x_{i,j}^{n+1} &\leftarrow x_{i,j}^{n+1}
-  
\frac{1}{D_{i,j}^{n}} 
\sum_{ll=-1}^{0} 
\left(
\frac{1}{2} (k^n_{i,j} + k^n_{i+d(ll+1),j+d(ll-0)} ) \frac{(x^n_{i,j} - x^n_{i+d(ll+1),j+d(ll-0)})}{(\Delta\xi_{i,j}^{d(ll+1),d(ll-0)})^2} \right. \notag\\
& \qquad \qquad \qquad \qquad  + \left. \frac{1}{2} (k^n_{i,j} + k^n_{i-d(ll+1),j-d(ll-0)} ) \frac{(x^n_{i,j} - x^n_{i-d(ll+1),j-d(ll-0)})}{(\Delta\xi_{i,j}^{-d(ll+1),-d(ll-0)})^2} \right) ,\\
y_{i,j}^{n+1} &\leftarrow y_{i,j}^{n+1}
-  
\frac{1}{D_{i,j}^{n}} 
\sum_{ll=-1}^{0} 
\left(
\frac{1}{2} (k^n_{i,j} + k^n_{i+d(ll+1),j+d(ll-0)} ) \frac{(y^n_{i,j} - y^n_{i+d(ll+1),j+d(ll-0)})}{(\Delta\xi_{i,j}^{d(ll+1),d(ll-0)})^2} \right. \notag\\
& \qquad \qquad \qquad \qquad  + \left. \frac{1}{2} (k^n_{i,j} + k^n_{i-d(ll+1),j-d(ll-0)} ) \frac{(y^n_{i,j} - y^n_{i-d(ll+1),j-d(ll-0)})}{(\Delta\xi_{i,j}^{-d(ll+1),-d(ll-0)})^2} \right) ,
\end{align}
in which $d(l)=1$ if $l=0$ else $d(l)=0$ and 
with
\begin{equation}
D_{i,j}^{n} = w_d 
\sum_{ll=-1}^{0}  
\left( \frac{1}{2} \frac{(k^n_{i,j} + k^n_{i+d(ll+1),j+d(ll-0)} ) }{(\Delta\xi_{i,j}^{d(ll+1),d(ll-0)})^2} + 
\frac{1}{2} \frac{(k^n_{i,j} + k^n_{i-d(ll+1),j-d(ll-0)} )}{(\Delta\xi_{i,j}^{-d(ll+1),-d(ll-0)})^2} \right) .
\end{equation}

\par\noindent
\textit{Controlling the element quality}

\begin{figure}[htbp]
\centering
\includegraphics[width=0.9\textwidth,trim=30mm 40mm 40mm 20mm,clip]{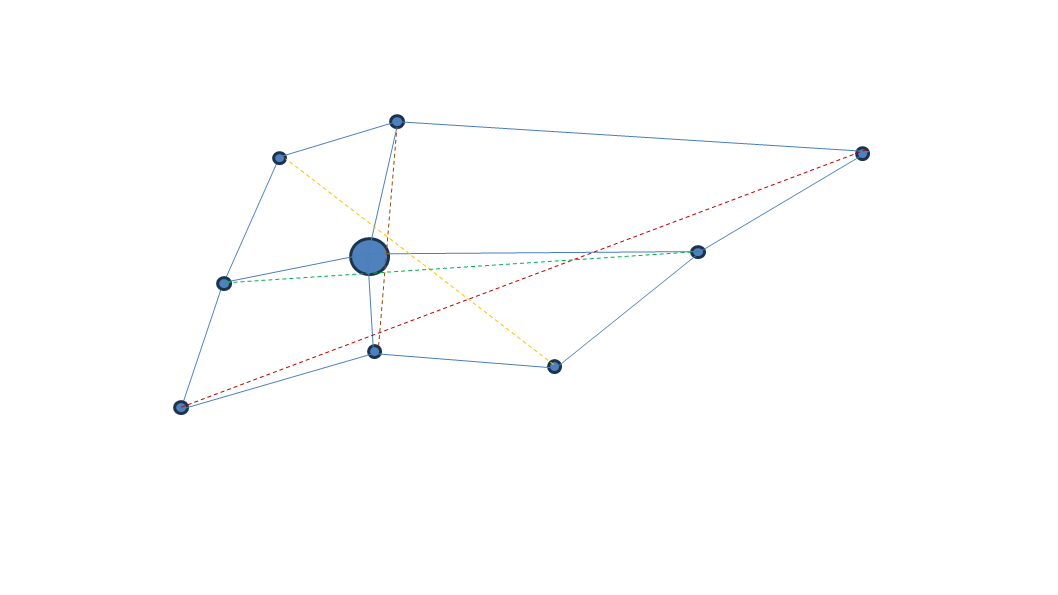}
\caption{Diagram showing how the grid or stencil lengths are used when combined with Equations~\eqref{xy-plane-color1} and~\eqref{xy-plane-color2} in order to determine the skewness of the grid or stencil. The equations are coloured to correspond to the lengths shown in the diagram.}
\label{grid-skew}
\end{figure}

The element quality is controlled through diagonal diffusion terms and in 2D there is one plane of two diagonal diffusion coefficients and thus 2 possible additional diffusion terms.  
The amendments of the diffusion coefficient are as follows (leaving out the superscript $n$ for clarity):  
\begin{equation}
\begin{split}
k_{i,j}^{1,1} \leftarrow k_{i,j}^{1,1} \quad\quad \quad\quad \quad\quad \quad\quad \quad\quad \quad\quad \quad\quad \quad\quad \quad\quad \quad\quad             \\
+ \alpha_1 \max\{k_{i,j}^{1,0}, k_{i,j}^{0,1} \} \min\{200,  \max\{ {\textcolor{red}{ d2^{i-1,j-1}_{i+1,j+1} }} / {\textcolor{orange}{ d2^{i-1,j+1}_{i+1,j-1} }} -4,0\}   \} \\
+ \alpha_2 \max\{k_{i,j}^{1,0}, k_{i,j}^{0,1} \} \min\{200,  \max\{ \textcolor{red}{ d2^{i-1,j-1}_{i+1,j+1} } / 
(2\max\{  \textcolor{green}{ d2^{i-1,j+0}_{i+1,j+0} }, \textcolor{brown}{ d2^{i+0,j-1}_{i+0,j+1} }\})  -1,0\}   \} ,
\end{split}
\label{xy-plane-color1} 
\end{equation}
\begin{equation}
\begin{split}
k_{i,j}^{1,-1} \leftarrow k_{i,j}^{1,-1} \quad\quad \quad\quad \quad\quad \quad\quad \quad\quad \quad\quad \quad\quad \quad\quad \quad\quad \quad\quad                \\
+ \alpha_1 \max\{k_{i,j}^{1,0}, k_{i,j}^{0,1} \} \min\{200,  \max\{\textcolor{orange}{d2^{i-1,j+1}_{i+1,j-1}} / \textcolor{red}{ d2^{i-1,j-1}_{i+1,j+1} }-4,0\}   \} \\
+ \alpha_2 \max\{k_{i,j}^{1,0}, k_{i,j}^{0,1} \} \min\{200,  \max\{\textcolor{orange}{d2^{i-1,j+1}_{i+1,j-1}} / 
(2\max\{   \textcolor{green}{d2^{i-1,j+0}_{i+1,j+0}}, \textcolor{brown}{d2^{i+0,j-1}_{i+0,j+1}} \})  -1,0\}   \} ,
\end{split}
\label{xy-plane-color2} 
\end{equation}
in which the distances squared across the grid stencil (say 3$\times$3 stencil) are defined: 
\begin{equation}
\begin{split}
d2^{i+li,j+lj}_{i+lii,j+ljj} = (x_{i+li,j+lj} - x_{i+lii,j+ljj} )^2  + (y_{i+li,j+lj} - y_{i+lii,j+ljj} )^2 
                                        . 
\end{split}
\end{equation}
The default values are $\alpha_1=\alpha_2=0.25$. The values $-4$ and $-1$ in the equations above are there to avoid the first term and the second term having too much influence when the aspect ratios are less than~$2$ and~$1$ respectively. The distances across a stencil are represented with different coloured terms within the equations above and are shown diagrammatically in 
Figure~\ref{grid-skew}. In the absence of a suitable boundary values of the diffusion coefficients that controls the skewness of the mesh we set all the additions to the diffusion coefficient to zero on the boundary and one node into the domain. 

\begin{algorithm}[H]
\caption{Static mesh adaptation via diffusion and interpolation error}
\label{alg:static-mesh-adaptation}
\begin{algorithmic}[1]
\State \textbf{Input:} Initial grid $\bm{x}^0$, solution history $\{\psi^n\}$, 
target error $\epsilon_0$, bounds $h_{\min}, h_{\max}$
\State \textbf{Initialise:} Set $\bm{x} \leftarrow \bm{x}^0$
\State \textbf{Step 1: Compute interpolation error $E_{i,j}^{li,lj}$}
\For{each node $(i,j)$ and stencil direction $(li,lj)$}
    \State Compute second derivative for three variables $\psi_{1}=h+H$, $\psi_{2}=(h+H)u$ and $\psi_{3}=(h+H)v$ %
    
    \begin{equation}
    \frac{\partial^2 \psi_{N}^{n}}{\partial \xi^2}
    \approx
    \frac{2\psi_{N\,i,j}^n - \psi_{N\,i+li,j+lj}^n - \psi_{N\,i-li,j-lj}^n}
    {(\Delta \xi^{li,lj})^2} \;\;\; \forall N \in\{1,2,3\} \;\;\;  \forall \, li,lj\in\{-1,0,1\},
    \nonumber
    \end{equation}
    \vspace{-1.5em}
    \State Compute maximum over time 
    \[
    E_{i,j}^{li,lj} = \left( \max_n \left\{\max_N \left|
    \frac{\partial^2 \psi_{N}^n}{\partial \xi^2}
    \right|\right\} \right)^{\frac{1}{R}} \;\;\; \forall N \in\{1,2,3\}
    \]
\EndFor
\vspace{-1.5em}
\State \textbf{Step 2: Compute desired mesh spacing $h_{i,j}^{li,lj}$}
\For{each node $(i,j)$ and direction $(li,lj)$}
    \[
    h_{i,j}^{li,lj} =
    \sqrt{
    \frac{(\Delta x_{i,j}^{li,lj})^2 \, \epsilon_0}
    {\gamma \, E_{i,j}^{li,lj}}
    } \;\;\; \forall \, li,lj\in\{-1,0,1\} \;\;\; h_{i,j}^{li,lj} \leftarrow \min\{\max\{h_{i,j}^{li,lj}, h_{\min}\}, h_{\max}\}
    \]
\EndFor
\vspace{-1.5em}
\State \textbf{Step 3: Define diffusion coefficient}
\For{each node $(i,j)$ and direction $(li,lj)$}
    \begin{equation}
    k_{i,j}^{li,lj} = \frac{1}{h_{i,j}^{li,lj}}
    \nonumber
    \end{equation}
\EndFor
\vspace{-1.5em}
\State \textbf{Step 4: Smooth diffusion coefficient}
\For{each node $(i,j)$}
    \[
    k_{i,j}^{li,lj} \leftarrow
    \sum_{ii=-1}^{1}\sum_{jj=-1}^{1}
    S_m^{ii,jj} \, k_{i+ii,j+jj}^{li,lj} \;\;\;  S_{m} \leftarrow M_{\xi}
    \] 
\EndFor
\vspace{-1.5em}
\State \textbf{Step 5: Interpolate $k$ onto physical grid}
\State Interpolate diffusion coefficients from computational grid to current grid positions
\State \textbf{Step 6: Solve diffusion equation (pseudo-time)}
\Repeat
    \For{each node $(i,j)$}
        \State Compute diffusion operator $(\nabla_\xi k \nabla_\xi \bm{x})_{i,j}$
        \State Update node positions $\bm{x}_{i,j} \leftarrow \bm{x}_{i,j} + \Delta t^* (\nabla_\xi k \nabla_\xi \bm{x})_{i,j}$
    \EndFor

    \State Update $\Delta t^*$ for stability if required
    \State Re-interpolate $k$ onto updated grid
\Until{convergence of $\bm{x}$}
\State \textbf{Output:} Adapted mesh $\bm{x}$

\end{algorithmic}
\end{algorithm}

\paragraph{Remarks}

This formulation ensures that regions with large solution curvature (e.g. near shocks, wetting–drying fronts, or strong velocity gradients) receive higher resolution, while smooth regions are coarsened. The resulting diffusion-driven mesh adaptation provides a physically consistent and computationally efficient mechanism for generating distorted structured grids.

\section{Results}
\label{results} 
\begin{figure}[H]
    \centering
    \begin{subfigure}[t]{0.5\textwidth}
    \includegraphics[width=\textwidth]{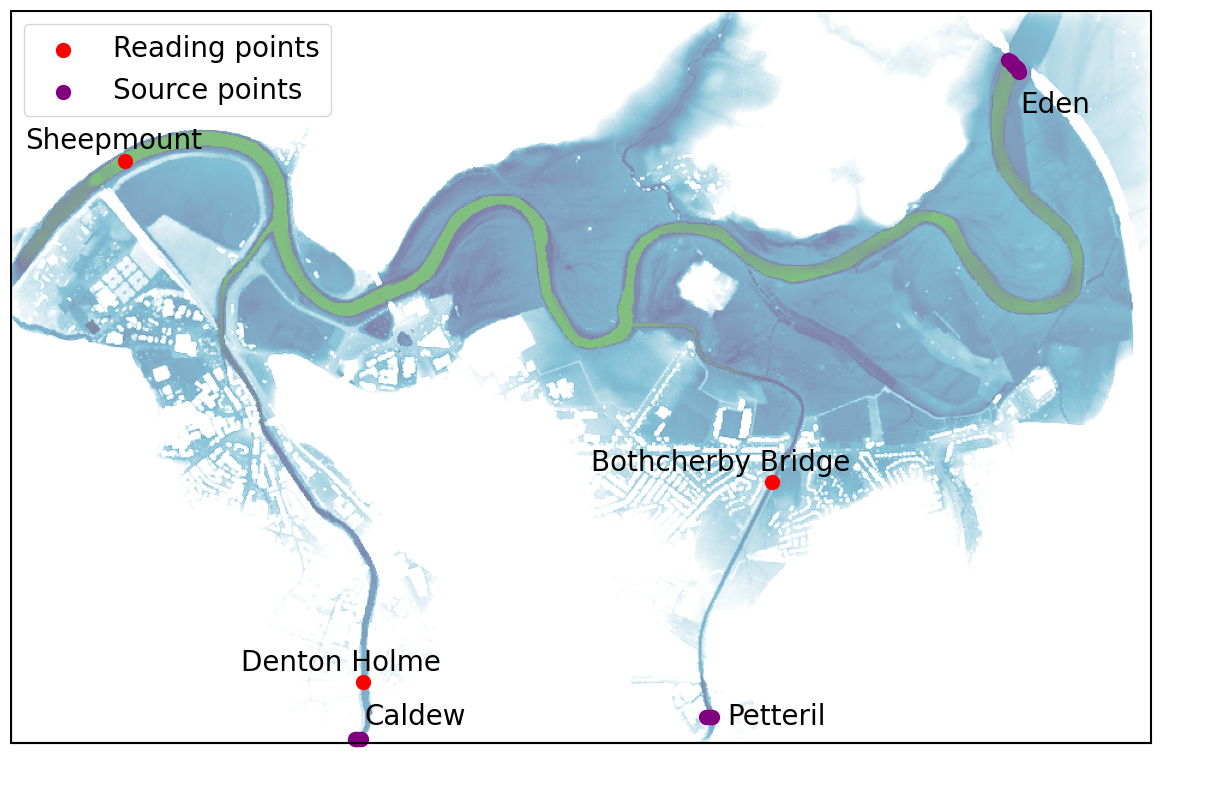}
    \caption{}
    \end{subfigure}
    \begin{subfigure}[t]{0.4\textwidth}
    \includegraphics[width=\textwidth]{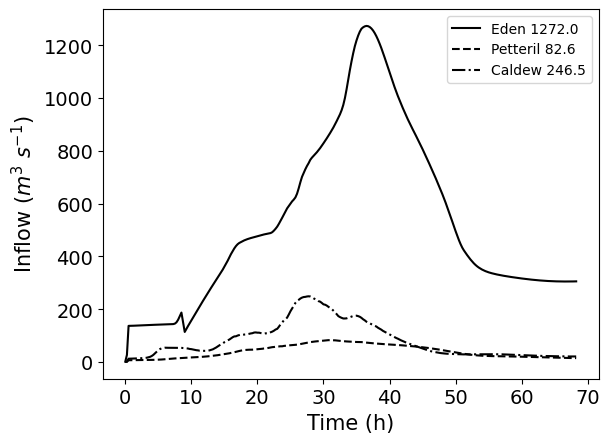}
    \caption{}
    \end{subfigure}
    \begin{subfigure}[t]{0.49\textwidth}
    \includegraphics[width=\textwidth]{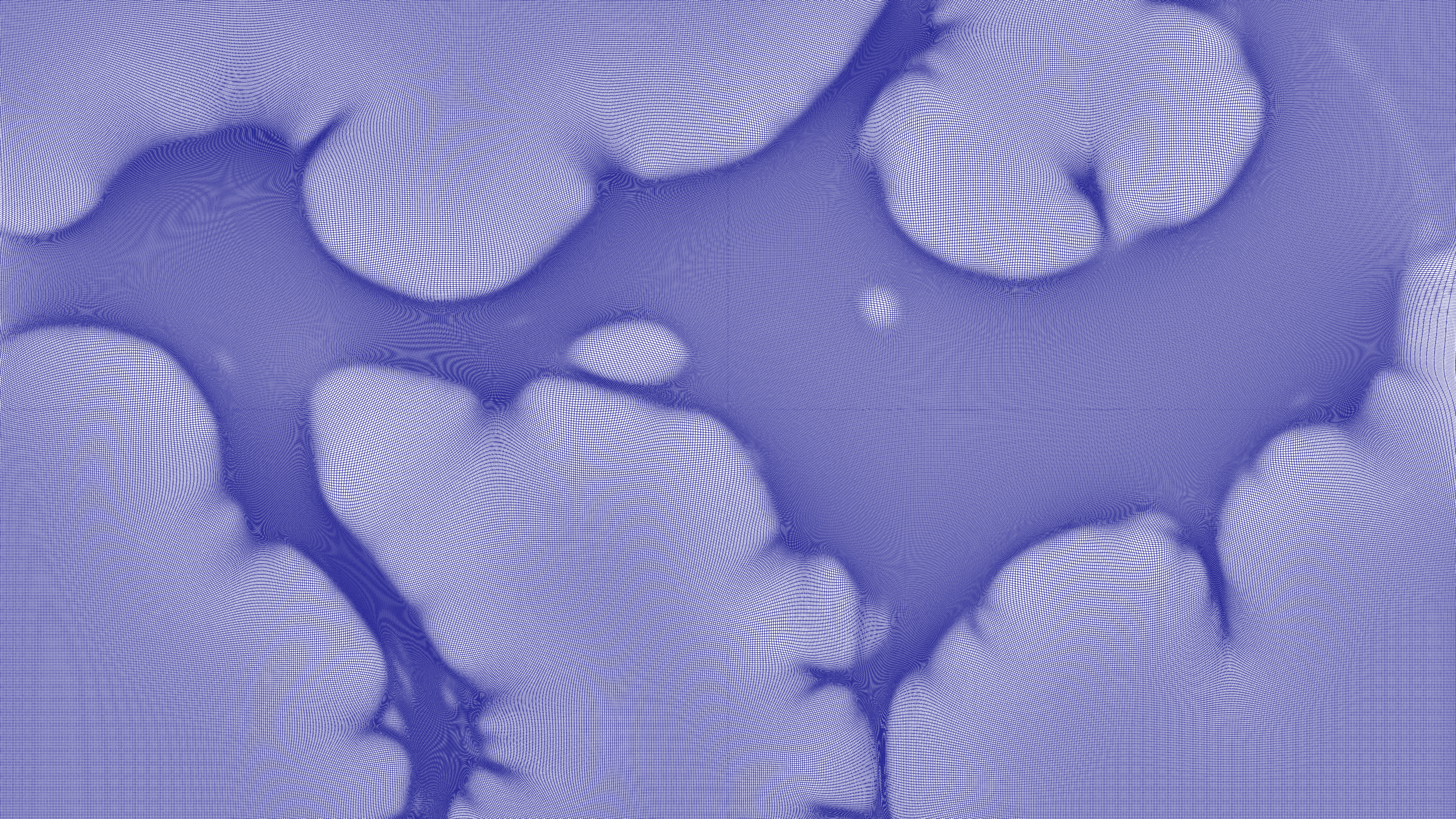}
    \caption{}
    \end{subfigure}
    \begin{subfigure}[t]{0.49\textwidth}
    \includegraphics[width=\textwidth]{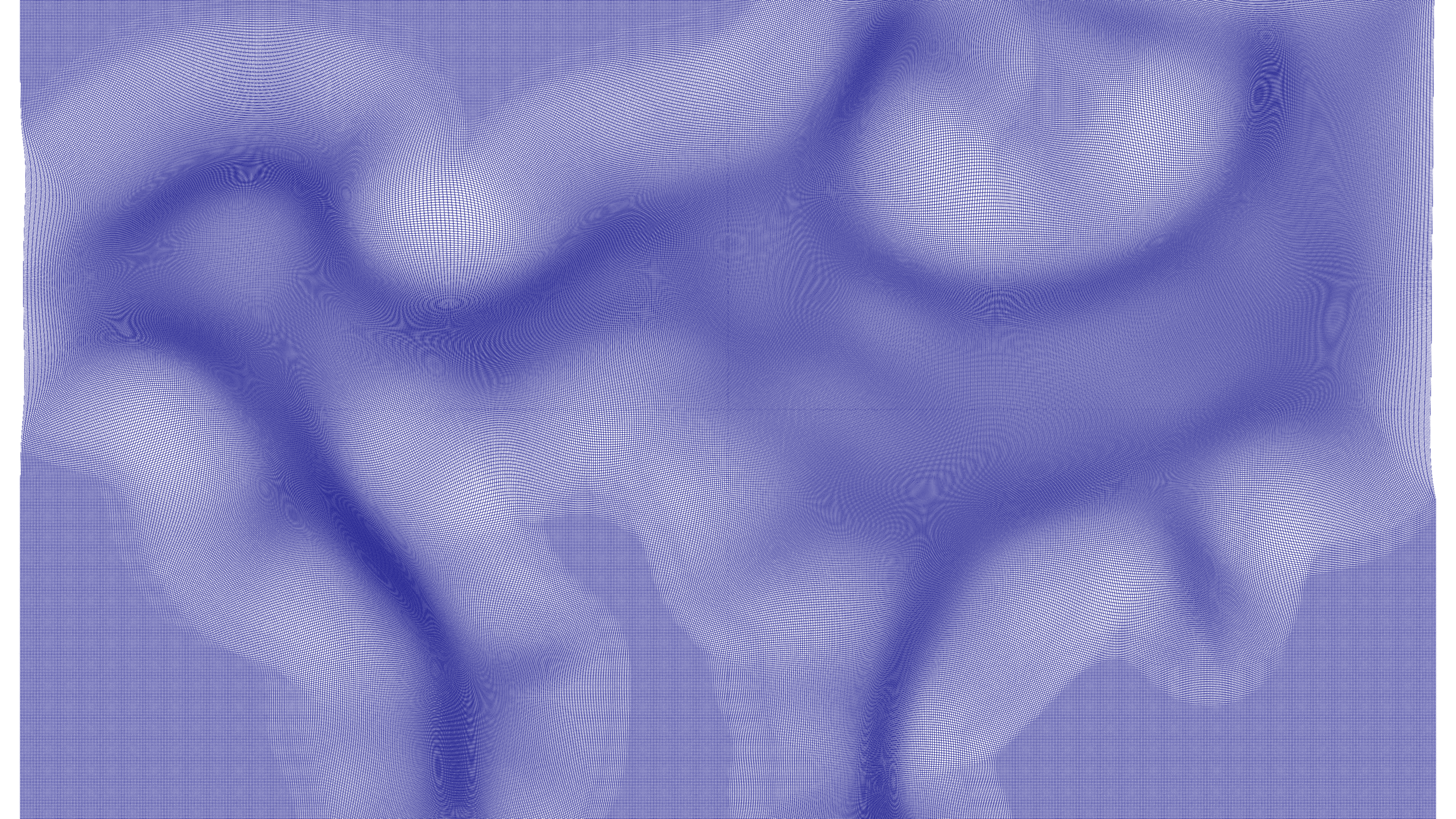}
    \caption{}
    \end{subfigure}
    \caption{Carlisle 2005 flooding: (a)~the flood extent when the Eden inlet was at its maximum flow rate (\SI{1200}{m/s}), with  positions and names of the sampling points indicated; (b)~the three rivers and topology of the area being modelled; (c) and (d)~distorted grid (Models C-F described in Table~\ref{tab:numerical_set_up})) generated by the moving the nodes according to a diffusion equation in response to the interpolation error.}
    \label{fig:DSWE_mesh}
\end{figure}

To validate the proposed distorted structured grid formulation within the \NP  framework, the Carlisle 2005 flooding event is simulated. The results are compared against observational data and numerical predictions obtained using \NP on regular grids. Figure~\ref{fig:DSWE_mesh} presents an overview of the case setup. Figure~\ref{fig:DSWE_mesh}~(a) shows the flood extent corresponding to time when the Eden inlet discharge was at a maximum (\SI{1200}{m^3/s}), which determines the  value of our particular mass source tensor $\bm{s_h}^{n+1}$ in Equation~\eqref{Petrov-h}, together with the locations of the sampling points (Sheepmount, Botcherby Bridge and Denton Holme). Figure~\ref{fig:DSWE_mesh}~(b) illustrates the inflow rates of the three rivers (Eden, Petteril and Caldew). Figure~\ref{fig:DSWE_mesh}~(c) and (d) display the two distorted structured grids generated using a diffusion equation to statically adapt the resolution of the structured grid. The distorted grid picks out the river channels and regions of strong hydraulic gradients, demonstrating the capability of the diffusion-based static mesh adaptation to concentrate resolution in important areas while preserving the bathymetry without loss of physical information. The details of the numerical setup and distorted mesh generation are summarised in Tables~\ref{tab:numerical_set_up} and \ref{tab:SWE_BCs}. Six models (C–-H) are tested on a range of distorted meshes, while two models (A and B) are evaluated on regular grids. All other numerical settings are kept identical to ensure a consistent comparison. In addition, simulations on a uniformly refined mesh are performed to provide reference (baseline) results.  The grid distortion will not change the original bathymetry (Geometry conservation) thus preserving the bathymetry without loss of physical information and holding this bathymetry information on a uniform structured grid and when needed this information is interpolated onto the distorted grid. 

\begin{table}[htbp]
    \footnotesize
    \centering
    \begin{tabular}{ccccccccc}
    \toprule
    \NP  & ConvFEM & Grid size & Mesh & Adaption ratio & Order of root & Grid points & \multicolumn{2}{c}{Time step (s)} \\
    \multicolumn{1}{c}{(Model)} & & \multicolumn{1}{c}{(m)} & & \multicolumn{1}{c}{($h_{max}/h_{min}$)} & \multicolumn{1}{c}{($R$)} & \multicolumn{1}{c}{($nx \times ny$)} & \multicolumn{1}{c}{Mesh} & \multicolumn{1}{c}{Fluid} \\
    \toprule
    Baseline & Linear & 2.5 & Regular & --- & --- & 1902$\times$1222 & --- & 0.125 \\
    \toprule   
    A & Linear & 5.0 & Regular & --- & --- & 951$\times$611 & --- & 0.25 \\
    B & Quadratic & 5.0 & Regular & --- & --- & 951$\times$611 & --- & 0.25 \\
    C & Linear & 5.0 & Distort & 5.0 & 3 & 951$\times$611 & 0.05 & 0.25 \\
    D & Quadratic & 5.0 & Distort & 5.0 & 3 & 951$\times$611 & 0.05 & 0.25 \\
    E & Linear & 5.0 & Distort & 5.0 & 2 & 951$\times$611 & 0.05 & 0.25 \\
    F & Quadratic & 5.0 & Distort & 5.0 & 2 & 951$\times$611 & 0.05 & 0.25 \\
    G & Linear & 5.0 & Distort & 5.0 & 1 & 951$\times$611 & 0.05 & 0.25 \\
    H & Quadratic & 5.0 & Distort & 5.0 & 1 & 951$\times$611 & 0.05 & 0.25 \\
    \toprule
    \end{tabular}
    \caption{Six \NP models used for the 2D Carlisle flooding simulation. The $R$ refers to the $n-th$ root as described in Algorithm~\ref{alg:static-mesh-adaptation} $E_{i,j}^{li,lj} = \left( \max_n \{\max_N \left|
    \frac{\partial^2 \psi_{N}^n}{\partial \xi^2}
    \right|\} \right)^{\frac{1}{R}}$.}
    \label{tab:numerical_set_up}
\end{table}

\begin{table}[htbp]
    \centering
    \begin{tabular}{lcc}
    \toprule
    Surface & Velocity  & Water depth  \\
            & \multicolumn{1}{c}{$(u,v)^T$} & \multicolumn{1}{c}{$\eta=h+H$}    \\
    \toprule
    West  & $u=0, \partial v / \partial x=0$ & $\eta=0$ \\
    East  & $u=0, \partial v / \partial x=0$ & $\eta=0$ \\
    North & $v=0, \partial u / \partial y=0$ & $\eta=0$ \\
    South & $v=0, \partial u / \partial y=0$ & $\eta=0$ \\
    \toprule
    \end{tabular}
    \caption{Boundary conditions applied in the 2D Carlisle flooding simulation, specifying the velocity components and water depth constraints imposed on each domain boundary.}
    \label{tab:SWE_BCs}
\end{table}

\begin{figure}[H]
    \centering
    \begin{minipage}{1.0\textwidth}
    \centering
    \begin{subfigure}[t]{0.5\textwidth}
    \includegraphics[width=\textwidth]{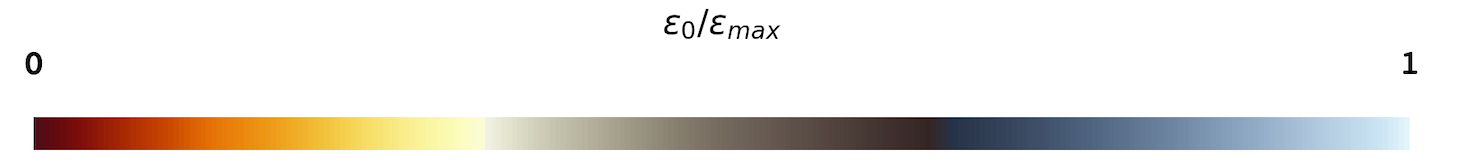}
    \end{subfigure}
    \end{minipage} 
    \begin{subfigure}[t]{0.3\textwidth}
    \includegraphics[width=\textwidth]{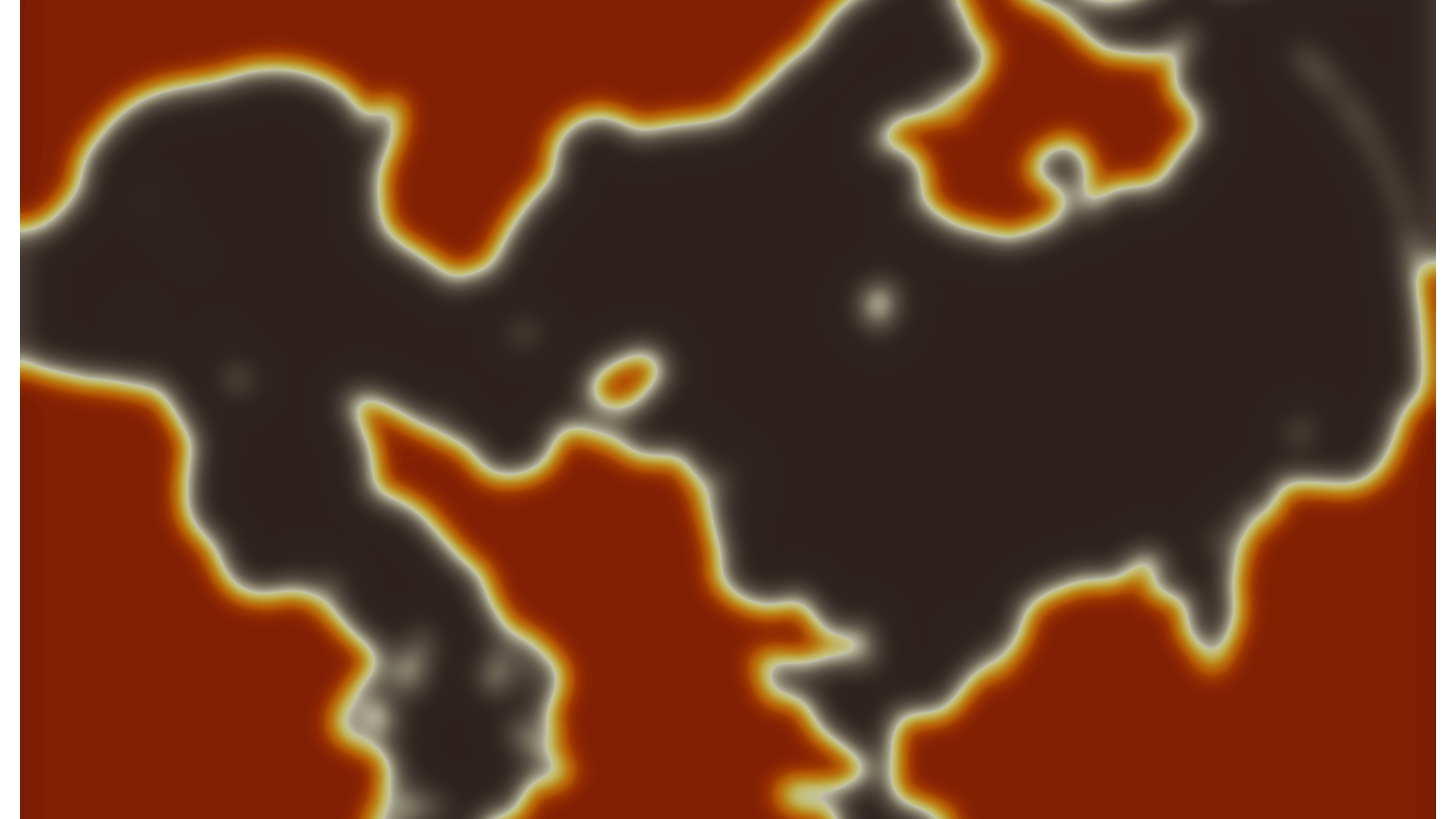}
    \caption{Static interpolate error $R=3$}
    \end{subfigure}
    \begin{subfigure}[t]{0.3\textwidth}
    \includegraphics[width=\textwidth]{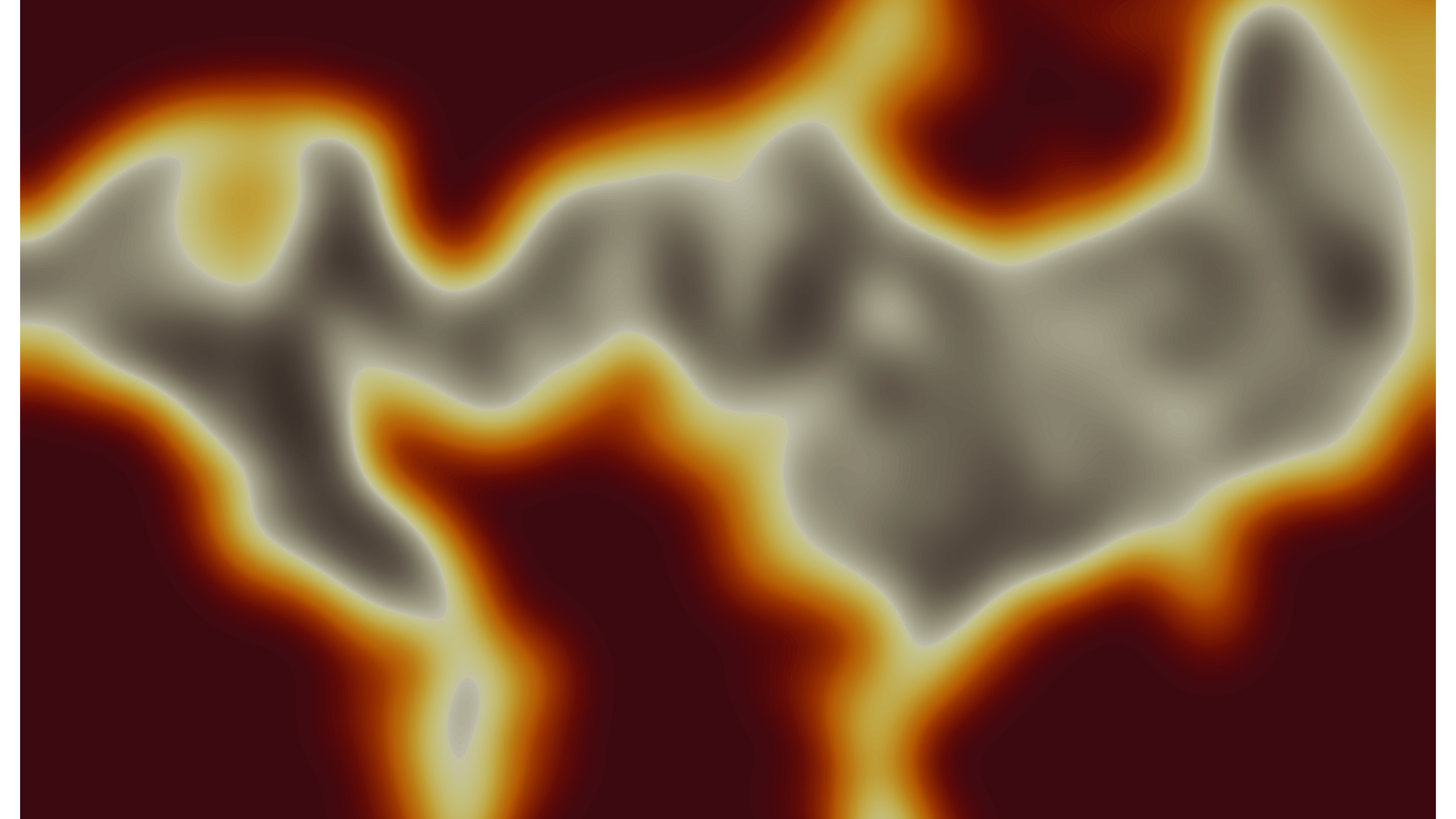}
    \caption{Static interpolate error $R=2$}
    \end{subfigure}
    \begin{subfigure}[t]{0.3\textwidth}
    \includegraphics[width=\textwidth]{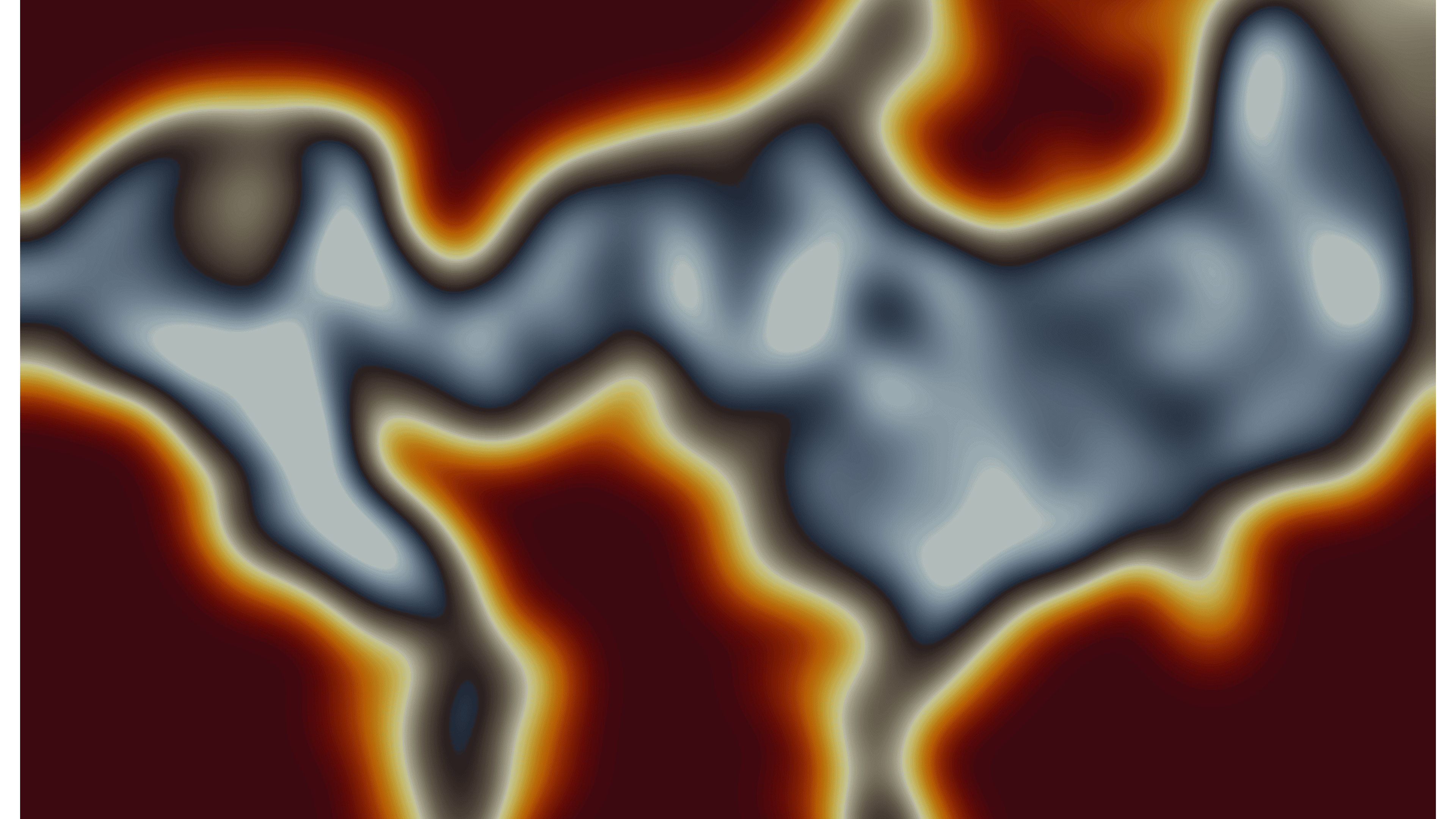}
    \caption{Static interpolate error $R=1$}
    \end{subfigure}

    \begin{subfigure}[t]{0.3\textwidth}
    \includegraphics[width=\textwidth]{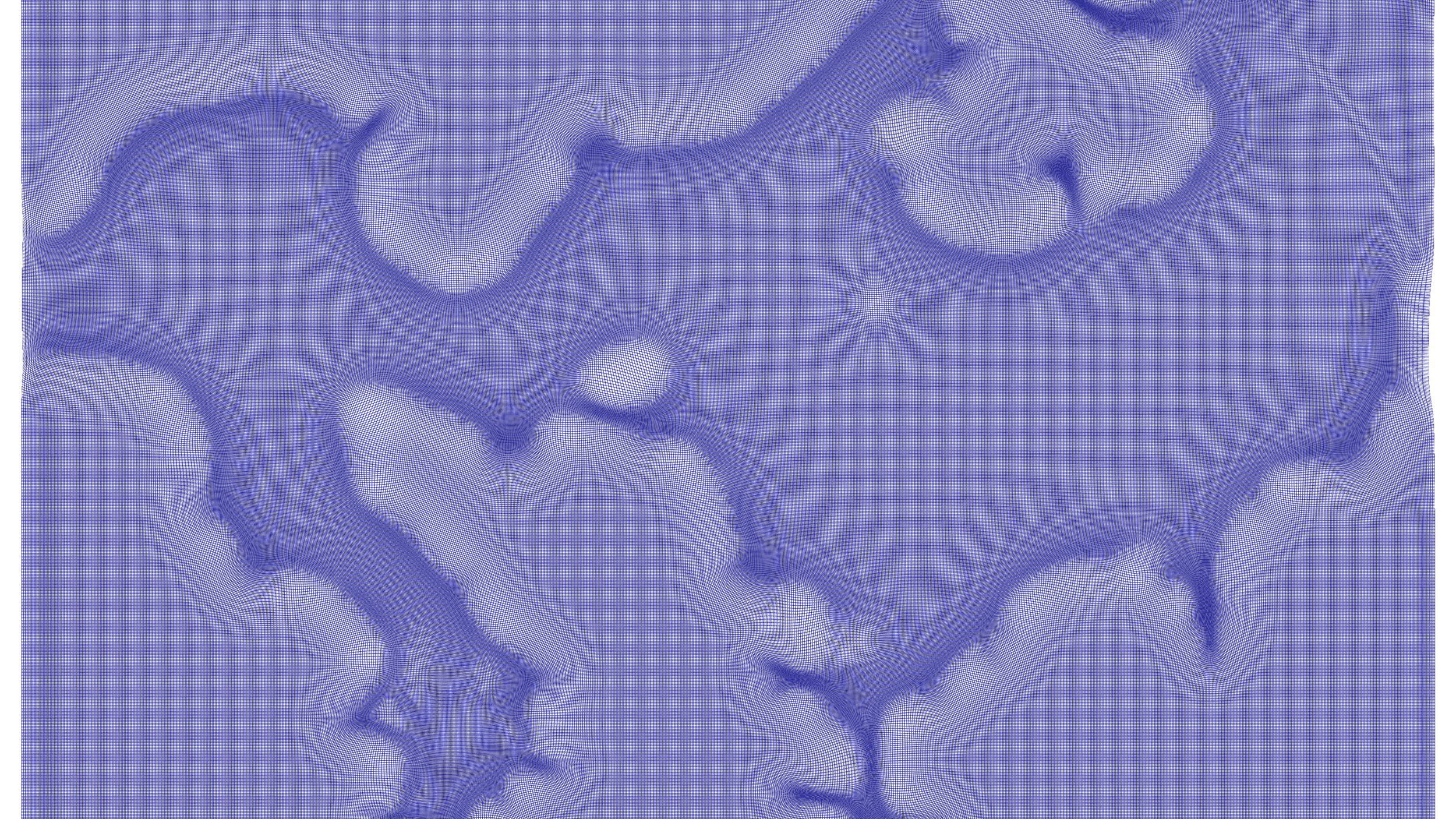}
    \caption{25,000 iterations}
    \end{subfigure}
    \begin{subfigure}[t]{0.3\textwidth}
    \includegraphics[width=\textwidth]{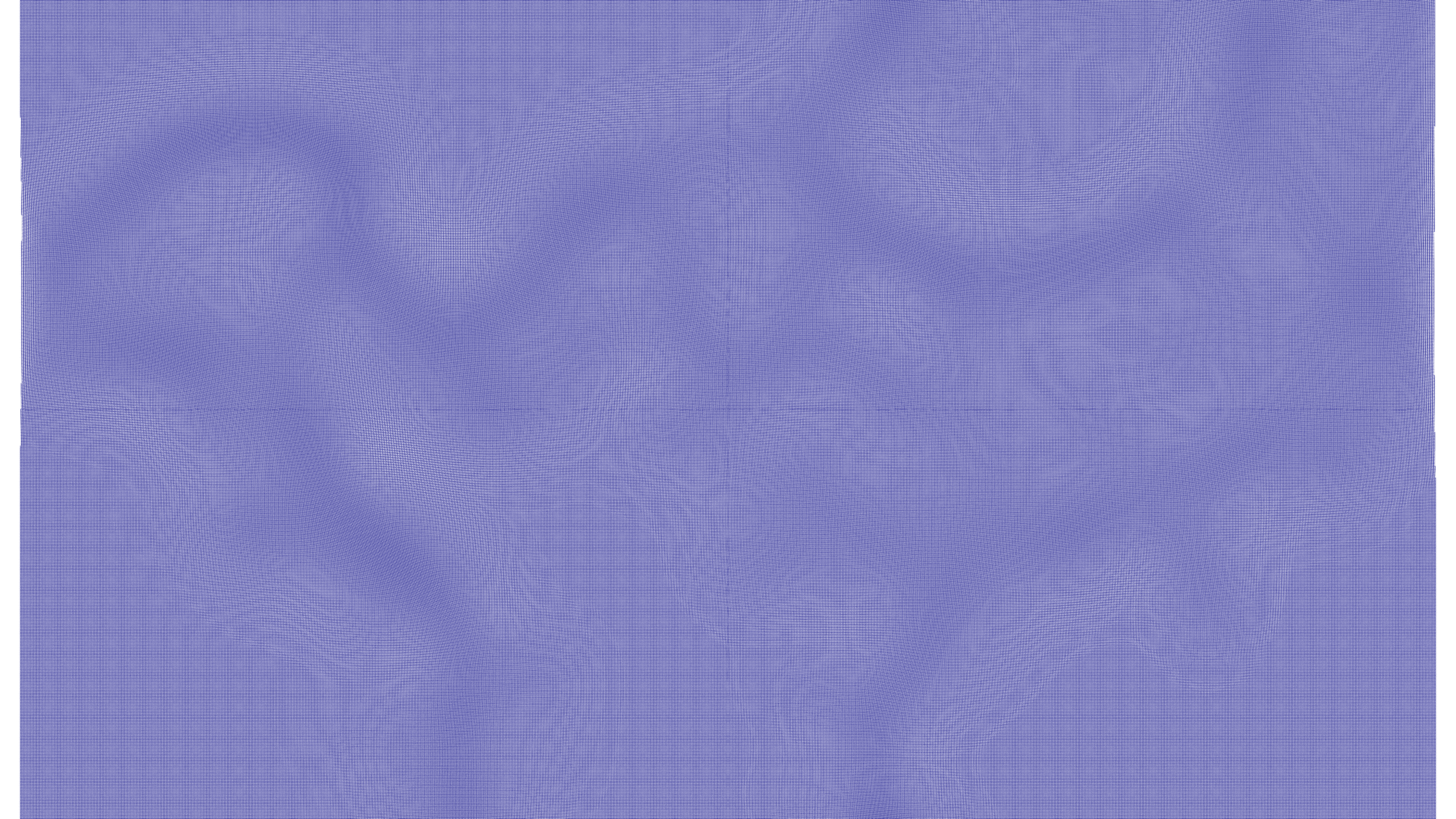}
    \caption{25,000 iterations}
    \end{subfigure}
    \begin{subfigure}[t]{0.3\textwidth}
    \includegraphics[width=\textwidth]{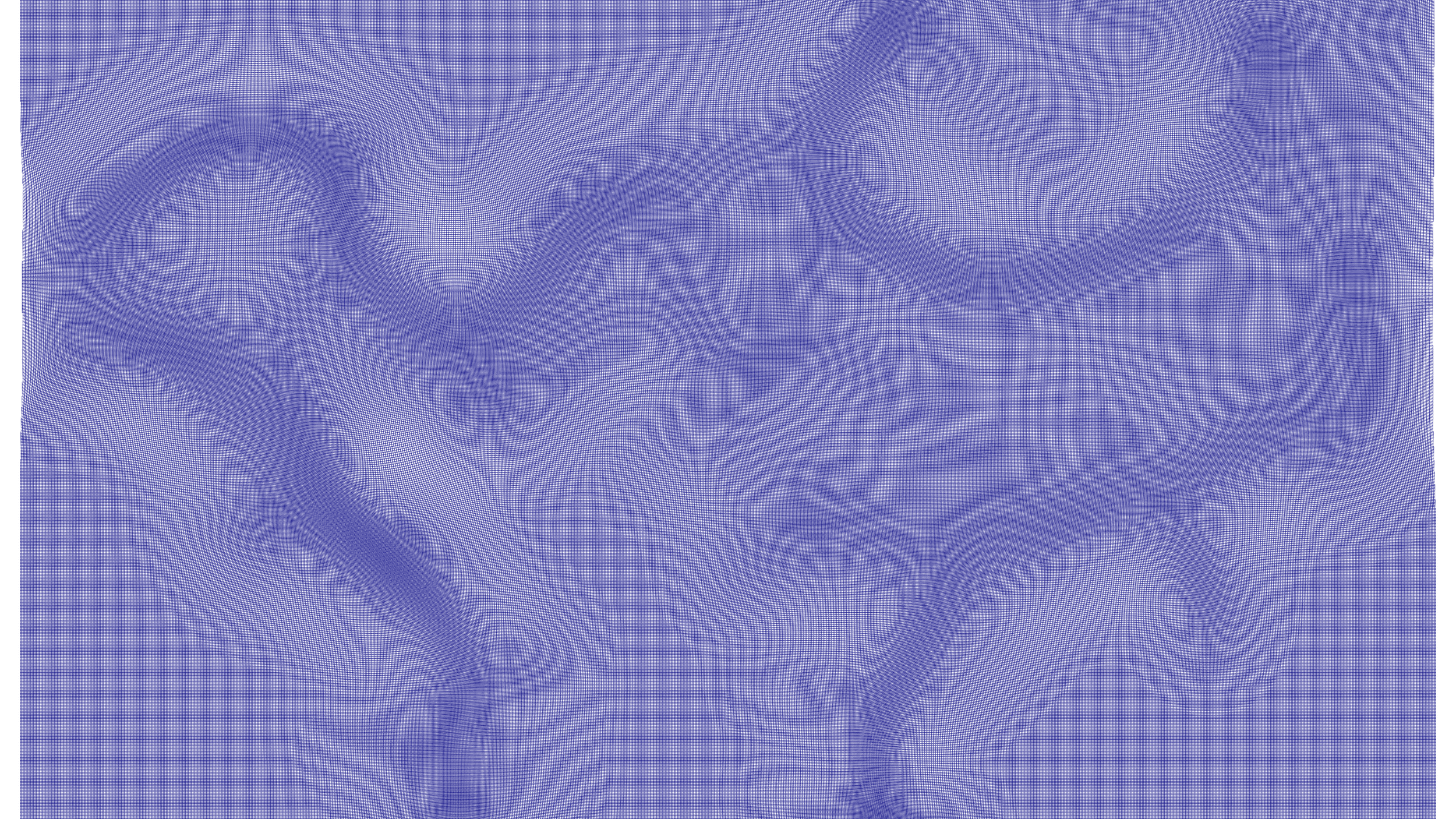}
    \caption{25,000 iterations}
    \end{subfigure}

    \begin{subfigure}[t]{0.3\textwidth}
    \includegraphics[width=\textwidth]{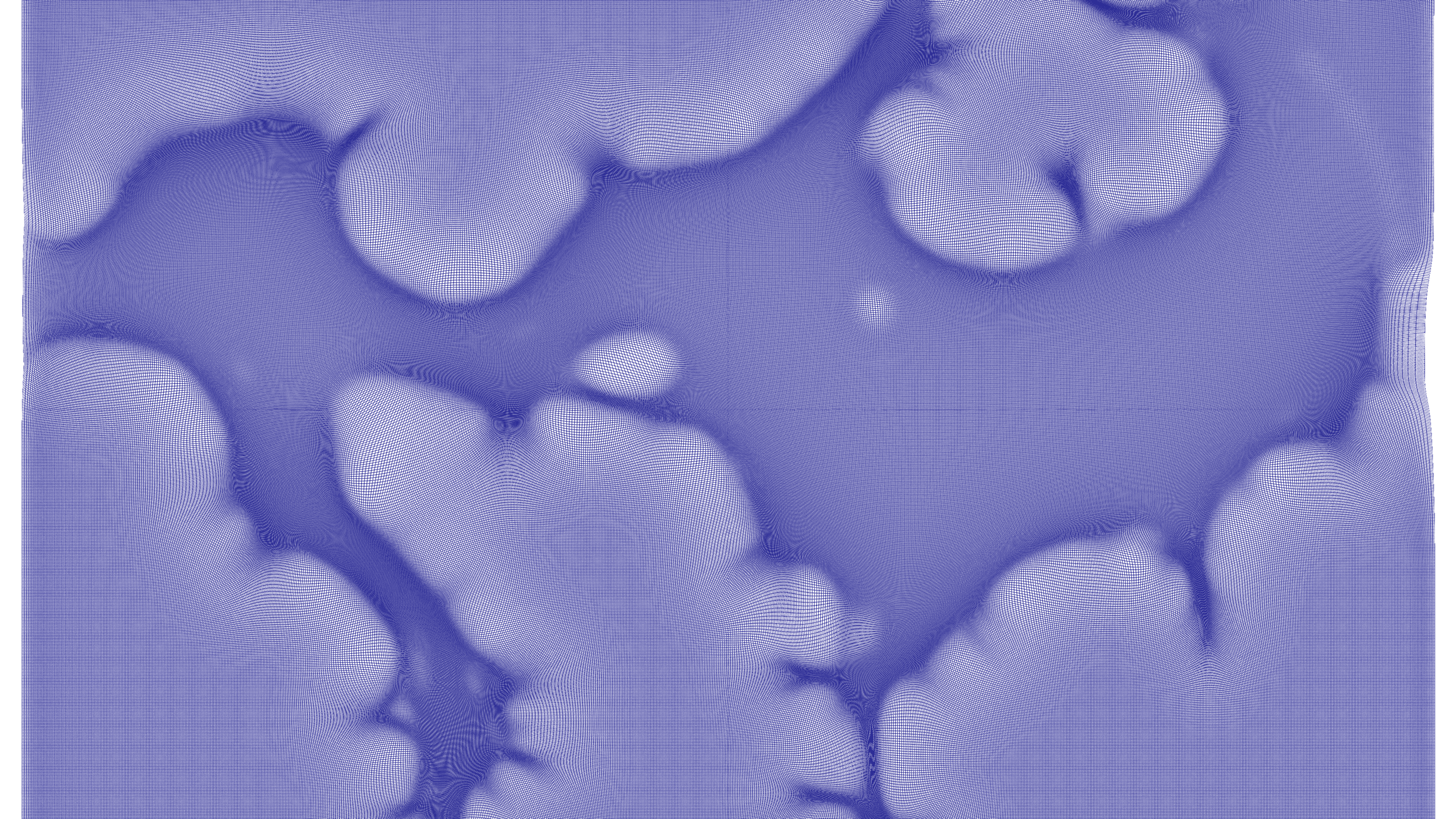}
    \caption{50,000 iterations}
    \end{subfigure}
    \begin{subfigure}[t]{0.3\textwidth}
    \includegraphics[width=\textwidth]{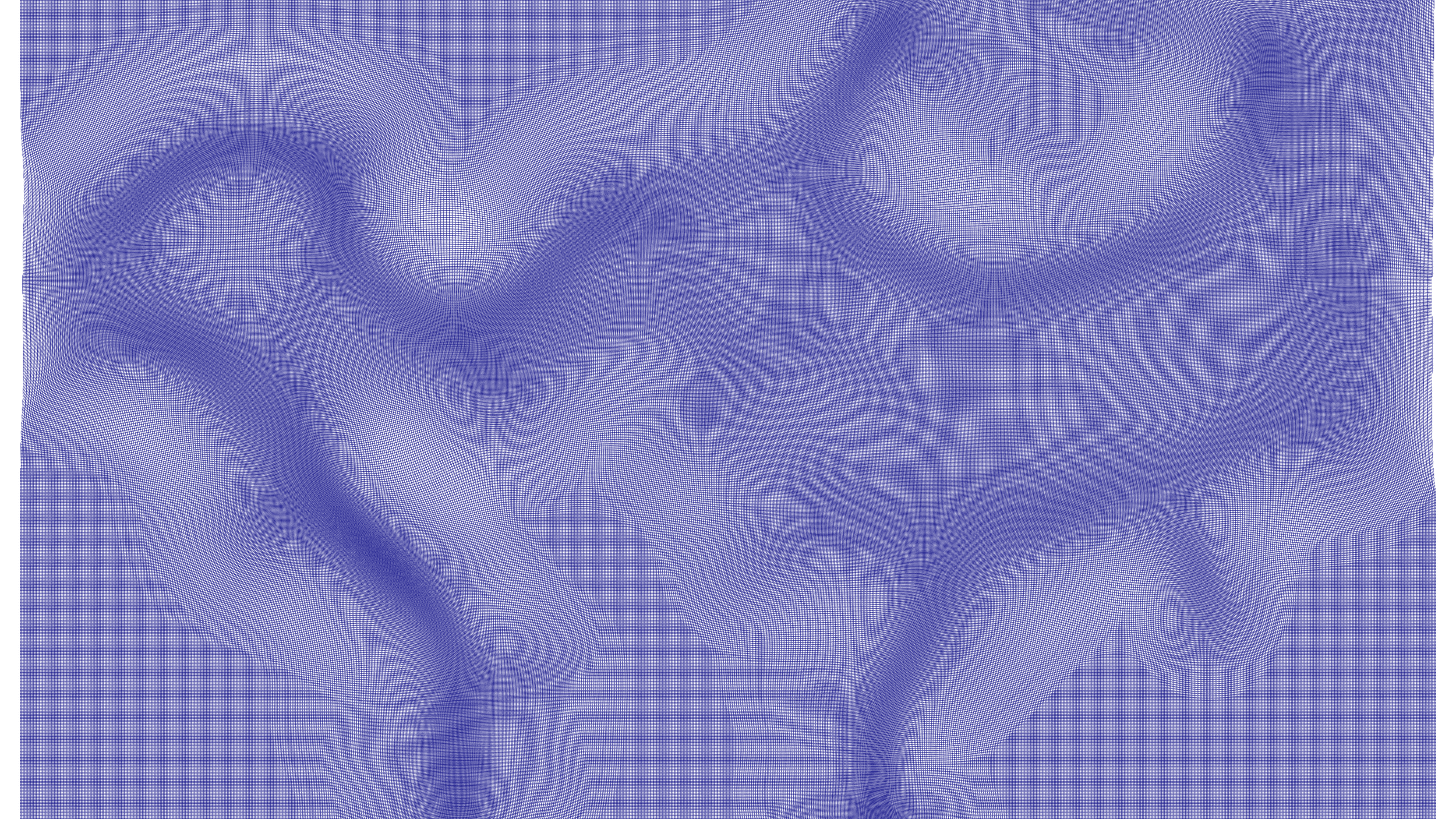}
    \caption{50,000 iterations}
    \end{subfigure}
    \begin{subfigure}[t]{0.3\textwidth}
    \includegraphics[width=\textwidth]{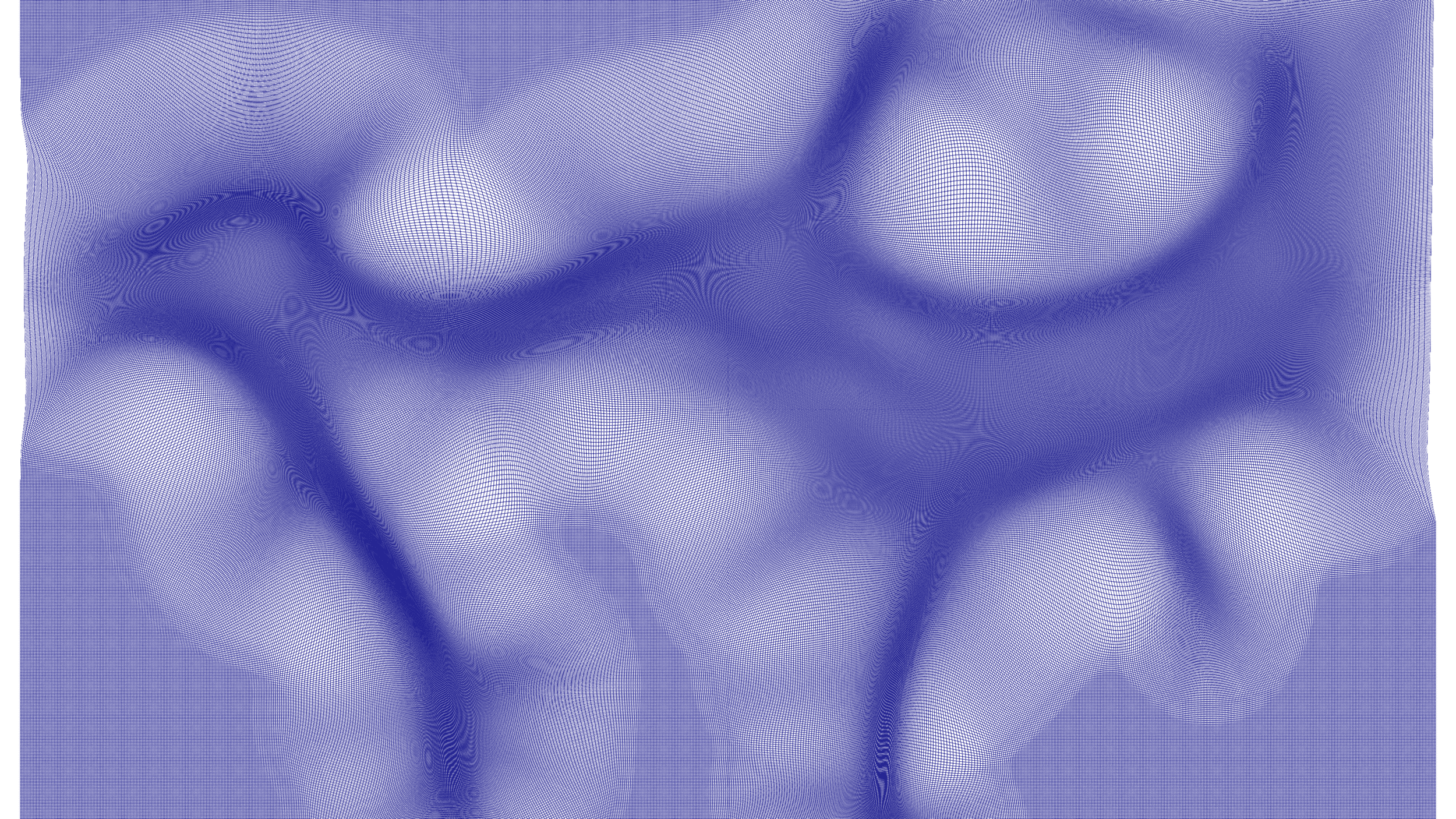}
    \caption{50,000 iterations}
    \end{subfigure}
    
    \begin{subfigure}[t]{0.3\textwidth}
    \includegraphics[width=\textwidth]{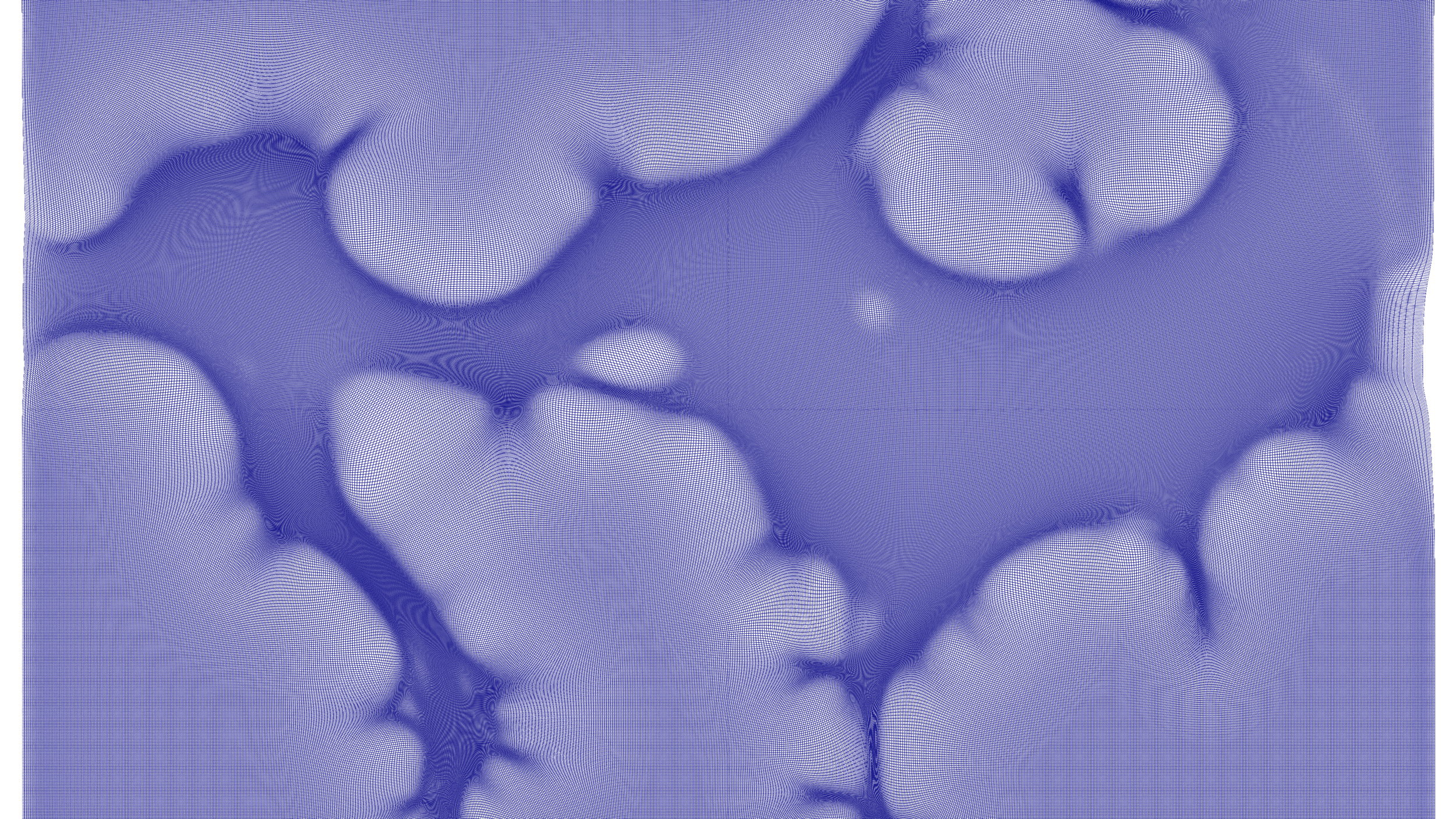}
    \caption{150,000 iterations}
    \end{subfigure}
    \begin{subfigure}[t]{0.3\textwidth}
    \includegraphics[width=\textwidth]{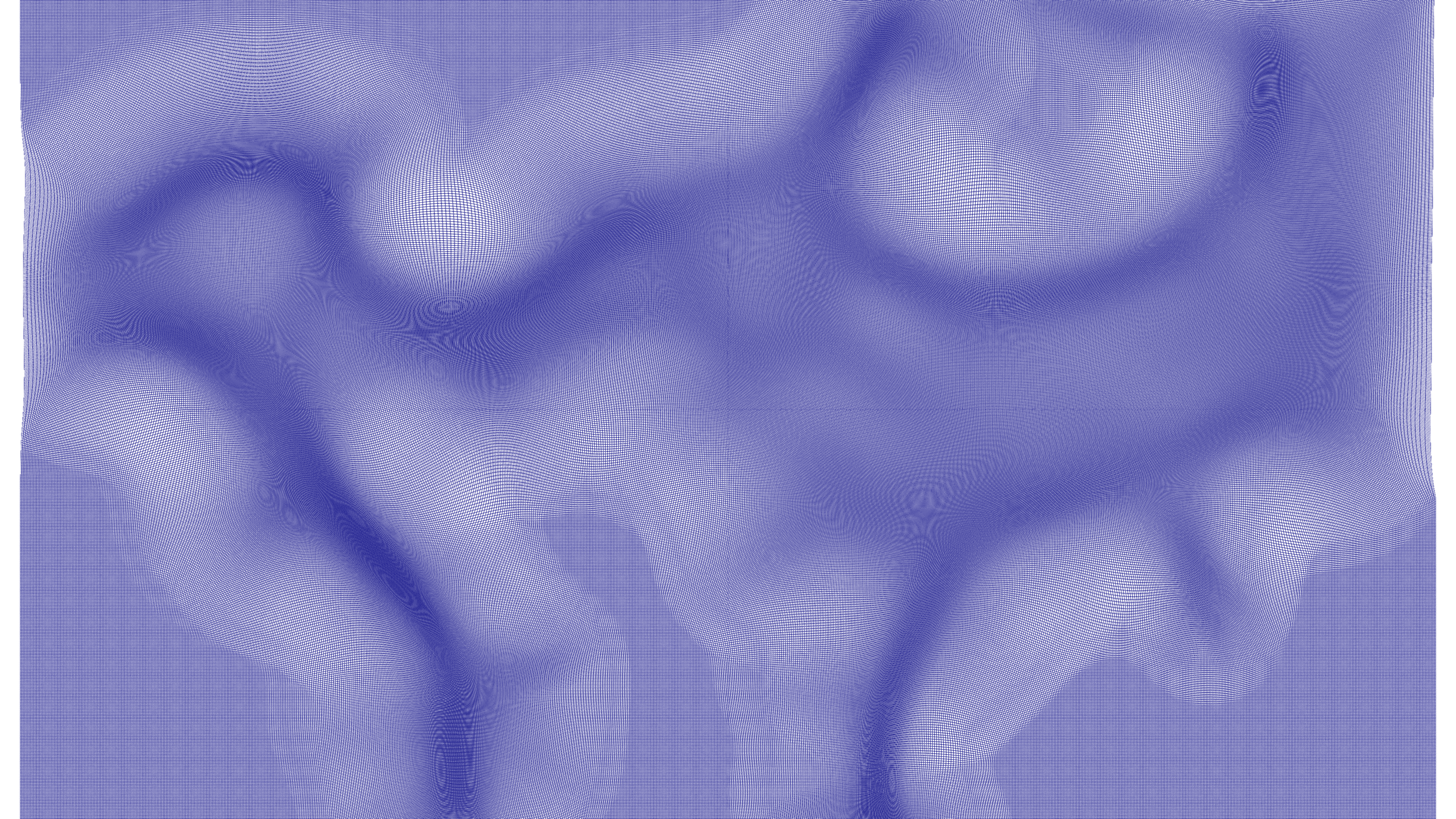}
    \caption{150,000 iterations}
    \end{subfigure}
    \begin{subfigure}[t]{0.3\textwidth}
    \includegraphics[width=\textwidth]{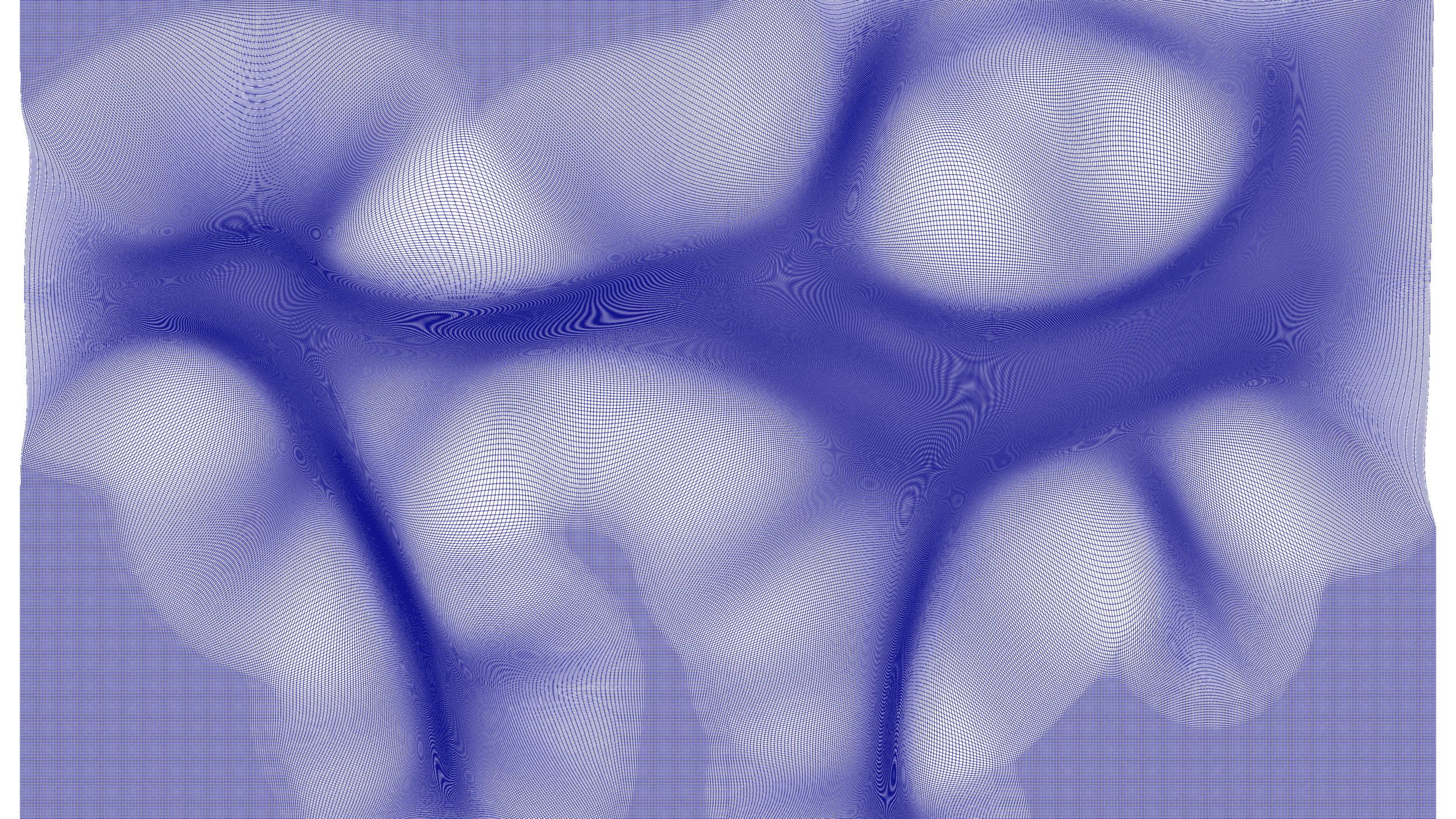}
    \caption{150,000 iterations}
    \end{subfigure}
    
    \caption{Static mesh distortion driven by the cumulative error field (a) obtained over the full simulation. (b)-(f), evolution of the distorted mesh after different Jacobi iterations (25,000 to 150,000) of solving diffusion equation for initial regular uniform coordinates. The mesh is used in Models C--H depicted in Table~\ref{tab:numerical_set_up}.}
    \label{fig:DSWE_mesh_details}
\end{figure}

Figure~\ref{fig:DSWE_mesh_details} presents Models C-H (See Table~\ref{tab:numerical_set_up}) respectively for the Carlisle flooding simulation. Sub-figures~(a)--(c) present the cumulative interpolation error obtained by integrating errors over the entire time of a structured grid simulation and taking different order of root (See Equation $E_{i,j}^{li,lj} = \left( \max_n \{\max_N \left| \frac{\partial^2 \psi_{N}^n}{\partial \xi^2} \right|\} \right)^{\frac{1}{R}}$ in Algorithm~\ref{alg:static-mesh-adaptation}). Areas of large error correspond to the location of rivers and their surrounding areas, which is where flooding will occur and the water velocities will be at their highest. There are mainly only two colours in Figure~\ref{fig:DSWE_mesh_details}~(a) because using a cube root (Models C and D $R=3$) for the production of the error measure may lead to a more uniform error distribution in inundated regions. With this error field, we apply the method of distorting the grid as described in Section~\ref{sec:method}. Sub-figures~\ref{fig:DSWE_mesh_details} (d)--(l) illustrate that as the diffusion algorithm progresses, the grid gradually deforms, concentrating resolution in regions of higher error while smoothing and coarsening in regions of lower error. This behaviour demonstrates that the proposed approach is able to identify areas requiring higher resolution in a physically consistent manner, suggesting that method is robust. The distorted grid shown in Sub-figures~\ref{fig:DSWE_mesh_details}~(j)--(l) is used to run the simulation again. Details of the numerical set up: time step, the structured grid size, the min and max node spacing in the distorted grid, and the mass source, are shown in Tables~\ref{tab:numerical_set_up} and \ref{tab:SWE_BCs}.

\begin{figure}[H]
    \centering
    \begin{subfigure}[t]{0.475\textwidth}
    \includegraphics[width=\textwidth]{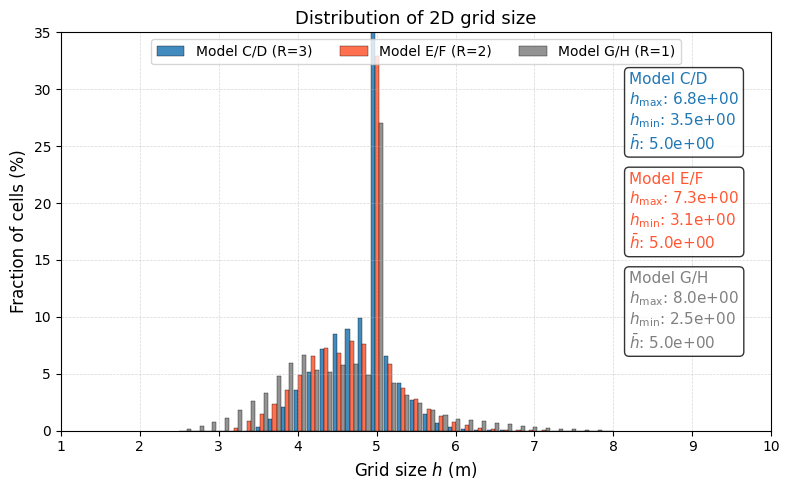}
    \caption{5,000 iterations}
    \end{subfigure}
    \begin{subfigure}[t]{0.475\textwidth}
    \includegraphics[width=\textwidth]{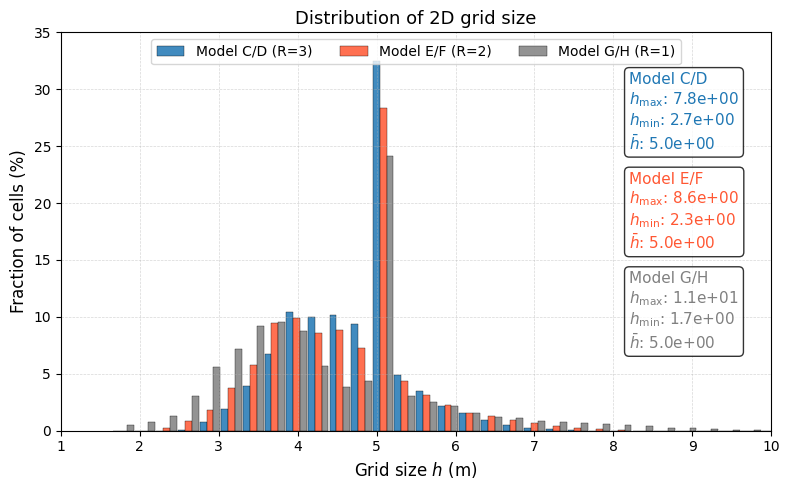}
    \caption{25,000 iterations}
    \end{subfigure}
    \begin{subfigure}[t]{0.475\textwidth}
    \includegraphics[width=\textwidth]{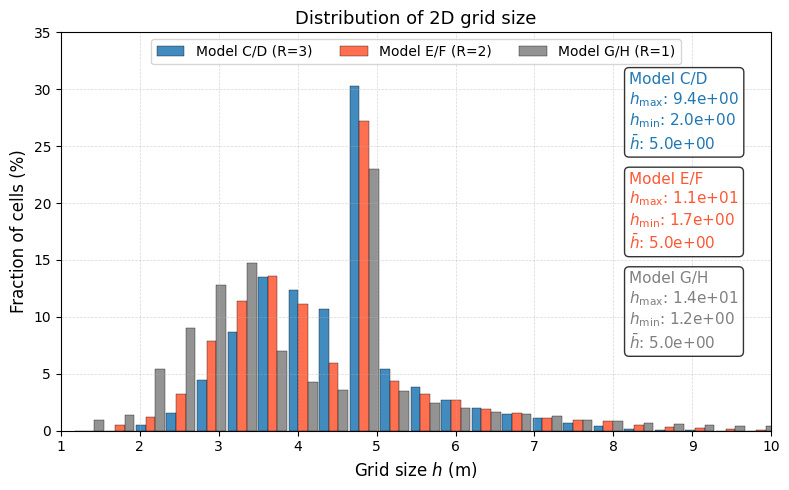}
    \caption{50,000 iterations}
    \end{subfigure}
    \begin{subfigure}[t]{0.475\textwidth}
    \includegraphics[width=\textwidth]{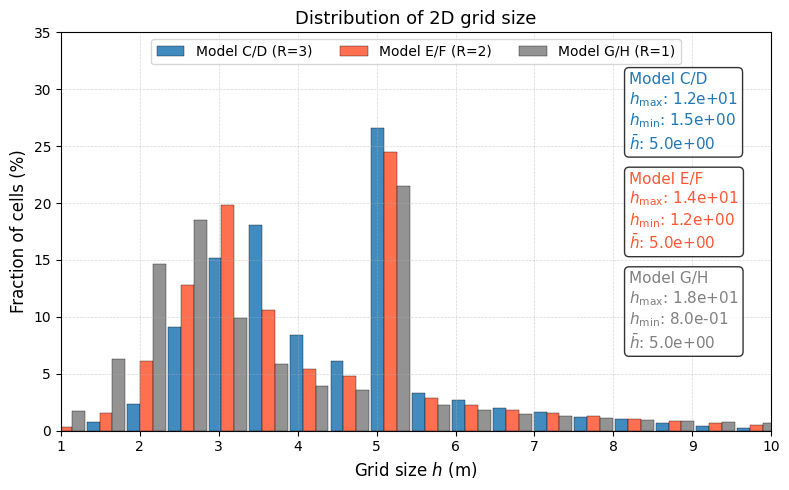}
    \caption{100,000 iterations}
    \end{subfigure}
    \caption{Histograms of grid size distribution during the mesh distortion process for the 2D Carlisle flooding simulation at (a) 5,000, (b) 25,000, (c) 50,000, and (d) 100,000 iterations. The results are obtained from Models C to H as depicted in Table~\ref{tab:numerical_set_up}.}
    \label{fig:DSWE_mesh_his}
\end{figure}

Figure~\ref{fig:DSWE_mesh_his} compares the evolution of the grid size distributions for Models C/D ($R=3$), Models E/F ($R=2$) and Models G/H ($R=1$) as listed in Table~\ref{tab:numerical_set_up}) after 5,000, 25,000, 50,000 and 100,000 iterations for the 2D Carlisle flooding simulation. In each figure, the width of bar $W_{b}$ is represented as $W_{b} = h_{max} - h_{min}$. At 5,000 iterations, both mesh configurations exhibit a narrow distribution concentrated around $h=\SI{5}{m}$, indicating that they remain close to the initial uniform grid. As the iterations increase, the distributions for both models progressively broaden, with an increasing proportion of both smaller and larger grid sizes. From 25,000 to 50,000 iterations, this trend becomes more significant, where the right tail develops towards larger grid size and the left side extends towards smaller cells. This indicates that the proposed distorted mesh capability can effectively redistribute cell sizes through local stretching and produce a wider range of spatial resolution across the domain, enhancing the accuracy of the numerical solution in capturing fine-scale features. At 100,000 iterations, the overall shape of the distribution becomes more stable in both models, suggesting that the solution of distorting mesh is converged. Importantly, despite the significant change in the shape of the mesh size distribution, the mean grid size remains consistently close to \SI{5}{m} at all stages. This demonstrates that the distortion is conservative with respect to the global mesh size, enabling mesh adaptation while maintaining the overall average resolution. 

On the other hand, the differences among the three error formulations are present but not significant. All three cases show the same overall trend, a progressive broadening of the distribution away from the initial peak at $h=\SI{5}{m}$ as the mesh distortion develops. Interestingly, at later iterations, Models G/H ($R=1$) exhibit a slightly wider spread towards both smaller and larger grid sizes, while Models E/F ($R=2$) show an intermediate behaviour and Models C/D ($R=3$) remain relatively more concentrated around the mean value. This suggests that reducing 
R could lead to a slightly stronger redistribution of mesh resolution, although the overall differences between the three formulations remain moderate. Overall, the results indicate that the choice of $R$ controls the final range of cell sizes and thus provides a practical means of adjusting the aggressiveness of the mesh adaptation procedure.


\begin{figure}[H]
    \centering
    \begin{subfigure}[t]{0.50\textwidth}
    \includegraphics[width=\textwidth]{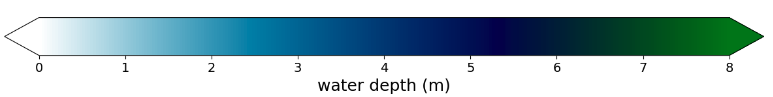}
    \end{subfigure}
    \begin{subfigure}[t]{0.99\textwidth}
    \includegraphics[width=\textwidth]{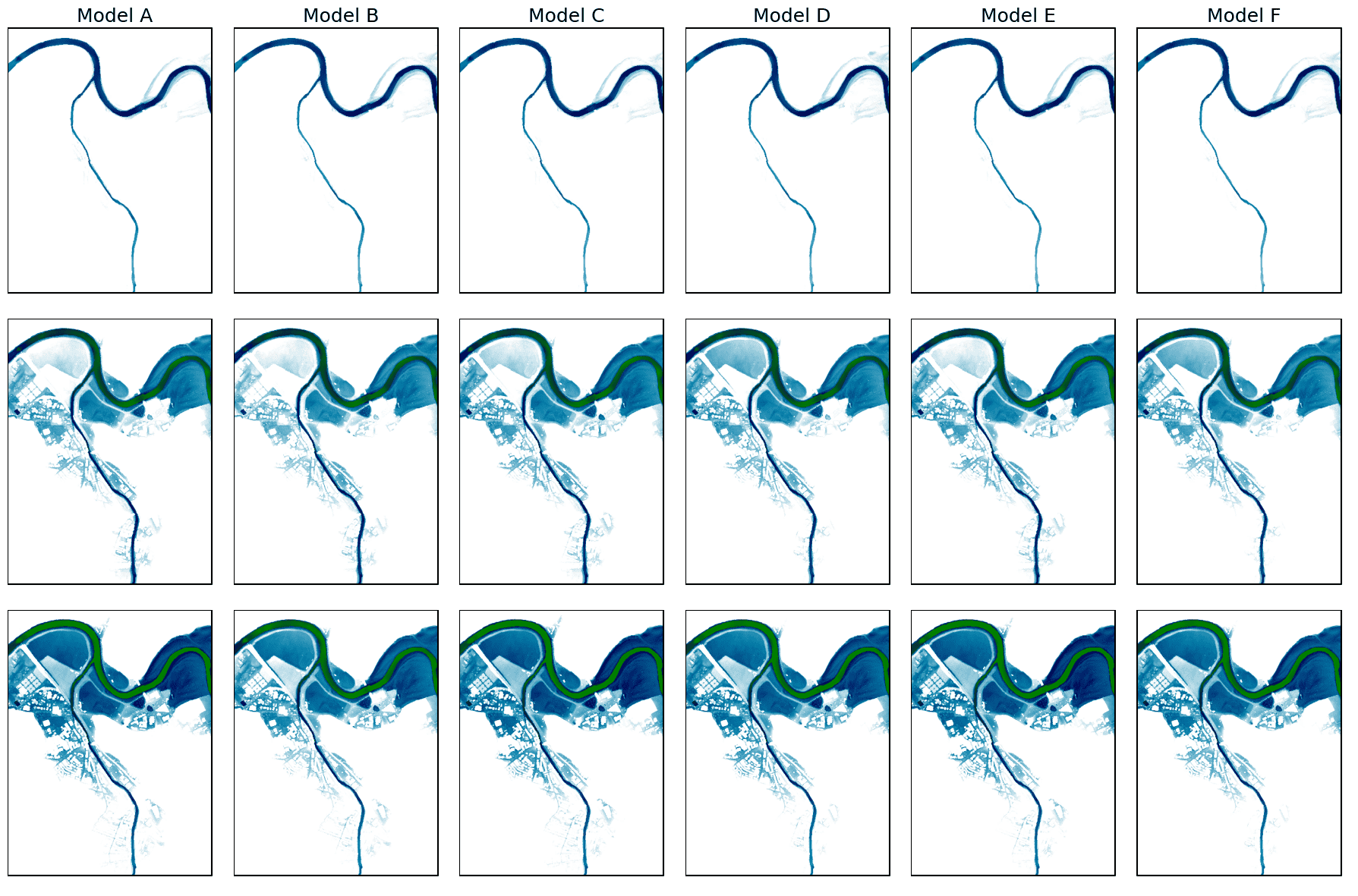}
    \end{subfigure}
    \caption{Simulated water depth in one focusing area (Denton Holme) on regular and distorted grid using Linear and Quadratic ConvFEM models. The results are shown at three different time levels, top, \SI{20}{h}; middle, \SI{40}{h}; bottom, \SI{60}{h}.}
    \label{fig:DSWE_water_depth_1}
\end{figure}

\begin{figure}[H]
    \centering
    \begin{subfigure}[t]{0.99\textwidth}
    \includegraphics[width=\textwidth]{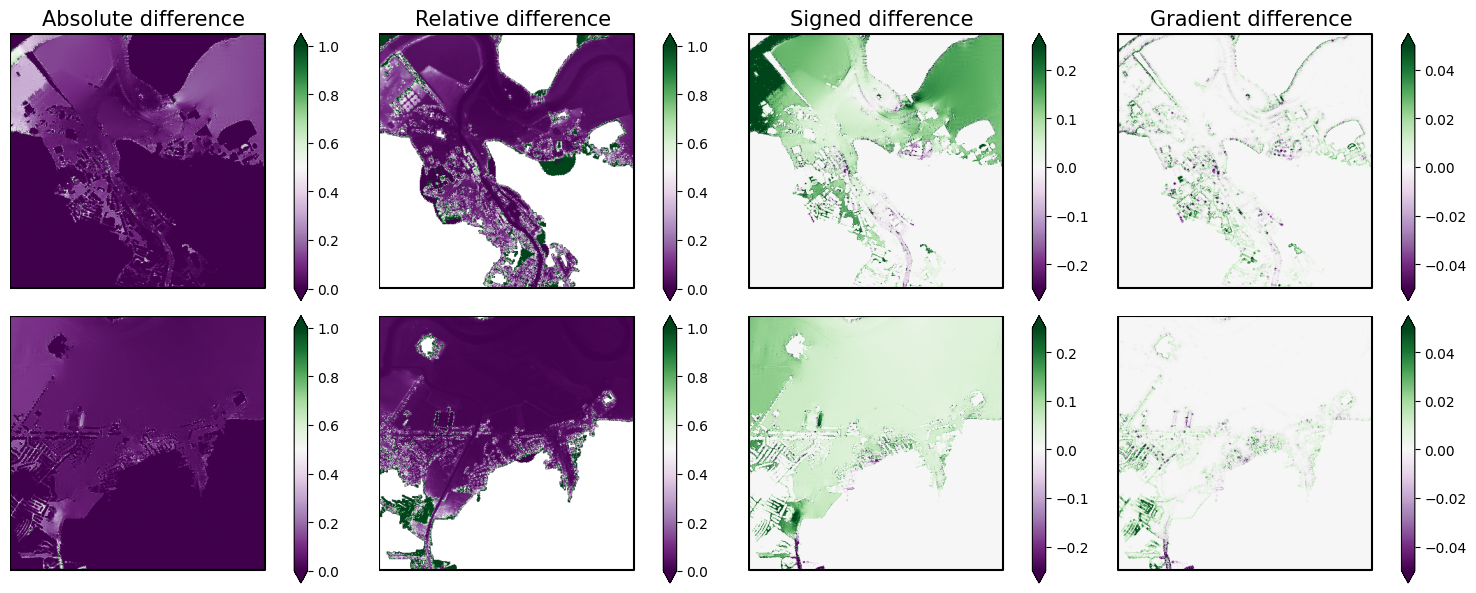}
    \caption{Linear ConvFEM}
    \end{subfigure}
    \begin{subfigure}[t]{0.99\textwidth}
    \includegraphics[width=\textwidth]{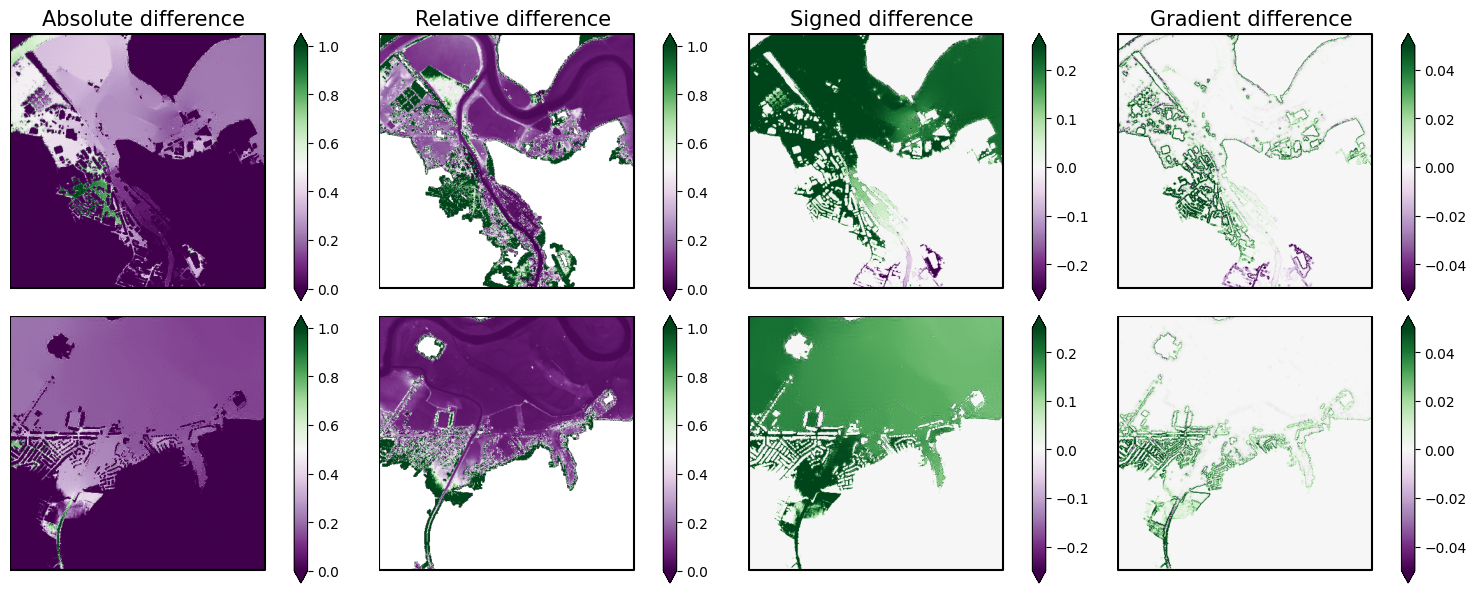}
    \caption{Quadratic ConvFEM}
    \end{subfigure}
    \caption{Spatial distribution of differences in water depth between regular and distorted grids using Linear and Quadratic ConvFEM models. Absolute, relative, signed, and gradient magnitude differences are shown for two representative regions, Denton Holme and Botcherby Bridge, after 60 hours of flooding.}
    \label{fig:DSWE_water_depth_difference}
\end{figure}

\begin{figure}[H]
    \centering
    \begin{subfigure}[t]{0.99\textwidth}
    \includegraphics[width=\textwidth]{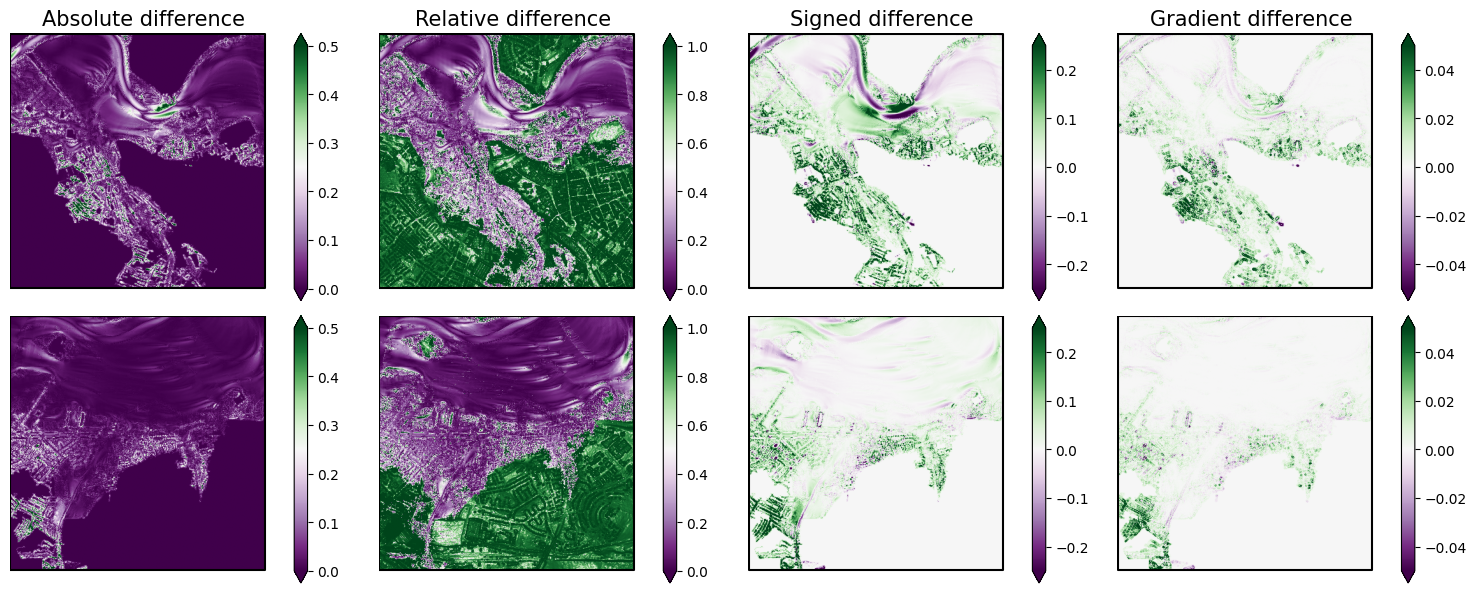}
    \caption{Linear ConvFEM}
    \end{subfigure}
    \begin{subfigure}[t]{0.99\textwidth}
    \includegraphics[width=\textwidth]{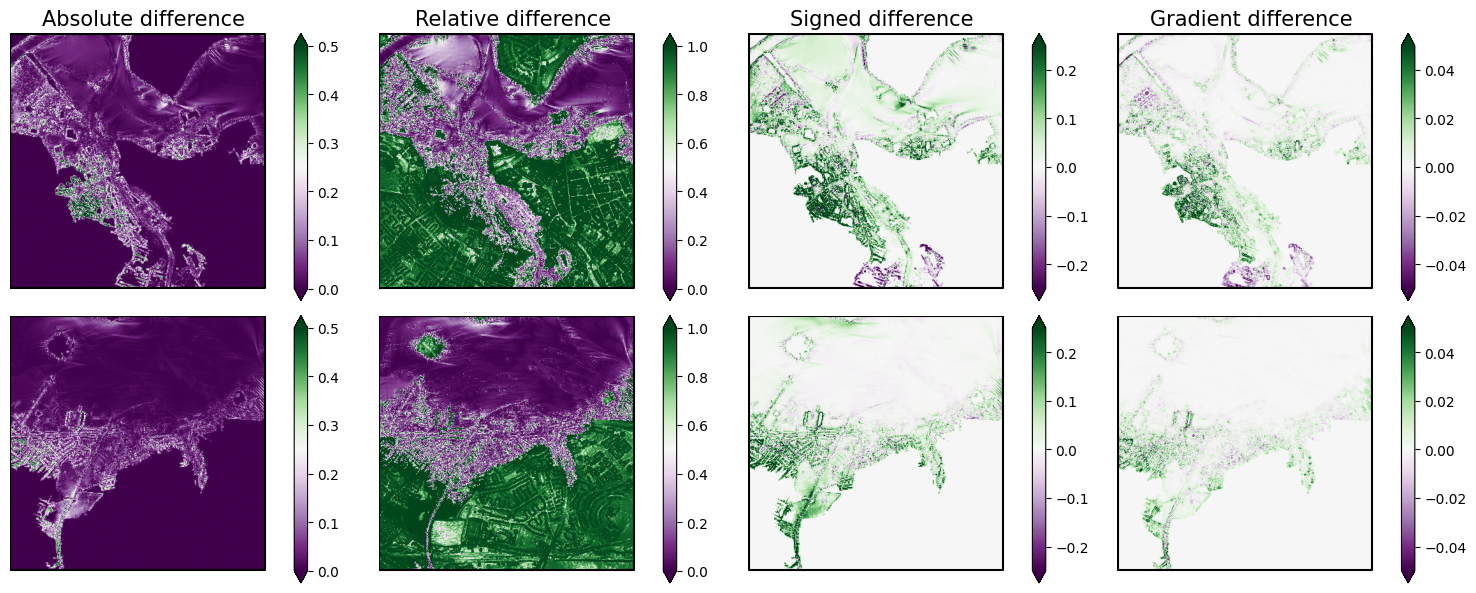}
    \caption{Quadratic ConvFEM}
    \end{subfigure}
    \caption{Spatial distribution of differences in water speed between regular and distorted grids using Linear and Quadratic ConvFEM models. Absolute, relative, signed, and gradient magnitude differences are shown for two representative regions, Denton Holme and Botcherby Bridge, after 60 hours of flooding.}
    \label{fig:DSWE_water_speed_difference}
\end{figure}

Figure~\ref{fig:DSWE_water_depth_1} compares simulated water depth in the Denton Holme region using six \NP models (Models A-F summarised in Table~\ref{tab:numerical_set_up}) on regular and distorted grids. The results are shown at three different time levels which are after 20~hours (top row), 40~hours (middle row) and 60~hours (bottom row) of flooding. Regarding the regular uniform grid (Models A and B), both linear and quadratic discretisations capture the overall flooding pattern, however, small-scale features near river banks and urban structures are less sharply resolved. This observation has been reported in~\cite{Chen2025}. To see the small scale features we have zoomed in on them in Figures~\ref{fig:DSWE_water_depth_difference} and \ref{fig:DSWE_water_speed_difference}. 
For the distorted grid solution (Models C to F), the water depth distribution exhibits improved spatial definition, particularly along narrow flow paths and wetting-drying interfaces. Moreover, the distorted grid leads to a more noticeable change in the predicted water depth near the river (particularly after 60 hours of flooding) when using the quadratic ConvFEM model (Models D and F) compared to the linear ConvFEM model (Models C and E).

\begin{figure}[H]
    \centering
    \begin{subfigure}[t]{0.50\textwidth}
    \includegraphics[width=\textwidth]{figures/SWE_RESULTS/legend_carlisle.png}
    \end{subfigure}
    \begin{subfigure}[t]{0.99\textwidth}
    \includegraphics[width=\textwidth]{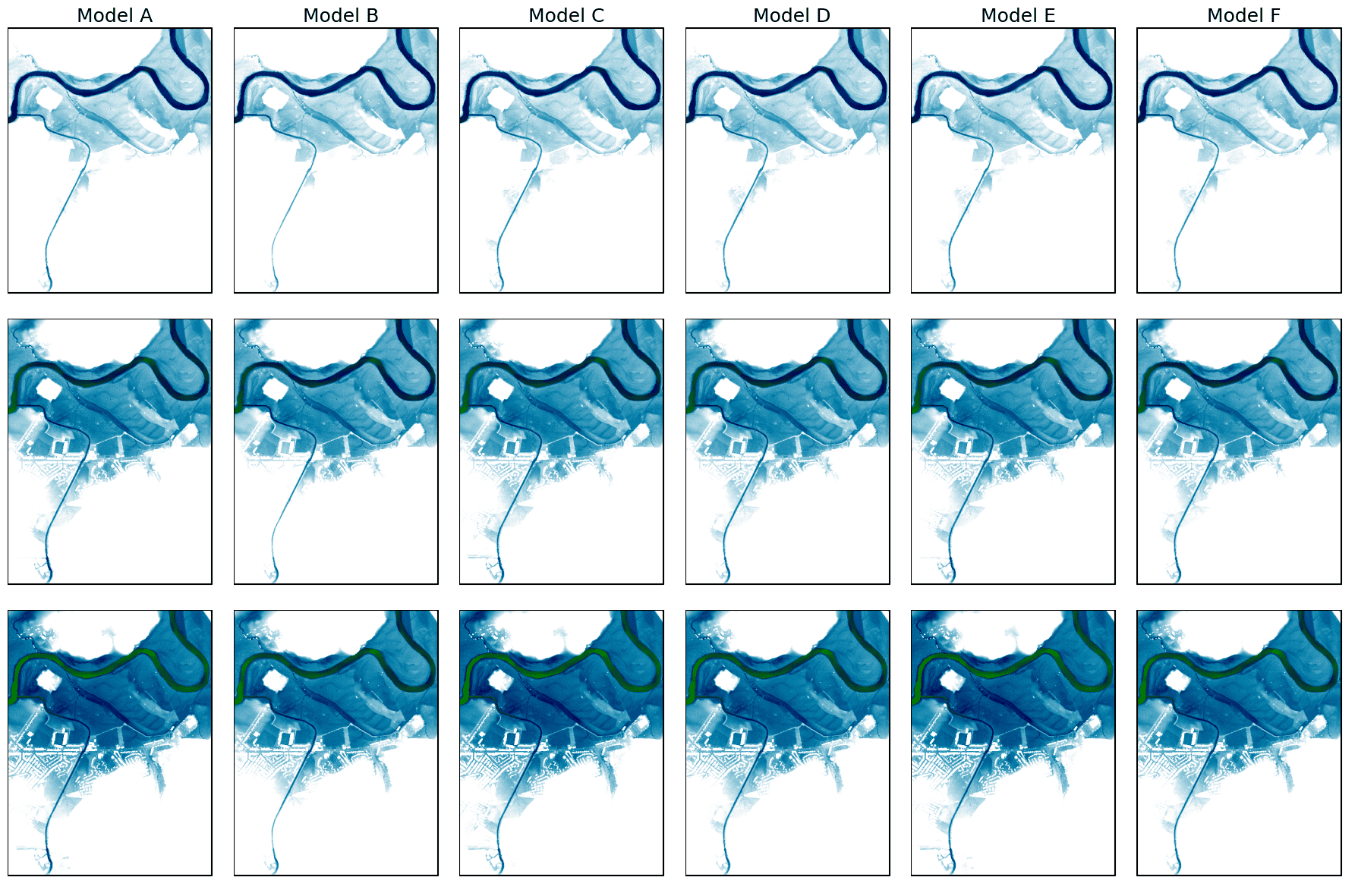}
    \end{subfigure}
    \caption{Simulated water depth in the area around Botcherby Bridge on regular and distorted grids using Linear and Quadratic ConvFEM models. The results are shown at three different time levels: top, \SI{20}{h}; middle, \SI{40}{h}; bottom, \SI{60}{h}.}
    \label{fig:DSWE_water_depth_2}
\end{figure}

A similar analysis is performed for another flooded area (Botcherby Bridge), as shown in Figure~\ref{fig:DSWE_water_depth_2}. The results present the water depth distributions obtained using the linear and quadratic ConvFEM on both regular and distorted grids. Three representative time levels (20 hours, 40 hours and 60 hours) are selected for comparison. The effect of the distorted grid on the predicted water depth near the river is more pronounced for the quadratic ConvFEM model (comparing Models D and F to Model B) than for the linear ConvFEM model (comparing Models C and E to Model A), particularly after 40 hours of flooding, resulting in relatively larger simulated water depths.

Figure~\ref{fig:DSWE_water_depth_difference} presents a detailed comparison of the water depth fields obtained using regular and distorted grids, discussed through absolute, relative, signed, and gradient magnitude differences for both Linear (Model A \textbf{$\mathbf{vs}$} Model C) and Quadratic (Model B \textbf{$\mathbf{vs}$} Model D) ConvFEM models. Two representative regions, Denton Holme and Botcherby Bridge, are selected to highlight localised behaviours after 60 hours of flooding. The absolute differences $|h_{dis}-h_{reg}|$ indicate that the overall discrepancies remain spatially localised, with the majority of the domain exhibiting minimal variation between the two grid configurations. Interestingly, larger differences are concentrated along river channels and flood fronts, suggesting that the distorted grid primarily influences dynamically active regions. The relative difference $\frac{|h_{dis}-h_{reg}|}{h_{reg}}$ further emphasises these localised effects, revealing that even where absolute differences are small, noticeable variations can arise in regions of shallow water depth. This highlights the sensitivity of low-depth areas to grid adaptation and suggests that the distorted grid enhances resolution in these regions. The signed difference $h_{dis}-h_{reg}$ shows a clear pattern of local redistribution of water depth with alternating positive and negative regions. These variations are spatially localised and do not exhibit a consistent domain-wide offset, suggesting that the distorted grid primarily modifies the solution locally while preserving the overall flooding pattern. The gradient difference $\nabla h_{dis}-\nabla h_{reg}$ provides additional insight into the underlying mechanism of these discrepancies. Differences are predominantly observed in regions with strong spatial gradients of water depth, such as flow boundaries and rapidly varying free-surface regions. This demonstrates that the distorted grid enhances the representation of sharp gradients and local flow features, which are otherwise smoothed on the regular grid. Comparing the Linear and Quadratic ConvFEM results, the Quadratic model exhibits more pronounced and structured differences, particularly in gradient-sensitive regions, indicating its increased sensitivity to mesh adaptation and its ability to capture finer-scale flow variations.

\begin{figure}[H]
    \centering
    \begin{subfigure}[t]{0.50\textwidth}
    \includegraphics[width=\textwidth]{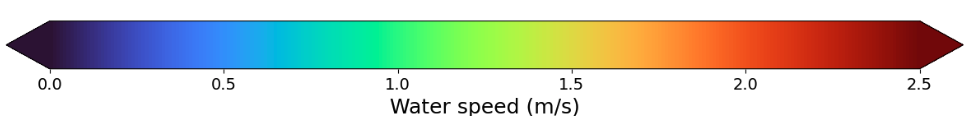}
    \end{subfigure}
    \begin{subfigure}[t]{0.99\textwidth}
    \includegraphics[width=\textwidth]{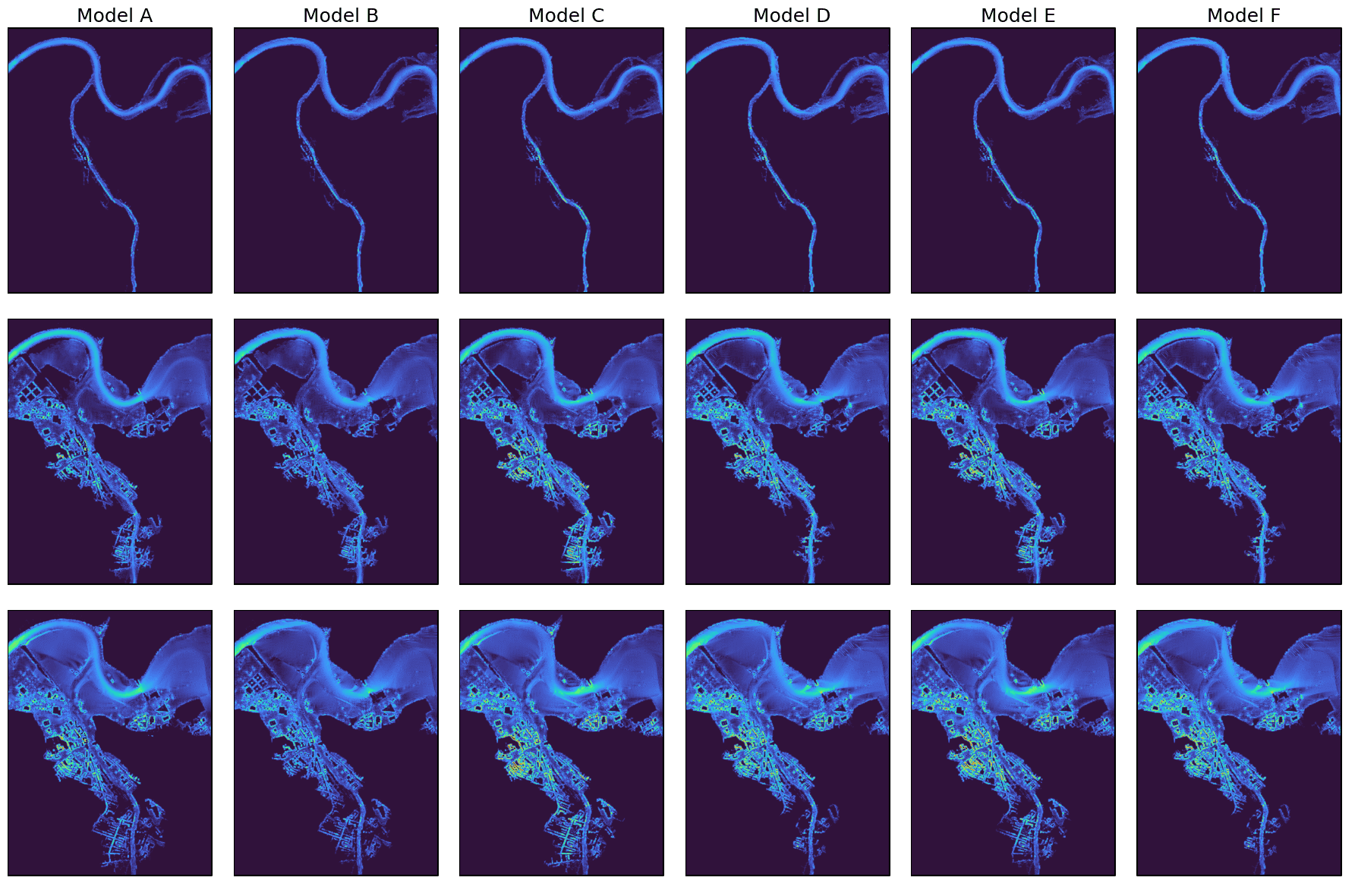}
    \end{subfigure}
    \caption{Results for water speed in the Denton Holme area on regular and distorted grids using Linear and Quadratic ConvFEM models. The results are shown at three different time levels, top, \SI{20}{h}; middle, \SI{40}{h}; bottom, \SI{60}{h}.}
    \label{fig:DSWE_water_speed_1}
\end{figure}

\begin{figure}[H]
    \centering
    \begin{subfigure}[t]{0.50\textwidth}
    \includegraphics[width=\textwidth]{figures/SWE_RESULTS/legend_carlisle_vel.png}
    \end{subfigure}
    \begin{subfigure}[t]{0.99\textwidth}
    \includegraphics[width=\textwidth]{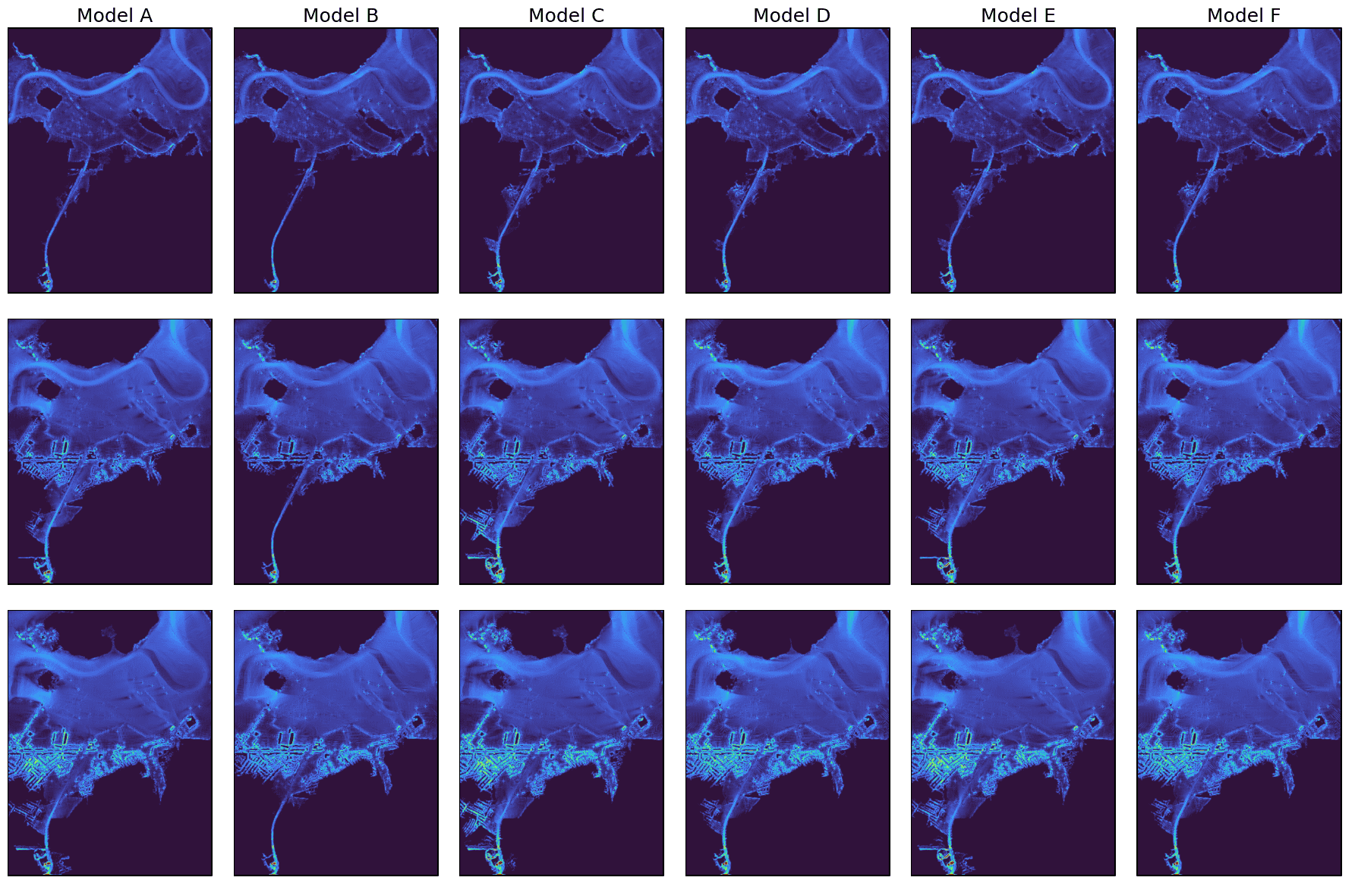}
    \end{subfigure}
    \caption{Water speed in the area surrounding Botcherby Bridge on regular and distorted grids using Linear and Quadratic ConvFEM models. The results are shown at three different time levels, top, \SI{20}{h}; middle, \SI{40}{h}; bottom, \SI{60}{h}.}
    \label{fig:DSWE_water_speed_2}
\end{figure}

Following the analysis of water depth, we examine the impact of the static distorted grid on the flow velocity. Figures~\ref{fig:DSWE_water_speed_1} and~\ref{fig:DSWE_water_speed_2} present the simulated water speed in two areas, Denton Holme and Botcherby Bridge. The same \NP  models and time levels are selected for comparison as in Figures~\ref{fig:DSWE_water_depth_1} and~\ref{fig:DSWE_water_depth_2} . Consistent with the results of the predicted water depth, the results obtained on the distorted grid indicate that using the distorted grid enables more detailed flow structures to be captured near in the areas near rivers.

Figure~\ref{fig:DSWE_water_speed_difference} presents the spatial distribution of differences in flow velocity between the regular and distorted grids, using the same set of metrics as for water depth, including absolute, relative, signed, and gradient magnitude differences. Two representative regions, Denton Holme and Botcherby Bridge, are analysed after 60 hours of flooding. Compared to the water depth results (Figure~\ref{fig:DSWE_water_depth_difference}), the velocity differences exhibit more pronounced and coherent spatial structures. The absolute differences $|U_{dis}-U_{reg}|$ are primarily concentrated along the main flow pathways, particularly within river channels, junctions, and regions of flow acceleration and deceleration. Although the magnitude of absolute differences is not directly comparable between water depth and velocity, the velocity field exhibits more coherent and spatially structured variations, indicating a stronger sensitivity to mesh distortion in dynamically active regions. The relative difference $\frac{|U_{dis}-U_{reg}|}{U_{reg}}$ highlights significant variations in low-velocity areas, where small absolute changes can lead to amplified relative discrepancies. This effect is particularly evident near floodplain regions and shallow flow zones, reflecting the increased sensitivity of velocity to local mesh adaptation. The signed difference $U_{dis}-U_{reg}$ reveals a pattern of local redistribution, with both positive and negative deviations distributed along flow structures. 
The results illustrate that the differences are primarily localised along flow structures, rather than forming a consistent domain-wide offset, suggesting that the distorted grid mainly introduces local redistribution of the velocity field. The gradient difference $\nabla U_{dis}-\nabla U_{reg}$ further highlights that discrepancies are concentrated in regions of strong velocity gradients, particularly along shear layers, channel edges, and zones of flow acceleration. These regions correspond to dynamically active areas where sharp spatial variations occur. The distorted grid produces more pronounced gradients in these locations, indicating an improved ability to resolve local flow features and sharp velocity transitions that appear more diffused on the regular grid. A comparison between the Linear and Quadratic ConvFEM models shows that the Quadratic formulation produces more structured and pronounced differences, particularly in gradient-sensitive regions. This suggests that higher-order discretisation benefits more from the adaptive characteristics of the distorted grid, leading to improved representation of fine-scale flow dynamics.

\begin{figure}[H]
    \centering
    \begin{subfigure}[t]{0.82\textwidth}
    \includegraphics[width=\textwidth]{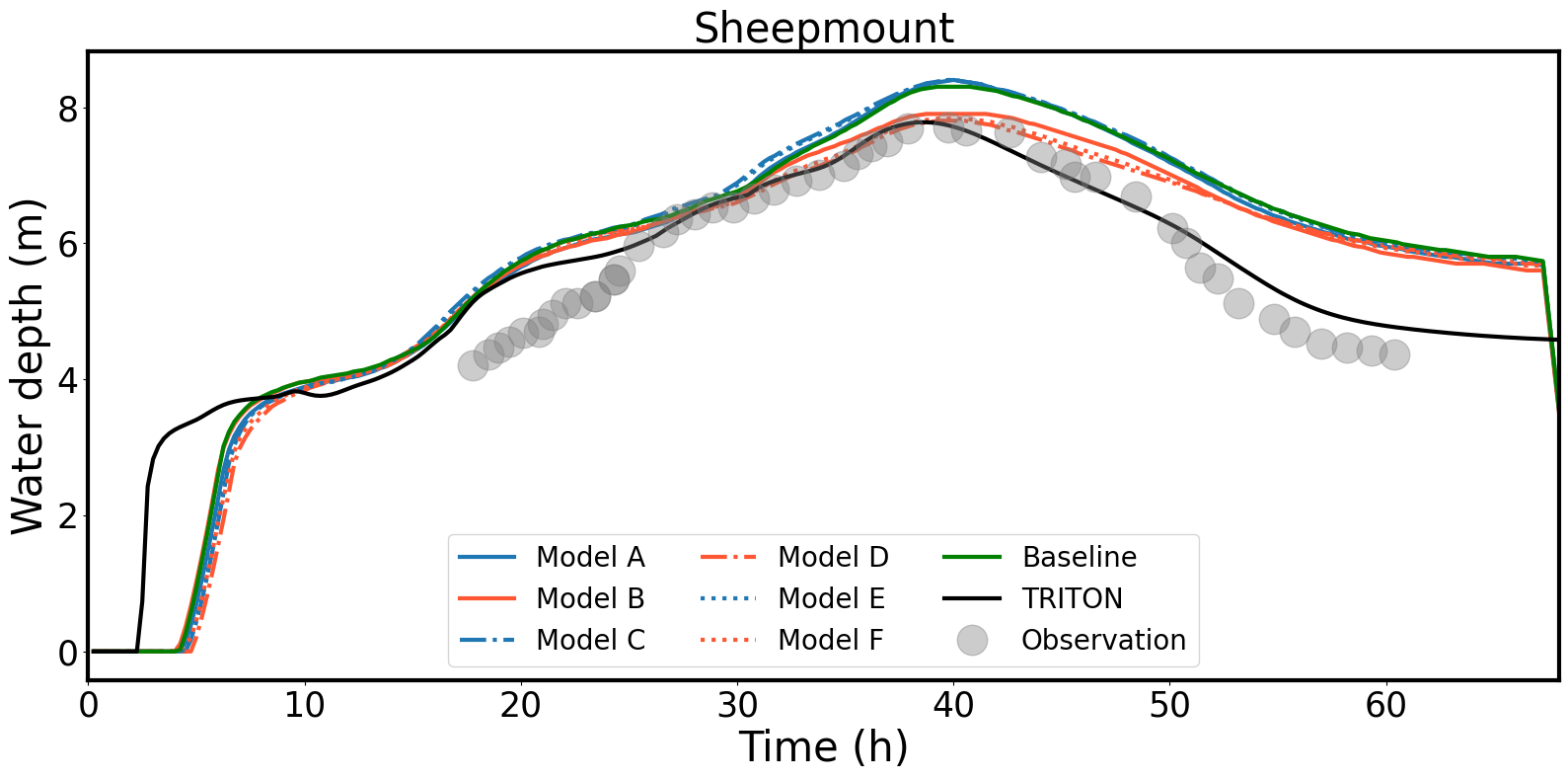}
    \end{subfigure}
    \begin{subfigure}[t]{0.82\textwidth}
    \includegraphics[width=\textwidth]{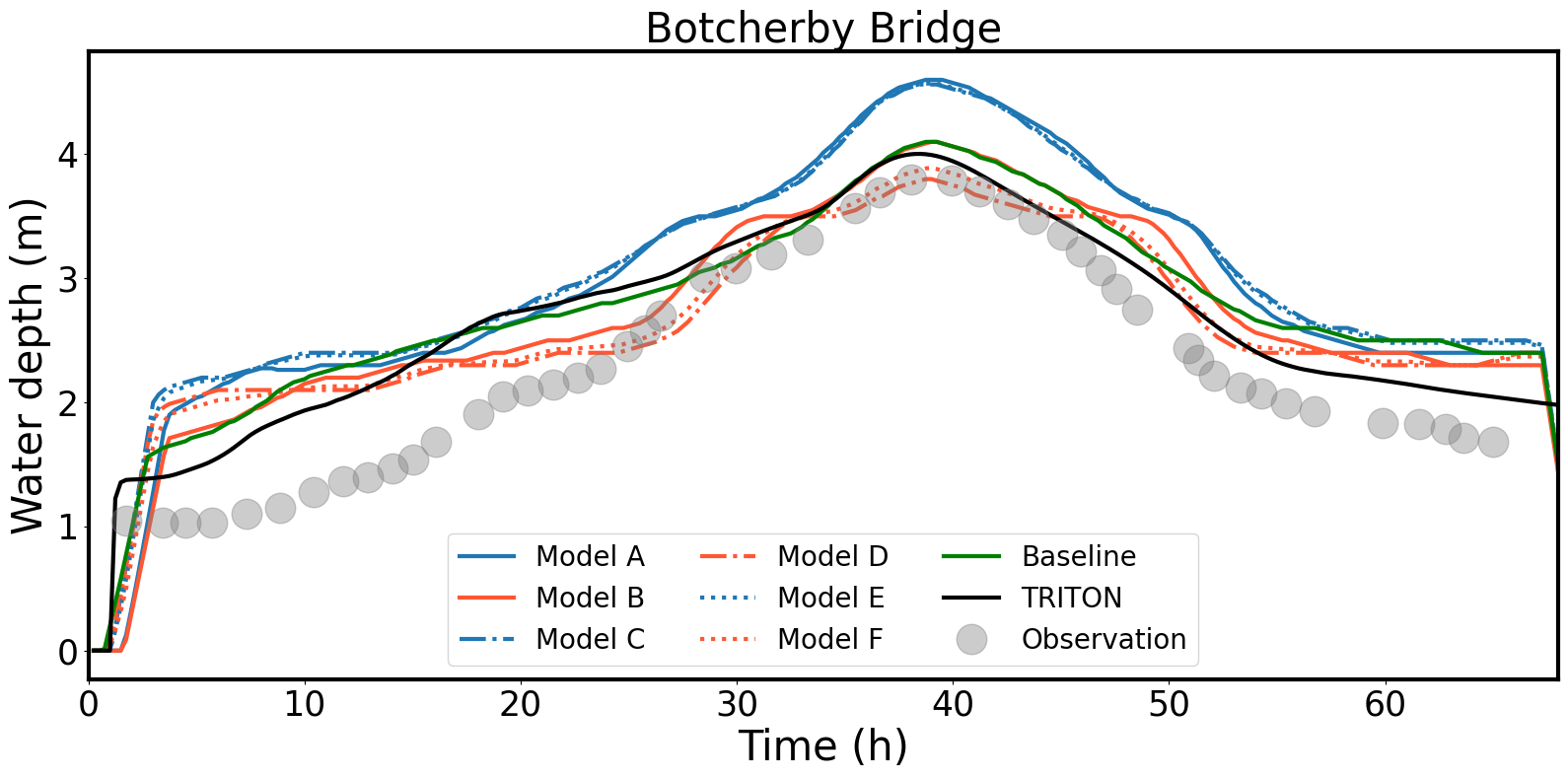}
    \end{subfigure}
    \begin{subfigure}[t]{0.82\textwidth}
    \includegraphics[width=\textwidth]{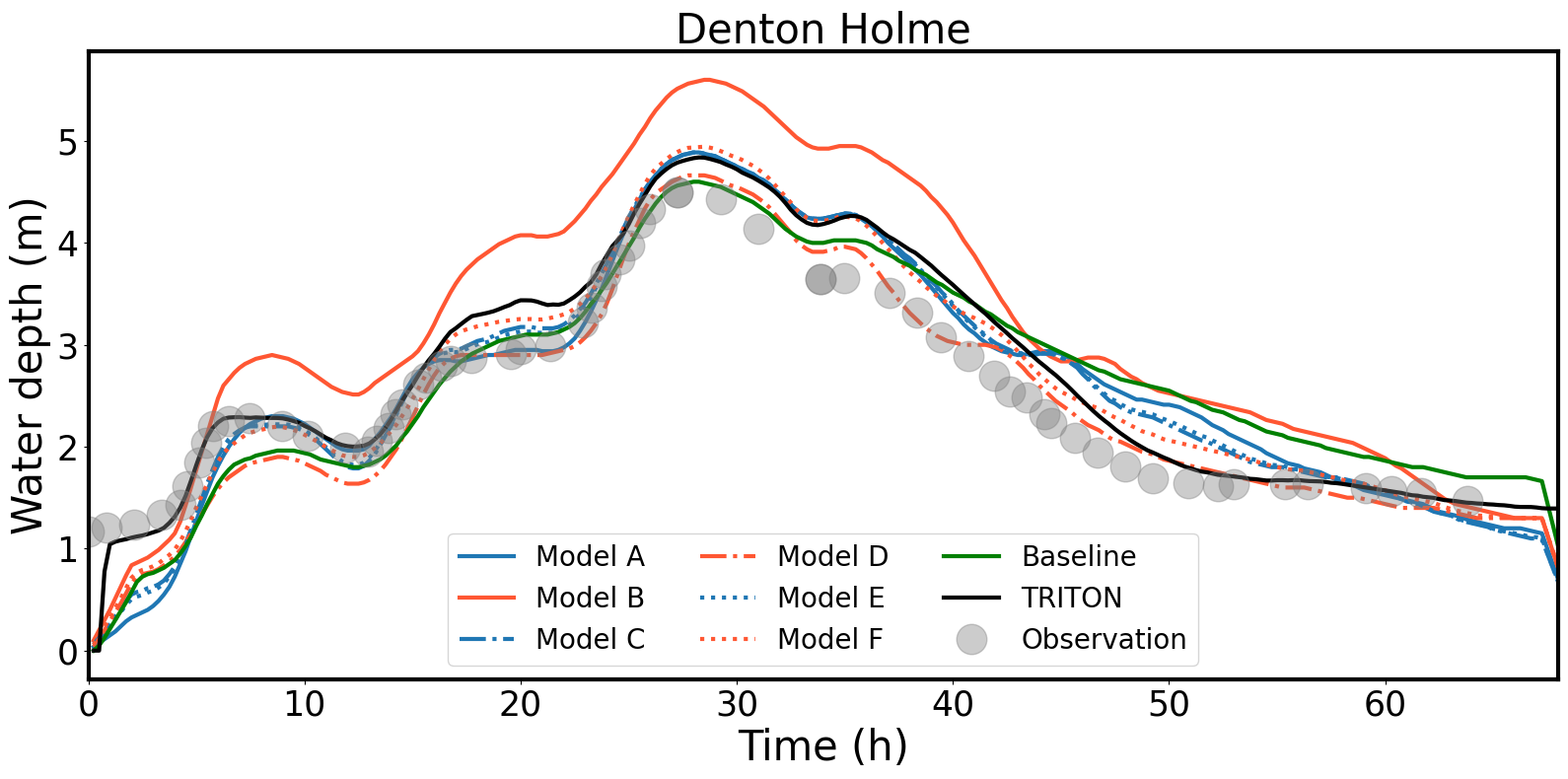}
    \end{subfigure}
    \caption{Numerical prediction of water depth versus time at three sampling points (Sheepmount, Botcherby Bridge and Denton Holme) on regular and distorted grid using Linear and Quadratic ConvFEM models. The details of the models considered are summarised in Table~\ref{tab:numerical_set_up}.}
    \label{fig:DSWE_validation}
\end{figure}

To quantify the improvement seen when using distorted grids, Figure~\ref{fig:DSWE_validation} compares the time series of predicted water depth (68 hours in total) at the three sampling locations (Sheepmount, Botcherby Bridge and Denton Holme) with observations and a solution generated by the TRITON model \cite{triton}. Two \NP (linear and quadratic ConvFEM) models are applied on regular and distorted grids, of which details are summarised in Table~\ref{tab:numerical_set_up}. At Sheepmount (top), all models capture the increase of the water level and the timing of the peak level accurately. The distorted grid simulations slightly reduce the overestimation of the peak seen in the regular grid and linear ConvFEM simulation. The quadratic ConvFEM models provide improved agreement with observations as the water level recedes. At Botcherby Bridge (middle), the differences between simulations become more significant during the peak flow period (from 35 hours to 45 hours). The distorted grid tends to moderate peak water-depth predictions and improves agreement with the observations. In addition, the quadratic discretisation reduces numerical oscillations and more accurately reproduces the shape of the observed water depth profile.

At Denton Holme (bottom), where complex urban topography strongly influences the flow dynamics, the advantages of the distorted grid become more evident. The distorted mesh reduces discrepancies either side of the peak water level and the quadratic distorted grid shows the closest overall agreement with the observations. These improvements are consistent with the mesh analysis presented earlier: the distorted grids redistribute resolution towards dynamically important regions, enabling better representation of local flow features such as rapid water level changes and peak dynamics. In general, the distorted structured grid maintains numerical stability and produces results comparable to or better than the regular grid, particularly when combined with quadratic ConvFEM discretisation.

\subsection{Computational Performance}

The computational cost of NN4PDE simulations scales approximately with the number of finite-element nodes and the number of Jacobi iterations used for solving the free-surface height. The total simulation time can therefore be approximated as proportional to the number of nodes, the number of time steps, and the iterations performed within each time step. This leads to the empirical model
\begin{equation}
\label{computational_speed}
\text{simulation duration} =
(1.3)^{p-1}\,
\mathcal{N_I}\,
\mathcal{T}\,
\mathcal{N}\,
\left(M_s + J\,F_s\right),
\end{equation}
where $p$ is the polynomial order ($p=1$ linear, $p=2$ quadratic), $\mathcal{N_I}$ is the number of nonlinear iterations per time step (typically $\mathcal{N_I}=2$), $\mathcal{T}$ is the number of time steps, $\mathcal{N}=N_x N_y$ is the number of FE nodes, and $J$ is the number of Jacobi iterations used for the free-surface solve. The constants $M_s$ and $F_s$ represent the computational cost per node for the main solver and the Jacobi update respectively, and depend on the GPU architecture. 
These coefficients are 
approximately: NVIDIA A100 GPU $(
M_s = 2.76 \times 10^{-9}\,\text{s}, \;
F_s = 9.2 \times 10^{-11}\,\text{s}
)$; NVIDIA H100 GPU 
$(
M_s = 1.20 \times 10^{-9}\,\text{s}, \;
F_s = 4.0 \times 10^{-11}\,\text{s}
)$;  NVIDIA H200 GPU 
$(
M_s = 1.00 \times 10^{-9}\,\text{s}, \;
F_s = 3.5 \times 10^{-11}\,\text{s}
)$. 
These coefficients were obtained by timing simulations of the Carlisle flooding benchmark using different numbers of Jacobi iterations and solving for the proportionality constants. For the Carlisle case considered here ($\mathcal{T}=306000$, $\mathcal{N}=951\times611=581061$, quadratic elements, $\mathcal{N_I}=2$, $J=1$), the model predicts a runtime of approximately \SI{24}{min} on an A100 GPU, compared with the measured runtime of \SI{21.3}{min}. 

\section{Discussion and Conclusions}
\label{dis-conclusions}
This study presents a convolution-based \NP  framework for solving the shallow water equations on distorted structured grids. By formulating the governing equations in a uniform computational space and incorporating metric terms through the Jacobian transformation, the method preserves conservation while enabling efficient structured-grid computation.

The advection operator is derived in both conservative and non-conservative forms, allowing flexibility depending on stability requirements. A nonlinear Petrov-Galerkin dissipation term provides adaptive stabilisation, particularly important in advection-dominated and distorted-grid regimes. Two approximations for the Laplacian operator were introduced to accelerate pressure and diffusion solves while retaining acceptable accuracy. Validation against the Carlisle 2005 flooding benchmark demonstrates that distorted meshes can achieve accuracy comparable to regular grids while enabling resolution concentration in regions of interest.

The convolutional structure of the discretised operators makes the framework naturally compatible with GPU acceleration and \NP implementations. Furthermore, the static mesh adaptation strategy based on diffusion principles provides a systematic approach to resolution control without sacrificing structured-grid efficiency.

Overall, this work demonstrates that distorted structured grids combined with convolutional operators provide a powerful and scalable methodology for geophysical flow simulation. The framework offers a promising foundation for future extensions to fully coupled atmospheric and oceanic systems, dynamic mesh adaptation, and hybrid physics-informed machine learning solvers.


\section*{Funding sources}
The authors would like to acknowledge the following UKRI grants: \textbf{GP4Streets}, ``Pioneering Climate Adaptation in Urban Environments'' (APP44894 / UKRI 1281); \textbf{AI4Urban-Health}, ``AI Solutions to Urban Health Using a Place-Based Approach'' (APP55547 / UKRI 1241); and AI-Respire, ``AI for personalised respiratory health and pollution (EP/Y018680/1). Support from Imperial-X's Eric and Wendy Schmidt Centre for AI in Science (a Schmidt Futures program) is gratefully acknowledged.


\bibliography{mybibfile}

\begin{thebibliography}{42}
\expandafter\ifx\csname natexlab\endcsname\relax\def\natexlab#1{#1}\fi
\providecommand{\url}[1]{\texttt{#1}}
\providecommand{\href}[2]{#2}
\providecommand{\path}[1]{#1}
\providecommand{\DOIprefix}{doi:}
\providecommand{\ArXivprefix}{arXiv:}
\providecommand{\URLprefix}{URL: }
\providecommand{\Pubmedprefix}{pmid:}
\providecommand{\doi}[1]{\href{http://dx.doi.org/#1}{\path{#1}}}
\providecommand{\Pubmed}[1]{\href{pmid:#1}{\path{#1}}}
\providecommand{\bibinfo}[2]{#2}
\ifx\xfnm\relax \def\xfnm[#1]{\unskip,\space#1}\fi
\bibitem[{Stoker(1957)}]{Stoker1957}
\bibinfo{author}{J.~J. Stoker}, \bibinfo{title}{Water Waves: The Mathematical
  Theory with Applications}, \bibinfo{publisher}{Interscience},
  \bibinfo{year}{1957}.
\bibitem[{Toro(2001)}]{Toro2001}
\bibinfo{author}{E.~F. Toro}, \bibinfo{title}{Shock-Capturing Methods for
  Free-Surface Shallow Flows}, \bibinfo{publisher}{Wiley},
  \bibinfo{year}{2001}.
\bibitem[{Teng et~al.(2017)Teng, Jakeman, Vaze, Croke, Dutta, and
  Kim}]{teng2017flood}
\bibinfo{author}{J.~Teng}, \bibinfo{author}{A.~Jakeman},
  \bibinfo{author}{J.~Vaze}, \bibinfo{author}{B.~F.~W. Croke},
  \bibinfo{author}{D.~Dutta}, \bibinfo{author}{S.~Kim},
\newblock \bibinfo{title}{{Flood inundation modelling: A review of methods}},
\newblock \bibinfo{journal}{Environmental Modelling \& Software}
  \bibinfo{volume}{90} (\bibinfo{year}{2017}) \bibinfo{pages}{201--216}.
\bibitem[{Zanardo and Salinas(2022)}]{Zanardo2022}
\bibinfo{author}{S.~Zanardo}, \bibinfo{author}{J.~L. Salinas},
\newblock \bibinfo{title}{{An Introduction to Flood Modeling for Catastrophe
  Risk Management}},
\newblock \bibinfo{journal}{WIREs Water} \bibinfo{volume}{9}
  (\bibinfo{year}{2022}) \bibinfo{pages}{e1568}.
  \DOIprefix\doi{10.1002/wat2.1568}.
\bibitem[{Spiekermann et~al.(2015)Spiekermann, Kienberger, Norton, Briones, and
  Weichselgartner}]{Spiekermann2015}
\bibinfo{author}{R.~Spiekermann}, \bibinfo{author}{S.~Kienberger},
  \bibinfo{author}{J.~Norton}, \bibinfo{author}{F.~Briones},
  \bibinfo{author}{J.~Weichselgartner},
\newblock \bibinfo{title}{{The Disaster-Knowledge Matrix – Reframing and
  Evaluating the Knowledge Challenges in Disaster Risk Reduction}},
\newblock \bibinfo{journal}{International Journal of Disaster Risk Reduction}
  \bibinfo{volume}{13} (\bibinfo{year}{2015}) \bibinfo{pages}{96--108}.
  \DOIprefix\doi{10.1016/j.ijdrr.2015.05.002}.
\bibitem[{Henonin et~al.(2013)Henonin, Russo, Mark, and
  Gourbesville}]{Henonin2013}
\bibinfo{author}{J.~Henonin}, \bibinfo{author}{B.~Russo},
  \bibinfo{author}{O.~Mark}, \bibinfo{author}{P.~Gourbesville},
\newblock \bibinfo{title}{{Real-Time Urban Flood Forecasting and Modelling: A
  State of the Art}},
\newblock \bibinfo{journal}{Journal of Hydroinformatics} \bibinfo{volume}{15}
  (\bibinfo{year}{2013}) \bibinfo{pages}{717--736}.
  \DOIprefix\doi{10.2166/hydro.2013.132}.
\bibitem[{Casulli(1990)}]{CASULLI199056}
\bibinfo{author}{V.~Casulli},
\newblock \bibinfo{title}{Semi-implicit finite difference methods for the
  two-dimensional shallow water equations},
\newblock \bibinfo{journal}{Journal of Computational Physics}
  \bibinfo{volume}{86} (\bibinfo{year}{1990}) \bibinfo{pages}{56--74}.
  \DOIprefix\doi{10.1016/0021-9991(90)90091-E}.
\bibitem[{Lundgren and Mattsson(2020)}]{lundgren2020efficient}
\bibinfo{author}{L.~Lundgren}, \bibinfo{author}{K.~Mattsson},
\newblock \bibinfo{title}{An efficient finite difference method for the shallow
  water equations},
\newblock \bibinfo{journal}{Journal of Computational Physics}
  \bibinfo{volume}{422} (\bibinfo{year}{2020}) \bibinfo{pages}{109784}.
  \DOIprefix\doi{10.1016/j.jcp.2020.109784}.
\bibitem[{LeVeque(2002)}]{LeVeque2002}
\bibinfo{author}{R.~LeVeque}, \bibinfo{title}{Finite Volume Methods for
  Hyperbolic Problems}, \bibinfo{publisher}{Cambridge University Press},
  \bibinfo{year}{2002}.
\bibitem[{Donea and Huerta(2003)}]{donea2003finite}
\bibinfo{author}{J.~Donea}, \bibinfo{author}{A.~Huerta}, \bibinfo{title}{Finite
  Element Methods for Flow Problems}, \bibinfo{publisher}{John Wiley \& Sons},
  \bibinfo{address}{Chichester}, \bibinfo{year}{2003}.
  \DOIprefix\doi{10.1002/0470013826}.
\bibitem[{Cockburn and Shu(2000)}]{Cockburn2000}
\bibinfo{author}{B.~Cockburn}, \bibinfo{author}{C.-W. Shu},
\newblock \bibinfo{title}{{The development of discontinuous Galerkin methods}},
\newblock \bibinfo{journal}{SIAM Review} \bibinfo{volume}{42}
  (\bibinfo{year}{2000}) \bibinfo{pages}{463--487}.
\bibitem[{Lee(2021)}]{lee2021discontinuous}
\bibinfo{author}{H.~Lee},
\newblock \bibinfo{title}{{Discontinuous Galerkin discretization of shallow
  water equations in implicit primal formulations for turbulent stresses}},
\newblock \bibinfo{journal}{Journal of Mechanical Science and Technology}
  \bibinfo{volume}{35} (\bibinfo{year}{2021}) \bibinfo{pages}{2471--2479}.
  \DOIprefix\doi{10.1007/s12206-021-0520-7}.
\bibitem[{LeFloch and Thanh(2011)}]{lefloch2011godunov}
\bibinfo{author}{P.~G. LeFloch}, \bibinfo{author}{M.~D. Thanh},
\newblock \bibinfo{title}{{A Godunov-type method for the shallow water
  equations with discontinuous topography in the resonant regime}},
\newblock \bibinfo{journal}{Journal of Computational Physics}
  \bibinfo{volume}{230} (\bibinfo{year}{2011}) \bibinfo{pages}{7631--7660}.
  \DOIprefix\doi{10.1016/j.jcp.2011.06.014}.
\bibitem[{Bernetti et~al.(2008)Bernetti, Titarev, and Toro}]{bernetti2008exact}
\bibinfo{author}{R.~Bernetti}, \bibinfo{author}{V.~A. Titarev},
  \bibinfo{author}{E.~F. Toro},
\newblock \bibinfo{title}{{Exact solution of the Riemann problem for the
  shallow water equations with discontinuous bottom geometry}},
\newblock \bibinfo{journal}{Journal of Computational Physics}
  \bibinfo{volume}{227} (\bibinfo{year}{2008}) \bibinfo{pages}{3212--3243}.
  \DOIprefix\doi{10.1016/j.jcp.2007.10.012}.
\bibitem[{Shaw et~al.(2021)Shaw, Kesserwani, Neal, Bates, and
  Sharifian}]{shaw2021lisflood}
\bibinfo{author}{J.~Shaw}, \bibinfo{author}{G.~Kesserwani},
  \bibinfo{author}{J.~Neal}, \bibinfo{author}{P.~Bates}, \bibinfo{author}{M.~K.
  Sharifian},
\newblock \bibinfo{title}{{LISFLOOD-FP 8.0: the new discontinuous Galerkin
  shallow-water solver for multi-core CPUs and GPUs}},
\newblock \bibinfo{journal}{Geoscientific Model Development}
  \bibinfo{volume}{14} (\bibinfo{year}{2021}) \bibinfo{pages}{3577--3602}.
  \DOIprefix\doi{10.5194/gmd-14-3577-2021}.
\bibitem[{Bunya et~al.(2009)Bunya, Kubatko, Westerink, and
  Dawson}]{bunya2009wetting}
\bibinfo{author}{S.~Bunya}, \bibinfo{author}{E.~J. Kubatko},
  \bibinfo{author}{J.~J. Westerink}, \bibinfo{author}{C.~Dawson},
\newblock \bibinfo{title}{{A wetting and drying treatment for the Runge--Kutta
  discontinuous Galerkin solution to the shallow water equations}},
\newblock \bibinfo{journal}{Computer Methods in Applied Mechanics and
  Engineering} \bibinfo{volume}{198} (\bibinfo{year}{2009})
  \bibinfo{pages}{1548--1562}. \DOIprefix\doi{10.1016/j.cma.2009.01.002}.
\bibitem[{Xia and Liang(2016)}]{xia2016gpu}
\bibinfo{author}{X.~Xia}, \bibinfo{author}{Q.~Liang},
\newblock \bibinfo{title}{{A GPU-accelerated Smoothed Particle Hydrodynamics
  (SPH) model for the shallow water equations}},
\newblock \bibinfo{journal}{Environmental Modelling \& Software}
  \bibinfo{volume}{75} (\bibinfo{year}{2016}) \bibinfo{pages}{28--43}.
  \DOIprefix\doi{10.1016/j.envsoft.2015.10.010}.
\bibitem[{Smith and Liang(2013)}]{smith2013towards}
\bibinfo{author}{L.~S. Smith}, \bibinfo{author}{Q.~Liang},
\newblock \bibinfo{title}{{Towards a generalised GPU/CPU shallow-flow modelling
  tool}},
\newblock \bibinfo{journal}{Computers \& Fluids} \bibinfo{volume}{88}
  (\bibinfo{year}{2013}) \bibinfo{pages}{334--343}.
  \DOIprefix\doi{10.1016/j.compfluid.2013.09.018}.
\bibitem[{Lacasta et~al.(2015)Lacasta, Morales-Hern{\'a}ndez, Murillo, and
  Garc{\'i}a-Navarro}]{lacasta2015gpu}
\bibinfo{author}{A.~Lacasta}, \bibinfo{author}{M.~Morales-Hern{\'a}ndez},
  \bibinfo{author}{J.~Murillo}, \bibinfo{author}{P.~Garc{\'i}a-Navarro},
\newblock \bibinfo{title}{{GPU implementation of the 2D shallow water equations
  for the simulation of rainfall/runoff events}},
\newblock \bibinfo{journal}{Environmental Earth Sciences} \bibinfo{volume}{74}
  (\bibinfo{year}{2015}) \bibinfo{pages}{7295--7305}.
  \DOIprefix\doi{10.1007/s12665-015-4705-7}.
\bibitem[{Berger and Colella(1989)}]{Berger1989}
\bibinfo{author}{M.~Berger}, \bibinfo{author}{P.~Colella},
\newblock \bibinfo{title}{Local adaptive mesh refinement for shock
  hydrodynamics},
\newblock \bibinfo{journal}{Journal of Computational Physics}
  \bibinfo{volume}{82} (\bibinfo{year}{1989}) \bibinfo{pages}{64--84}.
\bibitem[{Pain et~al.(2001)Pain, Umpleby, De~Oliveira, and Goddard}]{Pain2001}
\bibinfo{author}{C.~C. Pain}, \bibinfo{author}{A.~P. Umpleby},
  \bibinfo{author}{C.~R.~E. De~Oliveira}, \bibinfo{author}{A.~J.~H. Goddard},
\newblock \bibinfo{title}{{Adaptive unstructured mesh methods for CFD}},
\newblock \bibinfo{journal}{International Journal for Numerical Methods in
  Fluids} \bibinfo{volume}{34} (\bibinfo{year}{2001})
  \bibinfo{pages}{327--355}.
\bibitem[{Estrem et~al.(2020)Estrem, Moxey, Sherwin, and Peir{'o}}]{Estrem2020}
\bibinfo{author}{K.~Estrem}, \bibinfo{author}{D.~Moxey}, \bibinfo{author}{S.~J.
  Sherwin}, \bibinfo{author}{J.~Peir{'o}},
\newblock \bibinfo{title}{Efficient anisotropic mesh adaptivity for fluid
  dynamics simulations},
\newblock \bibinfo{journal}{Computer Methods in Applied Mechanics and
  Engineering} \bibinfo{volume}{361} (\bibinfo{year}{2020})
  \bibinfo{pages}{112761}. \DOIprefix\doi{10.1016/j.cma.2019.112761}.
\bibitem[{Huang and Russell(2011)}]{Huang2011}
\bibinfo{author}{W.~Huang}, \bibinfo{author}{R.~D. Russell},
\newblock \bibinfo{title}{Adaptive moving mesh methods},
\newblock \bibinfo{journal}{Applied Numerical Mathematics} \bibinfo{volume}{61}
  (\bibinfo{year}{2011}) \bibinfo{pages}{873--889}.
\bibitem[{Alauzet and Loseille(2010)}]{Alauzet2010}
\bibinfo{author}{F.~Alauzet}, \bibinfo{author}{A.~Loseille},
\newblock \bibinfo{title}{Parallel anisotropic mesh adaptation},
\newblock \bibinfo{journal}{Computers \& Fluids} \bibinfo{volume}{39}
  (\bibinfo{year}{2010}) \bibinfo{pages}{1760--1775}.
\bibitem[{Loseille and Alauzet(2011)}]{Loseille2011}
\bibinfo{author}{A.~Loseille}, \bibinfo{author}{F.~Alauzet},
\newblock \bibinfo{title}{{Anisotropic mesh adaptation for CFD computations}},
\newblock \bibinfo{journal}{Journal of Computational Physics}
  \bibinfo{volume}{230} (\bibinfo{year}{2011}) \bibinfo{pages}{1621--1645}.
\bibitem[{Kesserwani and Sharifian(2023)}]{Kesserwani2023}
\bibinfo{author}{G.~Kesserwani}, \bibinfo{author}{M.~K. Sharifian},
\newblock \bibinfo{title}{{(Multi)wavelet-based Godunov-type simulators of
  flood inundation: Static versus dynamic adaptivity}},
\newblock \bibinfo{journal}{Advances in Water Resources} \bibinfo{volume}{171}
  (\bibinfo{year}{2023}) \bibinfo{pages}{104357}.
  \DOIprefix\doi{10.1016/j.advwatres.2022.104357}.
\bibitem[{Thompson et~al.(1985)Thompson, Warsi, and Mastin}]{Thompson1985}
\bibinfo{author}{J.~F. Thompson}, \bibinfo{author}{Z.~U.~A. Warsi},
  \bibinfo{author}{C.~Mastin}, \bibinfo{title}{Numerical Grid Generation:
  Foundations and Applications}, \bibinfo{publisher}{North-Holland},
  \bibinfo{year}{1985}.
\bibitem[{Thompson et~al.(1998)Thompson, Soni, and Weatherill}]{Thompson1998}
\bibinfo{author}{J.~F. Thompson}, \bibinfo{author}{B.~K. Soni},
  \bibinfo{author}{N.~P. Weatherill}, \bibinfo{title}{Handbook of Grid
  Generation}, \bibinfo{publisher}{CRC Press}, \bibinfo{year}{1998}.
\bibitem[{Staniforth and Thuburn(2012)}]{Staniforth2012}
\bibinfo{author}{A.~Staniforth}, \bibinfo{author}{J.~Thuburn},
\newblock \bibinfo{title}{Horizontal grids for global weather and climate
  prediction models},
\newblock \bibinfo{journal}{Quarterly Journal of the Royal Meteorological
  Society} \bibinfo{volume}{138} (\bibinfo{year}{2012}) \bibinfo{pages}{1--26}.
\bibitem[{Miura and Skamarock(2013)}]{Miura2013}
\bibinfo{author}{H.~Miura}, \bibinfo{author}{W.~C. Skamarock},
\newblock \bibinfo{title}{An upwind-biased transport scheme using a quadratic
  reconstruction on spherical icosahedral grids},
\newblock \bibinfo{journal}{Monthly Weather Review} \bibinfo{volume}{141}
  (\bibinfo{year}{2013}) \bibinfo{pages}{832--847}.
  \DOIprefix\doi{10.1175/MWR-D-12-00048.1}.
\bibitem[{Phillips et~al.(2023)Phillips, Heaney, Chen, Buchan, and
  Pain}]{Phillips2023diffusion}
\bibinfo{author}{T.~R.~F. Phillips}, \bibinfo{author}{C.~E. Heaney},
  \bibinfo{author}{B.~Chen}, \bibinfo{author}{A.~G. Buchan},
  \bibinfo{author}{C.~C. Pain},
\newblock \bibinfo{title}{Solving the discretised neutron diffusion equations
  using neural networks},
\newblock \bibinfo{journal}{International Journal for Numerical Methods in
  Engineering} \bibinfo{volume}{124} (\bibinfo{year}{2023})
  \bibinfo{pages}{4659--4686}. \DOIprefix\doi{10.1002/nme.7321}.
\bibitem[{Chen et~al.(2026)Chen, Heaney, and Pain}]{Chen2026}
\bibinfo{author}{B.~Chen}, \bibinfo{author}{C.~E. Heaney},
  \bibinfo{author}{C.~C. Pain},
\newblock \bibinfo{title}{{Neural Physics: Using AI Libraries to Develop
  Physics-Based Solvers for Incompressible CFD}},
\newblock \bibinfo{journal}{Computers \& Fluids} \bibinfo{volume}{308}
  (\bibinfo{year}{2026}) \bibinfo{pages}{106981}.
  \DOIprefix\doi{10.1016/j.compfluid.2026.106981}.
\bibitem[{Chen et~al.(2024)Chen, Heaney, Gomes, Matar, and Pain}]{Chen2024CMA}
\bibinfo{author}{B.~Chen}, \bibinfo{author}{C.~E. Heaney},
  \bibinfo{author}{J.~L. M.~A. Gomes}, \bibinfo{author}{O.~Matar},
  \bibinfo{author}{C.~C. Pain},
\newblock \bibinfo{title}{Solving the discretised multiphase flow equations
  using machine learning libraries},
\newblock \bibinfo{journal}{Computer Methods in Applied Mechanics and
  Engineering} \bibinfo{volume}{426} (\bibinfo{year}{2024})
  \bibinfo{pages}{116974}. \DOIprefix\doi{10.1016/j.cma.2024.116974}.
\bibitem[{Chen et~al.(2025)Chen, Nadimy, Heaney, Sharifian, Estrem, Nicotina,
  Hilberts, and Pain}]{Chen2025}
\bibinfo{author}{B.~Chen}, \bibinfo{author}{A.~Nadimy}, \bibinfo{author}{C.~E.
  Heaney}, \bibinfo{author}{M.~K. Sharifian}, \bibinfo{author}{L.~V. Estrem},
  \bibinfo{author}{L.~Nicotina}, \bibinfo{author}{A.~Hilberts},
  \bibinfo{author}{C.~C. Pain},
\newblock \bibinfo{title}{{Solving the discretised shallow water equations
  using neural networks}},
\newblock \bibinfo{journal}{Advances in Water Resources} \bibinfo{volume}{197}
  (\bibinfo{year}{2025}) \bibinfo{pages}{104903}.
  \DOIprefix\doi{10.1016/j.advwatres.2025.104903}.
\bibitem[{Nadimy et~al.(2026)Nadimy, Chen, Chen, Heaney, and Pain}]{Nadimy2026}
\bibinfo{author}{A.~Nadimy}, \bibinfo{author}{B.~Chen},
  \bibinfo{author}{Z.~Chen}, \bibinfo{author}{C.~E. Heaney},
  \bibinfo{author}{C.~C. Pain},
\newblock \bibinfo{title}{Solving the discretised shallow water equations using
  non-uniform grids and machine-learning libraries},
\newblock \bibinfo{journal}{Environmental Modelling \& Software}
  \bibinfo{volume}{196} (\bibinfo{year}{2026}) \bibinfo{pages}{106752}.
  \DOIprefix\doi{10.1016/j.envsoft.2025.106752}.
\bibitem[{Huang et~al.(2026)Huang, Heaney, and van Reeuwijk}]{Huang2026}
\bibinfo{author}{R.~Huang}, \bibinfo{author}{C.~E. Heaney},
  \bibinfo{author}{M.~van Reeuwijk},
\newblock \bibinfo{title}{{Parameter estimation for land-surface models using
  Neural Physics}},
\newblock \bibinfo{journal}{Geoscientific Model Development}
  \bibinfo{volume}{19} (\bibinfo{year}{2026}) \bibinfo{pages}{8801--8814}.
  \DOIprefix\doi{10.5194/gmd-19-8801-2026}.
\bibitem[{Jiao et~al.(2026)Jiao, Zhang, Xue, Chen, Yang, Ba, See, Kang, Heaney,
  Pain et~al.}]{jiao2026np4buoyancy}
\bibinfo{author}{Y.~Jiao}, \bibinfo{author}{H.~Zhang},
  \bibinfo{author}{T.~Xue}, \bibinfo{author}{B.~Chen},
  \bibinfo{author}{J.~Yang}, \bibinfo{author}{T.~Ba}, \bibinfo{author}{S.~See},
  \bibinfo{author}{C.~W. Kang}, \bibinfo{author}{C.~E. Heaney},
  \bibinfo{author}{C.~C. Pain}, et~al.,
\newblock \bibinfo{title}{{NP4Buoyancy: An End-to-End Differentiable
  Neural-Physics Framework for Buoyancy-Driven Flows}},
\newblock \bibinfo{journal}{Journal of Computational Physics}
  (\bibinfo{year}{2026}) \bibinfo{pages}{115179}.
  \DOIprefix\doi{10.1016/j.jcp.2026.115179}.
\bibitem[{Phillips et~al.(2023)Phillips, Heaney, Chen, Buchan, and
  Pain}]{Phillips2024}
\bibinfo{author}{T.~R.~F. Phillips}, \bibinfo{author}{C.~E. Heaney},
  \bibinfo{author}{B.~Chen}, \bibinfo{author}{A.~G. Buchan},
  \bibinfo{author}{C.~C. Pain},
\newblock \bibinfo{title}{{Solving the Discretised Boltzmann Transport
  Equations Using Neural Networks: Applications in Neutron Transport}},
\newblock \bibinfo{journal}{arXiv preprint}  (\bibinfo{year}{2023})
  \bibinfo{pages}{2301.09991}. \DOIprefix\doi{10.48550/arXiv.2301.09991}.
\bibitem[{Vanzo et~al.(2016)Vanzo, Siviglia, and Toro}]{VANZO20161}
\bibinfo{author}{D.~Vanzo}, \bibinfo{author}{A.~Siviglia},
  \bibinfo{author}{E.~F. Toro},
\newblock \bibinfo{title}{{Pollutant Transport by Shallow Water Equations on
  Unstructured Meshes: Hyperbolization of the Model and Numerical Solution via
  a Novel Flux Splitting Scheme}},
\newblock \bibinfo{journal}{Journal of Computational Physics}
  \bibinfo{volume}{321} (\bibinfo{year}{2016}) \bibinfo{pages}{1--20}.
  \DOIprefix\doi{10.1016/j.jcp.2016.05.023}.
\bibitem[{Pavlidis et~al.(2016)Pavlidis, Gomes, Xie, Percival, Pain, and
  Matar}]{Pavlidis2016}
\bibinfo{author}{D.~Pavlidis}, \bibinfo{author}{J.~L. M.~A. Gomes},
  \bibinfo{author}{Z.~Xie}, \bibinfo{author}{J.~R. Percival},
  \bibinfo{author}{C.~C. Pain}, \bibinfo{author}{O.~K. Matar},
\newblock \bibinfo{title}{Compressive advection and multi-component methods for
  interface-capturing},
\newblock \bibinfo{journal}{International Journal for Numerical Methods in
  Fluids} \bibinfo{volume}{80} (\bibinfo{year}{2016})
  \bibinfo{pages}{256--282}. \DOIprefix\doi{10.1002/fld.4078}.
\bibitem[{Ronneberger et~al.(2015)Ronneberger, Fischer, and
  Brox}]{Ronneberger2015}
\bibinfo{author}{O.~Ronneberger}, \bibinfo{author}{P.~Fischer},
  \bibinfo{author}{T.~Brox},
\newblock \bibinfo{title}{{U-Net: Convolutional Networks for Biomedical Image
  Segmentation}},
\newblock in: \bibinfo{booktitle}{Medical Image Computing and Computer-Assisted
  Intervention (MICCAI)}, volume \bibinfo{volume}{9351} of
  \textit{\bibinfo{series}{LNCS}}, \bibinfo{publisher}{Springer},
  \bibinfo{year}{2015}, pp. \bibinfo{pages}{234--241}.
  \DOIprefix\doi{10.48550/arXiv.1505.04597}.
\bibitem[{tri(2022)}]{triton}
\bibinfo{title}{{TRITON: Two-dimensional Runoff Inundation Toolkit for
  Operational Needs}}, \bibinfo{organization}{Oak Ridge National Lab.},
  \bibinfo{year}{2022}. \bibinfo{note}{Accessed 01-December-2023}.

\end{thebibliography}

\end{document}